\documentclass[acmsmall]{acmart}
\usepackage{wasysym}
\usepackage{pdflscape}
\usepackage{hyperref}
\usepackage{multirow}
\usepackage{array}
\usepackage{cancel}
\usepackage{tikz}

\def\checkmark{\tikz\fill[scale=0.4](0,.35) -- (.25,0) -- (1,.7) -- (.25,.15) -- cycle;}
\def\BibTeX{{\rm B\kern-.05em{\sc i\kern-.025em b}\kern-.08emT\kern-.1667em\lower.7ex\hbox{E}\kern-.125emX}}

\acmJournal{CSUR}

\definecolor{Changecolor}{HTML}{00006B}

\begin{document}

\title{Smart Home Personal Assistants: A Security and Privacy Review}

\author{Jide~S.~Edu}
\affiliation{%
\institution{King's College London}
  \streetaddress{Department of Informatics, Faculty of Natural and Mathematical Science, Strand campus}
  \city{London}
   \country{UK}}
\email{jide.edu@kcl.ac.uk}

\author{Jose~M.~Such}
\affiliation{%
\institution{King's College London}
  \streetaddress{Department of Informatics, Faculty of Natural and Mathematical Science, Strand campus}
  \city{London}
   \country{UK}}
\email{jose.such@kcl.ac.uk}

\author{Guillermo~Suarez-Tangil}
\affiliation{%
\institution{King's College London}
  \streetaddress{Department of Informatics, Faculty of Natural and Mathematical Science, Strand campus}
  \city{London}
   \country{UK}}
\email{guillermo.suarez-tangil@kcl.ac.uk}


\renewcommand{\shortauthors}{J. Edu et al.}

\newcommand{\fixme}[1]{\textcolor{red}{\textbf{FIXME:} #1}}
\newcommand{\todo}[1]{\textcolor{red}{\textbf{TODO:} #1}}

 \newcommand{\change}[1]{ \textcolor{Changecolor}{{#1}}}
\newcommand{\review}[1]{{\bf \textcolor{Changecolor}{#1}}}

\newif\ifcomment
\commentfalse
\commenttrue
\ifcomment
\newcommand{\guillermo}[1]{{\bf \textcolor{blue}{GST: #1}}}
\newcommand{\jose}[1]{{\bf \textcolor{red}{JMS: #1}}}
\newcommand{\jide}[1]{{\bf \textcolor{purple}{JSE: #1}}}
\else
\newcommand{\guillermo}[1]{}
\newcommand{\jose}[1]{}
\newcommand{\jide}[1]{}
\fi

\begin{abstract} 
Smart Home Personal Assistants (SPA) are an emerging innovation that is changing the means by which home users interact with technology. However, several elements expose these systems to various risks: i) the open nature of the voice channel they use, ii) the complexity of their architecture, iii) the AI features they rely on, and iv) their use of a wide range of underlying technologies. This paper presents an in-depth review of SPA's security and privacy issues, categorizing the most important attack vectors and their countermeasures. Based on this, we discuss open research challenges that can help steer the community to tackle and address current security and privacy issues in SPA. One of our key findings is that even though the attack surface of SPA is conspicuously broad and there has been a significant amount of recent research efforts in this area, research has so far focused on a small part of the attack surface, particularly on issues related to the interaction between the user and the SPA devices. 
To the best of our knowledge, this is the first article to conduct such a comprehensive review and characterization of the security and privacy issues and countermeasures of SPA.
\end{abstract} 



\maketitle

\section{Introduction}   
Human-computer interaction (HCI) has traditionally been conducted in the form of different types of peripheral devices such as the keyboard, mouse, and most recently tactile screens. This has been so because computing devices could not decode the meaning of our word, let alone understand our intent. Over the last few years, however, the paradigm has shifted, as we witnessed the rapid development of voice technology in many computing applications. Since voice is one of the most effective and expressive communication tools, voice technology is changing the way in which users interact with devices and the manner they consume services. Currently, the most significant innovations that use voice technology are Smart Home Personal Assistants (SPA). SPA are intelligent assistants that take instructions from users, process them, and perform the corresponding tasks. They offer hands-free and eye-free operations, allowing users to perform diverse activities using voice commands while concentrating elsewhere on other tasks. Besides offering users the benefit of a quick interaction --- humans speak faster than they type~\cite{ruan_wobbrock_liou_ng_landay_2018}, using voice for HCI can be considered more natural~\cite{Kamm_Candace_1995} when compared to other interfaces like keyboard, and mouse. Not to mention the stronger social presence offered to users when they hear synthesized speeches very much like their own as responses from this technology~\cite{doi:10.1207/S1532785XMEP0701}.

SPA are rapidly becoming common features in homes and are increasingly becoming integrated with other smart devices~\cite{alexa-Toni_Reid}. It is believed that 10\% of the world consumers own SPA devices~\cite{OVUM}. According to a recent survey by Voicebot, over 50 million Alexa Echo devices have been sold to date in the US alone~\cite{alexa-Reality}.
There are a number of features that contribute to the popularity of SPA. 
SPA are quite different from early voice-activated technologies that could only work with small inbuilt commands and responses. 
Instead, SPA use Internet services and benefits from recent advances in Natural Language Processing (NLP), which allow them to handle a wide range of commands and questions. They enable a playful interaction, making their use more engaging~\cite{Luger:2016:LRB:2858036.2858288}. 
They are assigned a name and a gender, which encourages users to personify them and therefore interact with them in a human-like manner~\cite{doi:10.1111/j.1559-1816.1999.tb00142.x}.
They are used to maintain shopping and to-dos lists, purchase goods, and food, play audio-books, play games, stream music, radio and news, set timers, alarms and reminders~\cite{alexa-theverge}, get recipe ideas, control large appliances~\cite{alexa-Cnet}, send messages, make calls~\cite{alexa-CNN} and many more depending on their usage context~\cite{alexa-guardian, hoy_2018}. With the continuous proliferation and the rapid growth of SPA, we are now approaching an era when SPA will not only be maneuvering our devices at home but also replacing them in many cases. For instance, many SPA are now able to make phone calls, which positions them as a communicating device, and a likely alternative to landlines phones in the future, and some SPA are also equipped with display interface for watching videos/movies and smart home cameras directly in the SPA devices~\cite{alexa-display}. 
 
As these devices become increasingly popular~\cite{OVUM}, the most sought-after features expose SPA to various risks.  
Some of those features are the open nature of the voice channel they use, the complexity of their architecture, the AI features they rely on, and the use of a wide range of different technologies.
It is paramount to understand the underlying risks behind their use and fathom how to mitigate them. While most of these devices have incorporated some security and privacy mechanisms in their design, there is still a significant number of security and privacy challenges that need to be addressed. This is all the more important because SPA carry out distinct roles and perform various functions in single and multi-user environments, particularly in an intimate domain like homes. Since users co-locate with this technology, it also has an impact on the changes in their neighboring environment~\cite{Purington:2017:AMN:3027063.3053246}. In fact, there have already been reported security and privacy incidents in the media involving SPA, such as the case of an Amazon Alexa recording an intimate conversation and sending it to an arbitrary contact~\cite{alexa-guardian}. 
Users are concerned about these devices' security and privacy ~\cite{Fruchter:2018:CAT:3170427.3188448,Easwara_Moorthy_2015}.
In the absence of better technical security and privacy controls, users are implementing workarounds like turning off the SPA when they are not using it~\cite{Noura_Abdi_2019,Lau:2018:AYL:3290265.3274371}. 
Unfortunately, several mitigating techniques proposed in various studies fall short in addressing these risks. 
For instance, authors in~\cite{Lei_Xinyu} propose a presence-based access control system that does not support an extensive set of use cases. 
Furthermore, other solutions, such as the one in~\cite{feng_fawaz_shin_2017}, affect the usability of the SPA. 
  
Despite the fast-growing research on SPA's security and privacy issues, the literature lacks a detailed characterization of these issues. This paper offers the first comprehensive review of existing security and privacy attacks and countermeasures in smart home personal assistants and categorizes them. For this, we first provide an overview and background of the architectural elements of SPA, which is vital to understand both potential weaknesses and countermeasures. In addition, and based on our analysis and categorization of risks, attacks, and countermeasures, this paper presents a roadmap of future research directions in this area. We found out that while the attack surface of SPA is distinctly broad, the research community has focused only on a small part of it. In particular, recent works have focused mostly on issues related to the direct interaction between a user and their SPA. While those problems are indeed very important and further research is needed for effective countermeasures, we also found that research is needed to address other issues related to authorization, speech recognition, profiling, and the technologies integrated with SPA (e.g., the cloud, third-party skills, and other smart devices).

\subsection{Research Questions}
We focus on the following main research questions: 
\begin{itemize}
    \item RQ1---What are the main security and privacy issues behind the use of SPA?
    \item RQ2---What are the features that characterize the known attacks to SPA?
    \item RQ3---What are the main limitations of the existing countermeasures, and how can they be improved?
    \item RQ4---What are the main open challenges to address the security and privacy of SPA?
\end{itemize}

\subsection{Research Method}
We used a systematic literature review (SLR) approach~\cite{hannay_sjoberg_dyba_2007, kitchenham_pearl} to assess existing literature on the security and privacy of SPA. The primary search process involved searching for keywords related to the study (smart home personal assistants, voice assistants, privacy, security) through databases like  ACM Digital Library, Web of Science, IEEE Xplore Digital Library, and ScienceDirect. The secondary search process consisted of searching publications manually in the relevant research area for completeness. Regarding the inclusion and exclusion criteria for the papers we found through the search process above, we included in this review papers that describe research on SPA or research that is of direct relevance or application to SPA. The papers are reviewed with respect to their techniques, years, criteria, metrics, and results. We exclude position papers or short papers that do not describe any results.

\subsection{Review Structure}

The rest of this article is structured as follows: Section~\ref{sec:background} offers an introduction to SPA, their architecture. In Section~\ref{sec:Issues}, we describe the different security and privacy issues in the SPA. Known attacks on SPA are discussed in Section~\ref{sec:Attacks}. Section~\ref{sec:countermeasures} describes existing countermeasures, and Section~\ref{sec:discussion} provides a summary and some discussions on future research directions. Finally, Section~\ref{sec:conclusions} draws the conclusion.

\section{Background}
\label{sec:background}

SPA have a complex architecture (see details in Section \ref{background-arch}).  As a general introduction, and despite the fact that different SPA across different vendors have a few distinctive characteristics, all SPA perform similar functions and share some common features. In particular, SPA's architectures usually include, together with other architectural elements such as cloud-based processing and interaction with other smart devices, the following: i) a \emph{voice-based intelligent personal agent} such as Amazon's Alexa, Google's Assistant, Apple's Siri, and Microsoft's Cortana~\cite{alexa-Statista}; and ii) a \emph{smart speaker} such as Amazon's Echo family, Microsoft's home speaker, Google's home Speaker, and Apple's HomePod. Note that, while we focus on SPA as one full instantiation and ecosystem based on voice-based personal assistants, some of the issues mentioned in this review may apply to other non-SPA voice-based personal assistants, as there are parts of their architecture that may be similar, especially those parts not related to the smart speaker.

SPA decode users' voice input using NLP to understand users' intent. Once the intent is identified, it delegates the requests to a set of \emph{skills}\footnote{Note that, for ease of exposition, we adopt Amazon's terminology of \emph{Skills}, but these may be called differently in other SPA platforms. For instance, in Google's Assistant and Google Home, skills are called \emph{Actions} instead.} from where it obtains answers and recommendations. Conceptually, skills are similar to mobile apps, which interface with other programs to provide functionality to the user. The entire skills ecosystem provides an environment that offers the user the ability to run more complex functions such as calendar management, shopping, music playback, and other home automation tasks. There are two types of skills, namely: {\em native skills} and {\em third-party skills}. The former are skills given by the SPA provider that perform basic functions and leverage providers' strengths in areas such as productivity (Microsoft Cortana), search (Google Assistant), and e-commerce (Amazon Alexa)~\cite{White:2018:SDV:3289258.3185336}. The latter are skills built by third-party developers using \textit{skill kits}~\cite{alexa_developer, Google_developer}, which are development frameworks with a set of APIs offered by the SPA provider to perform basic operations. There are currently thousands of SPA skills hosted online, although the numbers keep growing daily. For example, Amazon's skill market now has over 70,000 Alexa skills worldwide~\cite{alexa-Voicebot.ai} and the Google Assistant skill market has over 2,000 skills~\cite{google_Voicebot.ai}. These skills are classified into different categories such as \textit{home control skills, business and finance skills, health and fitness skills, games and trivia skills, news skills, social skills, sports skills, utilities skills}, etc. As further support to the skills, SPA often have the ability to learn information about users' preferences such as individual language usages, words, searches, and services using Machine Learning (ML) techniques~\cite{Naik2018} to make them smarter over time.

\subsection {Smart Home Personal Assistants Architecture}  
\label{background-arch}

\begin{figure}[ht]
\centering
\includegraphics[width=0.9\columnwidth, trim=15 50 2 80, clip]{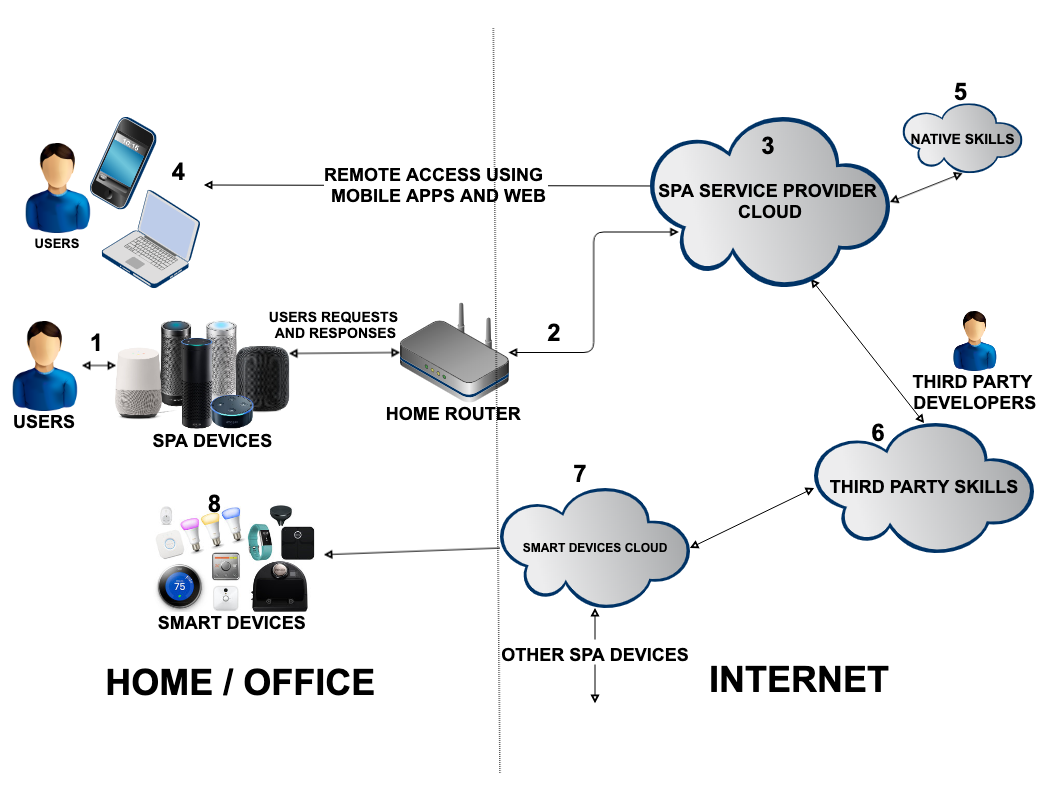}
\Description[SPA System]{The SPA device puts itself into recording mode to record and send the user's speech to the SPA cloud for processing.}
\caption{SPA architecture and its key components.~\cite{Google-SDK-Preview, Understand_Skills}}  
\label{fig:Capture.png}
\end{figure}

SPA are Internet-based systems with a regular iteration of updates. One benefit of this is that its capabilities are wide-ranging and dynamic --- they will evolve along with the proliferation of new Internet services. Figure~\ref{fig:Capture.png} shows the key components in the SPA system architecture. Each component is a potential attack point for an adversary. How some of them may be exploited is discussed in Section~\ref{sec:Attacks}. 

Point 1 represents the point of interaction between the users and the SPA devices. 
SPA devices such as Amazon Echo are equipped with powerful microphones, and the device itself consists of a voice interpreter that records users' utterances. To make use of the SPA, the voice interpreter needs to be activated. Many of the voice interpreters are often pre-activated and run in the background. After the voice interpreter is activated, it then waits for the wake-up word to be triggered~\cite{2800-18}. Once it receives the wake-up keyword, it puts the SPA into recording mode. In recording mode, any user utterances are processed and sent through the home router (Point 2) to the SPA cloud (Point 3)~\cite{Understand_Skills} for further analysis. Only the wake-up command is executed locally, while all other commands are sent to the cloud. Hence, the SPA must always be online.   

The captured utterances are decoded using NLP in the SPA cloud as we detail in Section \ref{subsec:background-nlp} below. It must overcome the issue of background noise, echo, and accent variation in the process of extracting the intent~\cite{2800-18}. Once the intent is extracted, it is then used to determine which skill to invoke. There are two ways to invoke a skill. First, they can be explicitly invoked by using their activation name: for example, where a skill name is ``Tutor Head,'' it can be triggered by saying the words: ``talk to Tutor Head.'' Explicit invocation can be extended to use a deep link connection, as detailed here~\cite{GOOGLE} for Google Assistant. For instance, ``talk to Tutor Head to find the next course'' where the next course is a predefined action under the ``Tutor Head'' skill. Second, skills can be implicitly invoked by an intent's query composition without explicitly using their invocation name. If a query does not directly match with a skill, the SPA will either inform the user or match the query to another similar skill when appropriate.

By default, the SPA provider will try to find a native skill to process the request invoked by the user \citep{alexa_developer}. In this case, the SPA cloud service then sends the intent to its native skill, which processes the request in the cloud of the SPA (Point 5) and sends a response back to the SPA device. When there are no native skills available, the request is sent to third-party skills (Point 6). These are typically hosted in a remote web service host controlled by the developer of the third-party skill. Once the request is processed, the third-party skill returns the answers to the SPA cloud service, which sometimes asks for more information before the request is finalized. In the case where the intent is meant to control other smart devices, the relevant information is forwarded to their respective cloud service (at Point 7), and from there, the instructions are relayed to the target smart device (at Point 8).

\subsection {Natural Language Processing in SPA}
\label{subsec:background-nlp}
\begin{figure}[htb]  
\centering
\includegraphics[width=0.85\columnwidth]{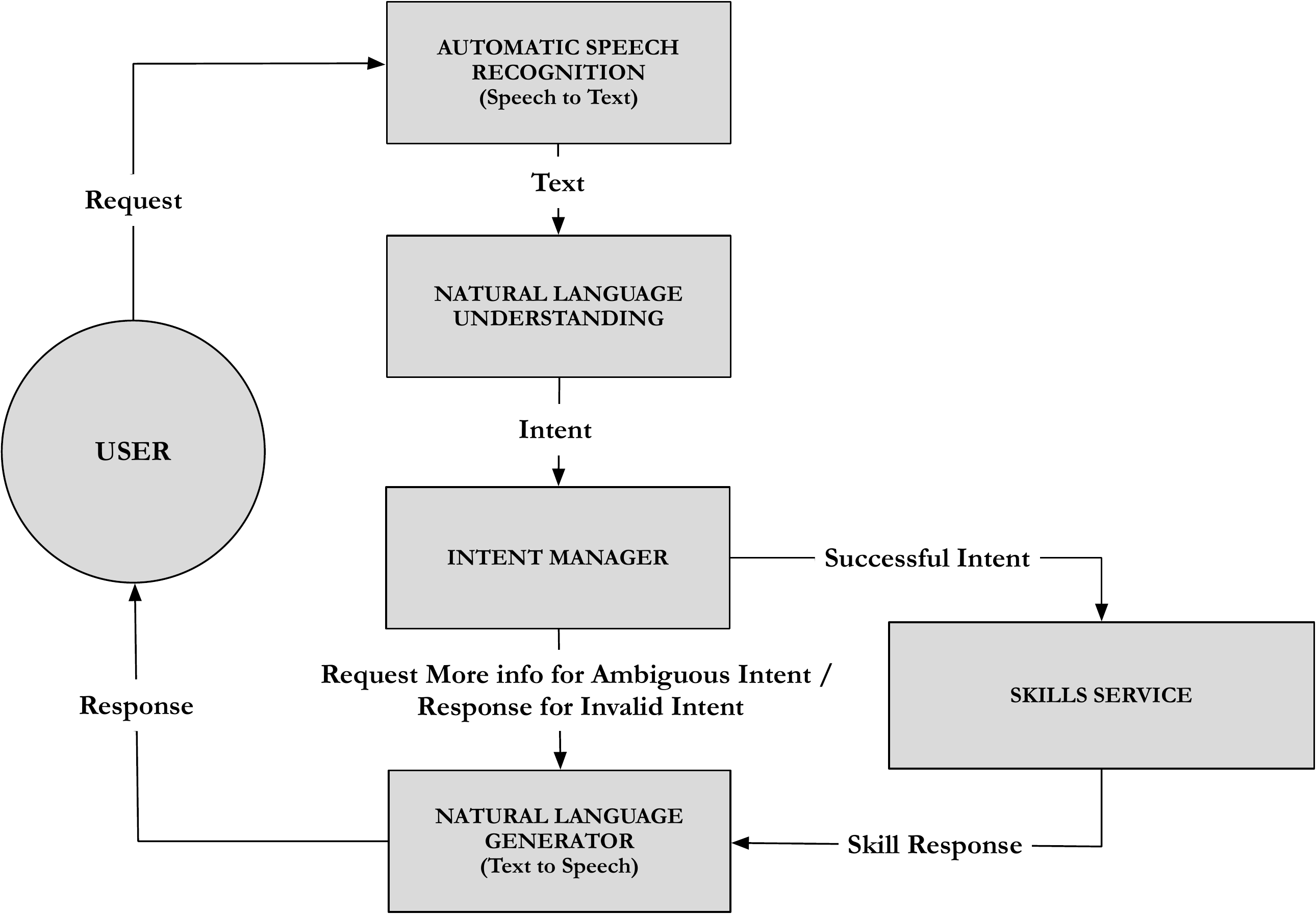}
\Description[NLP Speech System]{A diagram depicting how NLP system converts the users' speech to text in order to extract their intent and fulfill their request.}
\caption{NLP speech system from {\em Speech To Text} (ASR) to {\em Text To Speech}}.
\label{fig:nlp-to-tts}  
\end{figure}

SPA benefit from recent advances in Natural Language Processing (NLP), which allow them to handle a wide range of commands and questions. The NLP improvements are attributed to: i) a number of novel advances in ML, ii) a better knowledge of the construction and use of the human language, iii) an increase in the computing power, and iv) the availability of sizable labeled datasets for training speech engines~\cite{hirschberg_manning_2015}. Processing user speech includes a complex procedure that involves audio sampling, feature extraction, and speech recognition to transcribe the requests into text. Since humans speak with idioms and acronyms, it takes an extensive analysis of natural language to get correct outputs. For instance, issuing a command to an SPA asking it to remind you about a meeting at a specific time can be done in several ways. While some parts of this command are more specific than others and can easily be understood, such as the day of the week, other words that support them can be dynamic. This implies that understanding an intention as simple as a meeting reminder might require non-trivial interactions. Figure~\ref{fig:nlp-to-tts} illustrates the process involved in understanding a user's intent and generating responses.

Intent recognition starts with signal processing, which offers the SPA a number of chances to make sense of the audio by cleaning the signal. The idea is to enhance the target signal, which implies recognizing the surrounding noise to reduce it~\cite{mosner_wu_raju_krishnan}. That is one reason why most SPA devices are equipped with multiple microphones to roughly ascertain where the signal is coming from so that the device can concentrate on it. Once the original signal is identified, acoustic echo cancellation~\cite{yang_2018} is then used to subtract the noise from the received signal so that only the vital signal remains.
Typically, most speech recognition systems work by converting the sound waves from the user's utterances into digital information~\cite{Garimella2015RobustIB}. This is further analyzed in order to extract features from the user's speech, such as frequency and pitch. Primarily, Automatic Speech Recognition (ASR) comprises of two steps: features extraction and pattern classifiers using ML~\cite{londhe_ahirwal_lodha_2016}. There are several feature extraction methods, with Mel frequency cepstral coefficient (MFCC) being one of the most popular since it is believed to mimic the human auditory system~\cite{zolnay_kocharov_schlüter_ney_2007}. These features are then fed into an acoustic model trained using ML techniques to match the input audio signal to the correct text \cite{45601}. For instance, ML models based on Hidden Markov Model (HMM)\cite{Garimella2015RobustIB} often compare each part of the waveform against what comes previously and what comes next, and against a dictionary of waveforms to discover what is being said.

Once the SPA cloud has the text that transcribes what the user has said, it employs Natural Language Understanding (NLU), a key component of Natural Language Processing (NLP), to understand what the user intends to do. This is done using discreet to discreet mapping, with some instances relying on statistical models or ML techniques like deep learning to assume the likely intent. The more data available to the NLP system from regular usage, the better the prediction of the user's intent. After the NLU extracts the intent, the intent manager then decides whether more information is needed to provide an accurate answer before forwarding the intent to the skill service for processing. After the intent is processed, the generated skill response is sent to the Natural Language Generation (NLG), where it is converted into natural language representation. It is then communicated back to the user, and it is typically (e.g., Amazon Echo) played by a smart speaker.

\subsection{Assets in the SPA Architecture}
\label{Subsec:Assets}
Next, we discuss the assets in the SPA architecture from SPA users' point of view, and why users consider the assets to be important or sensitive, to understand what is at stake, and what should be protected.

\subsubsection{SPA and Smart Devices}
The SPA device and other peripheral smart devices are essential assets in this domain. 
There are different types of SPA devices attending to where the personal assistant interacts with the user. 
SPA devices can be integrated into smart speakers like Amazon Echo, Google Home, and Apple HomePod. 
As illustrated in Figure \ref{fig:Capture.png}, SPA also interact with other smart home devices~\cite{Noura_Abdi_2019, 238321} such as smart heating and cooling devices (e.g., Nest, or Ecobee 4), Smart security (e.g., Scout, or Abode), smart lighting devices (e.g., Philip Hue, or LIFX), Smart kitchen (e.g., GE+ Geneva) and surveillance cameras (e.g., Cloud Cam, Netgear Arlo Q). 
All these assets are generally characterized by the hardware they are built on.

\subsubsection{Personal Data}
Personal data is one of the most valuable assets in the SPA ecosystem because of the amount and variety in which personal data is collected, shared, and processed. Therefore, many of the security issues explained below also impact users' privacy, even though this may affect users differently depending on what they value based on their perceptions and preferences~\cite{tabassum2019investigating, Lau:2018:AYL:3290265.3274371, Noura_Abdi_2019}. All this may in turn be defined by the user's understanding of the data flows in SPA~\cite{Noura_Abdi_2019} and what they have experienced in other computing contexts~\cite{238321}. We give more details below of examples of particular types of personal data in the SPA ecosystem.

\paragraph{User voice records (audio clips and transcripts):}

SPA need to continuously learn from past computations for reliable speech recognition. To achieve this, SPA need a large training dataset of user conversations. Users are known to have concerns about the storage of the recordings of those conversations in some cases and, particularly, about what they may be used for~\cite{malkin2019privacy}.

\paragraph{User account data:}
Users also have data as part of their account with the SPA provider. For instance, in Amazon Alexa, this includes users~\emph{location, mobile number, email address, name, device address, payment information, and shopping lists~\cite{Amazon_Permissions}.}
Note, however, this data is not restricted to the SPA provider, and skills can request permission to access data from the user's account with the SPA provider.

\paragraph{Skill interaction data:}
Skills can potentially ask users for any personal data through their conversations with the users. 
In fact, there is research evidence that skills collect personal data during voice interaction without asking for any permissions regarding user account data~\cite{10.1145/3319535.3363274}. Birthdate, age, and blood type are examples of the data they may ask for according to this research.

\paragraph{Smart devices data:}
The integration between an SPA and other smart devices brings the smart home into one verbally controlled system and offers the SPA the privilege to manage the services of other connected smart devices. This integration enables access to home sensors that generate valuable personal data.

\paragraph{Behavioural data:}
Apart from the raw data mentioned so far, other sensitive data can be \emph{inferred} from user actions with the SPA or by processing the raw data. This includes predicting users' behavioral characteristics like user interests, usage patterns and sleeping patterns as shown in~\cite{DBLP:journals/corr/abs-1803-00466,chung_park_lee_2017},
where the authors demonstrate how personal information can be inferred from data stored by the SPA provider by using a forensic toolkit that extracts valuable artifacts from Amazon Alexa by taking advantage of the Alexa unofficial API.

\subsubsection{Other Assets}

There are other assets such as reputation, financial well-being, physical well-being, emotional well-being, and relationships, all of which could be valued differently by users. For instance, if an attacker successfully breaks into an SPA,  users could be affected financially if there are unauthorized purchases, emotionally from shame or embarrassment, as well as suffer damage to their reputation if an adversary uses SPA to impersonate them. In fact, some SPA users restrict the use they make of SPA to avoid impacts on these assets, e.g., many SPA users avoid purchasing through SPA because they do not think the process is secure or trustworthy enough~\cite{Noura_Abdi_2019}.

\color{black}

\section{Security and Privacy Issues}
\label{sec:Issues}
In this section, we present a classification of the main security and privacy issues of SPA. We use this classification to later map current attacks and countermeasures in Sections~\ref{sec:Attacks} and~\ref{sec:countermeasures}.

\subsection{Weak Authentication}  
Here, we discuss issues related to how SPA verify users and how an adversary can exploit such a process.  

\subsubsection{Wake-up Words}
\label{sec:issue-wake-up-words}
By design, SPA authentication is done using wake-up words that are recognized locally in the device. A user has the option to select a wake-up word from a set of predefined options, having one by default. It is therefore very easy for an attacker to infer the wake-up word of the user. In addition to the wake-up word, SPA have no additional ways of authenticating the user. The device will accept any command succeeding the wake-up keyword. Hence, it is easy for anyone in proximity to issue commands to the SPA. Authors in~\cite{Lei_Xinyu,alepis_patsakis_2017,zhang_yan_ji_zhang_zhang_xu_2017,2018arXiv180501525Z} have shown how this weak authentication can be used as a proxy to more elaborated security and privacy attacks. 

\subsubsection{Always On, Always Listening}  
As mentioned, the voice command interpreter constantly listens to the user utterances while waits for the wake-up word. 
Having a device permanently on and always listening poses important security and privacy concerns. 
Accidentally saying the wake-up word or any other phonetically similar words will put the assistants to record. 
Consequently, any conversation that follows is uploaded to the Internet. This issue could affect the users' privacy in a situation where private or confidential conversations are accidentally leaked, or where an attacker can retrieve sensitive information from these devices. Likewise, it could also affect the device security as an adversary can issue an unauthorized command to compromise such devices and use them to target other connected smart devices. Recently, due to this feature, a private conversation of a couple was accidentally recorded and sent to a random contact with the Echo device~\cite{The_Telegraph}. This example shows that the users are not in total control of their voice data.

\subsubsection{Synthesized Speech}  

SPAs are known to listen to audio playback. Just recently, a Tv commercial by Burger King prompted Google Home to read information to the user from Wikipedia about the Whopper hamburger~\cite{alexa-CNBC}. While major SPAs like Alexa and Google have now figured out how to filter out background media \cite{Papayiannis2018, Amazon_Acoustic_cancellation}, they are still vulnerable to synthesizing audio that exploits side channels or adversarial examples. For instance, they are vulnerable to inaudible sound reproduced at ultrasonic frequencies~\cite{zhang_yan_ji_zhang_zhang_xu_2017, 211283}, and synthesized speech transmitted through electromagnetic radiation. In particular, laser-powered ``light commands'' \cite{Sugawara2019LightCL}. Since the SPA wake-up word can be readily guessed, and the SPA has no means of detecting if a user is in close proximity, there is little or no limit to which speech can be supplied to them and by whom, provided it is meaningful and can be matched with an intent. Synthesized speech (like ultrasonic/inaudible attacks) could offer an adversary a covert channel to issue a malicious command. An attacker could even distribute these speeches over channels like TV and radio to attack multiple targets at once.

\subsection{Weak Authorization} 
In this part, we evaluate the issues regarding how the SPA manages the level of access to data, and the mechanisms users have to control that. 

\subsubsection{Multi-user Environment} 
The absence of proper functional role separation prevents users from correctly defining what and how resources should be accessed. It is challenging to specify who has access to which resources and how such access should be granted. By default, in a multi-user environment --- which many households are, any user can put the SPA into recording mode and issue out instructions to it. 
Even though the primary user can specify certain access controls for secondary users, the level of granularity is generally coarse and not extensive. 
For instance, any member of an Amazon household (a feature that allows sharing of contents with family members) can modify the device set-up such as the network connection, sound, and many more without the primary user consent. 

\subsubsection{Weak Payment Authorization}  
SPA systems are increasingly supporting online ordering. Implementing proper security controls challenges usability. For instance, Amazon Alexa users have the option to set a 4-digit PIN code to confirm purchases. At the time of writing, this option is not enabled by default. Even when such an option is turned on, it is vulnerable due to weak lockout\footnote{Lockout is a security mechanism that locks an application for some time before a reattempt is allowed.} implementation~\cite{haack_williams}. This is because Alexa allows two PIN tries before an ordering process lockout, after which the user has to restart the ordering process from the beginning. However, there is no restriction on how many times a user can try to order after every lockout~\cite{haack_williams}. 
Following this, vendors have tried to implement alternative countermeasures against misuse in the ordering process. We next show two cases of this. First, some vendors have prevented changes to the shipping address during ordering. Preventing any change to the shipping address during this process is not enough when dealing with ``insiders'' (i.e., unauthorized users who have access to the premises where the SPA are installed).
The case described in~\cite{mom.me} shows how a kid recently made an unauthorized order worth of about \$300 using her mother's Amazon account~\cite{mom.me}. 
Second, other vendors have tackled this weak authorization problem by providing prompt notification to the users about orders.
This poses a problem to users who do not frequently check their phones or emails, or who may not understand what is happening.

\subsubsection {External Party} 
One important concern is how SPA providers, skills developers, developers of integrated smart home devices, and those that have direct access to any of the points of the SPA architecture secure users from external parties that do not have access to any of these points. Like in every other cloud service, the question remains on how data gathered by those involved in the SPA system is shared with third parties, particularly regarding what kind of controls and mechanisms can be implemented to provide more control to users. Informed decisions can sometimes be taken when third parties provide privacy policies and terms of use~\cite{hrw}. 
However, it is currently uncertain what the scope of those terms might imply and how they are enforced.

\subsection{Profiling}
Beyond authorization, i.e., deciding who has access to what data, there is also the problem of data inference --- traditionally known as information processing~\cite{solove06}. 
Data inference has a particularly dangerous incarnation in SPA in the form of profiling. 
Profiling identifies, infers, and derives relevant personal information from data collected from users. Profiled data can be related to the interests, behaviors, and preferences of the targeted users~\cite{Ayse_Cufoglu}. In this subsection, we look into how SPA data can be used to profile users.

\subsubsection{Traffic Analysis} 
\label{sec:issues:en-route}
A good instance of an en-route type of profiling is traffic analysis.
An attacker can take advantage of SPA traffic's improper concealment to profile a user as shown in~\cite{DBLP:journals/corr/ApthorpeRF17}. 
In particular, attackers can leverage en-route profiling to infer a user's presence. 
This can be further used to conduct more sophisticated attacks. En-route profiling attacks can be made even when the network traffic is encrypted. 
While there are obfuscation techniques that can be used to hinder these types of attacks, they have not been adopted in SPA. In this scenario, the most plausible adversary would be a dishonest or unethical Internet service provider. Governments or other global adversaries with access to the user network traffic can also exploit this weakness. The practicality of this threat to encrypted SPA traffic is shown in~\cite{DBLP:journals/corr/ApthorpeRF17}. While authors in~\cite{DBLP:journals/corr/ApthorpeRF17} perform traffic analysis without even needing an in-depth inspection of the network packages, MiTM techniques --- such as SSL-stripping~\cite{10.1007/978-3-642-33469-6_30} --- might be used to perform profiling over plain-text.

\subsubsection{Uncontrolled Inferences} 
Profiling, in this case, is about inferences made by any of the parties in the SPA ecosystem (third-party skill developers, SPA providers, etc.) from data they collect with the consent of the user. This includes some of the personal data mentioned in Section \ref{Subsec:Assets} (conversations, account data, interaction data, etc.). That is, the starting point is data about the user that the user may have consented to share. This data is then used to infer \emph{new} data about the user that the user had not shared. An example would be the behavioral data mentioned in Section \ref{Subsec:Assets}. 
Therefore, the problem is that even when users can choose whether they share some data, they have no control over what the parties can do with the data, or what kind of inferences or aggregations they could make to derive other new personal information about the user, e.g., users' tastes or predilections. 
Note that in some cases, collusion between the parties might be possible to be able to conduct more powerful inferences. For instance, malicious skills may collude to aggregate personal data from multiple skills similar to what we have seen in smartphone apps~\cite{memon2015colluding}. Here, skill connection pairing~\cite{alexa-Bret_Kinsella} may be leveraged to create colluding skills aiming at getting more elaborated profiling. Uncontrolled inferences are especially critical as advances in data analysis enable automated techniques to make sense of unstructured data at scale.

\subsection {Adversarial AI}
\label{sec:issues:ai}
As described in Section \ref{subsec:background-nlp}, for an SPA to fulfill the user's request, it needs to first understand what the user's said, understand what the user wants, and before selecting the best skill to fulfill the request. For these, the speech recognition system uses AI techniques like NLP and ML. However, these techniques can introduce the issues discussed below.
 
\subsubsection {Adversarial ML}

ML in SPA system is used for many tasks, including speech recognition. Conventionally, ML is designed based on the notion that the environment is safe, and there is no interference during training and testing of the model~\cite{papernot2016towards}. However, such an assumption indirectly overlooks cases where adversaries are actively meddling with the learning process~\cite{papernot2016towards}. 
ML is known to be vulnerable to specially-crafted inputs, described as adversarial examples, which are usually derived by slightly modifying legitimate inputs~\cite{DBLP:journals/corr/SzegedyZSBEGF13}. These perturbations typically remain unknown to the person supervising the ML task but are wrongly classified by already trained ML models. Examples can be used to manipulate what the SPA system understands from spoken user commands~\cite{zhang_yan_ji_zhang_zhang_xu_2017}. This could then be used to generate a denial of service attack, invoke an incorrect skill~\cite{191968}, or to reduce the ML model quality and performance~\cite{DBLP:conf/sp/Carlini018}. Most ML models that perform the same task tend to be affected by similar adversarial inputs even if they use different architectures and are trained on different datasets~\cite{DBLP:journals/corr/PapernotMG16}. This allows the attacker to easily craft adversarial inputs with little knowledge about the target ML model. Research has also shown that speech recognition models often find it challenging to differentiate words with similar phonemes~\cite{217575}, e.g.: distinguish between ``Cat'', ``Pat'', and Fat, which can come handy when crafting adversarial inputs. Commonly exploited ML vulnerabilities are not the only type of examples that may apply.
For instance, to predict the best skill to process the user's request, most SPA continuously learn from the user interactions and regularly retrain their ML models with new data. Attackers could insert adversarial samples into the training dataset to corrupt the ML models (poisoning attack). Another example would be targeting the ML models to extract valuable information (membership inference attack), e.g.: the accent of the speakers in speech recognition models \cite{7958568}.

\subsubsection{NLP Vulnerabilities}
\label{sec:issues:NLP}
Although adversarial ML has a direct effect on the NLP system in SPA as it underpins many NLP tasks used for speech recognition, there are also other parts of the NLP system in SPA that do not directly use ML but that may also be exploited. Following the example of skill invocation given in the previous subsection, the adversarial NLP problem appears once user utterances have already been transcribed into text and the system needs to decide which skill to invoke from the text (note the difference with the problem of translating into text two words with similar pronunciation). In particular, Amazon's Echo and Alexa seem to use the lengthiest string match when deciding which skill is called~\cite{2018arXiv180501525Z}. For example, the text ``talk to \textit{tutor head} for me please'' will trigger the skill ``\textit{tutor head for me}'' rather than the skill ``\textit{tutor head}.'' In a similar way to adversarial ML, an attacker could use such difficulty to trick users into invoking a malicious skill intentionally. This can be achieved by registering a skill with the same name (but longest possible string match) than a legitimate skill. Besides, there is currently no restriction on the number of skills that can be registered, hence, an adversary can register as many skills as possible to increase the possibility of getting their skills called.

\subsection{Underlying and Integrated Technologies}
To broaden SPA capabilities and offer ubiquitous services, SPA rely on skills and other existing infrastructures like cloud services and smart devices. This means that they can potentially inherit or be subject to issues and vulnerabilities present in or arising from these technologies.

\subsubsection{Third-party Skills}

An attacker could take advantage of lax enforcement of the skill implementation policies and exploit the interaction between the user and the SPA system. For example, by faking the hand over process, a malicious skill can pretend to hand over control to another skill and deceive users into thinking that they are interacting with a different skill (Voice Masquerading attack) in order to eavesdrop on user conversations and collect sensitive information. After all, it is difficult for the user to determine if they are taking to the right skill at a particular period of time because of the vagueness of voice command \cite{natatsuka_iijima_watanabe_akiyama_sakai_mori_2019}. Likewise, a malicious skill can fake or ignore the skill termination command and continue to operate stealthily \cite{SRLabs_Squatting}. Furthermore, the existing SPA architecture support only permission-based access control on sensitive data. It is insufficient at controlling how skills use data once they get access~\cite{inproceedings}. This could create privacy concerns, especially in over-privileged skills as it does not allow users to specify the intended data flow patterns once a skill has permissions to access data. 
In fact, authorizing a malicious skill to access confidential information may result in leaking sensitive information to unwanted parties. In the SPA ecosystem, the end-user does not have any kind of access to the skills, which is rather different from the apps in smartphones that will be running in your phone, so protection mechanisms in the smartphone can be used to target apps. In contrast, users don't have a way to install any protection mechanisms beyond those the SPA provider can put in place for skills. A user must rely on the SPA provider to ensure that such services are as secure as they need to be. However, even if the SPA provider would provide a vetting process, related works have shown that they can be successfully evaded~\cite{2018arXiv180501525Z, SRLabs_Squatting}.

\subsubsection{Smart Home Devices} 

While SPA integration with other smart home devices brings the smart home into one verbally controlled system, it also creates a single key point of interest to attackers. 
Attackers can take advantage of this in two ways. On the one hand, breaching the SPA can allow attackers to control a wide range of connected devices. 
More so, privacy issues could emerge from data accumulation, data acquisition, and integration as discussed in~\cite{madaan_ahad_sastry_2018,RomanIoT18}, where the authors perform a comprehensive review of privacy threats of Information Linkage from data integration in IoT ecosystems. On the other hand, vulnerabilities in connected smart devices could be used as an intermediate step to attack the SPA~\cite{8283484,Denning:2013:CSM:2398356.2398377,surveySmart}. 
Attacks in connected smart home devices have been investigated in numerous works, including: 1) snooping attack where an adversary listens to the smart home traffic to read confidential data~\cite{Denning:2013:CSM:2398356.2398377}, 2) privilege escalation where attackers use design and configuration flaws in smart home devices to elevate privileges and access confidential home users information, 3) insecure interactions between apps that are used for controlling peripheral devices and third-party counterpart apps which could open channels for remote attackers, and 4) other direct compromises of various smart home devices \cite{Denning:2013:CSM:2398356.2398377,7546527}. For instance, the API service on Google Home before mid-July 2018 was reported to be vulnerable to DNS rebinding attacks, which allow remote attackers to initiate a denial of service attack, extract information about the Wi-Fi network or accurately locate this device~\cite{CVE201812716}.
It is important to note that some of the issues we identify in this review are not specific to SPA alone. They are also present in other smart home and IoT devices, since the SPA and other IoT devices conduct information exchange and communications in a similar way, and are often co-located within the same environment. Nonetheless, the SPA ecosystem is quite unique, e.g., the speech and intent recognition steps, which determine the actual third-party skill that is to serve a user command may lead to specific adversarial AI issues as mentioned above.

\subsubsection{Cloud} 

While the cloud offers the advantage of having readily available and virtually unlimited resources, they also present attackers with new opportunities~\cite{modi_patel_borisaniya_patel_rajarajan_2012}. 
On the one hand, they are data-rich environments that are centrally located in a single point, and in particular in SPA architectures, they keep most of the personal data mentioned in Section \ref{Subsec:Assets}. If this element is breached, attackers may get access to valuable and sensitive information. This is the most concrete and frequently mentioned threat by users regarding smart home data~\cite{238321}. 
On the other hand, they usually offer multiple remote ways of accessing the data (e.g., web or app-enabled access) and facilitate online configuration, thereby widening the attack surface.
The SPA provider cloud (point \#3 in Figure \ref{fig:Capture.png}) is therefore subject to these issues. Most importantly, data in the cloud are subjected to insider attacks (i.e., abuse of authorized access)~\cite{alexa-CBS-Human, Google-theverge-human}. For instance, some SPA providers may let employees listen to recorded conversations as they view this process as a critical part of evaluating their SPA speech recognition system~\cite{Google-theverge-human} and a way of improving customer experience~\cite{alexa-CBS-Human}. This is a critical issue when their privacy statements fail to mention this type of usage or whether conversations are used anonymously~\cite{Assistants-Capecodtoday}. Likewise, the SPA provider cloud could also suffer from incomplete data deletion~\cite{ramokapane2016assured}. This situation may enable SPA providers to retain (intentionally or accidentally) private data even after being deleted (assuming users manage to find the way to delete information from the cloud, which is not always easy for them~\cite{ramokapane2017feel}). For instance, it is known that Amazon could keep transcripts of users' voice interactions with Alexa even after the recordings are deleted~\cite{alexa-delete}.

\section{Attacks}
\label{sec:Attacks}
This section offers a review of known attacks on the SPA system and examines the vulnerabilities they exploit w.r.t. the issues described in Section~\ref{sec:Issues} and the point they target in the architecture in Section \ref{sec:background}. Table~\ref{table:1} shows an overview of the most relevant attacks mapped to the vulnerability(ies) they exploit and the affected points in the architecture. 
We found that most of the attacks target the following elements of the architecture depicted in Figure~\ref{fig:Capture.png}: 

\begin{enumerate}
\item { User to SPA device} (\#1): There is a wide range of attacks targeting this point of the architecture. In particular, we identify related works i) exploiting weak authentication, and ii) attacking underlying and integrated technologies.

\item { SPA device to SPA service provider cloud} (\#2): There is an attack reported in the literature that targets this point of the architecture and exploits improper concealment of SPA traffic.

\item { SPA service provider cloud} (\#3): Several attacks are also found at this point of the architecture targeting the SPA cloud components. We identify works exploiting i) ML Vulnerabilities, and ii) underlying technologies.

\item { Third-party Web skills} (\#6): Attacks targeting this point of the architecture exploit user misconceptions about the SPA system, and in particular about the skill. We show related works exploiting NLP subsystem vulnerabilities.

\end{enumerate}

We could not find any attacks targeting architectural elements \#4 (remote access via mobile and Web), \#5 (native Web skills), \#7 (smart device cloud), and \#8 (connected smart devices).
However, this does not mean that attacks targeting those architectural elements are not possible. In fact, some of the threats outlined in~\cite{Denning:2013:CSM:2398356.2398377} and the attacks demonstrated by researchers in~\cite{Chouhan_2016} could possibly exploit \#8. Besides, some of the vulnerabilities that exist in \#3 might also be found in \#7 as they are both cloud technology. Likewise, attacks targeting \#6, such as voice squatting and voice masquerading~\cite{2018arXiv180501525Z}, might also be possible in \#5 since both are skill services. Nevertheless, as far as we know, they have not been exploited yet. We discuss this more in detail later on in Section \ref{sec:discussion}.

We next describe the attacks we found in related literature by types (or categories) of attacks, particularly looking at the vulnerabilities (described in Section \ref{sec:Issues}) that they exploit and the assumptions they make on the environment.

\subsection{Side Channel Attacks}
\label{sec:User-SPA-attack}
This includes attacks that are based on information gained from the way an SPA is implemented, rather than vulnerabilities in the SPA itself. 
The \emph{always on, always listening} and the \emph{lack of arbitrary wake-up words}  within the \emph{weak authentication} category are the most exploited vulnerabilities in this class of attack.

Lei Xinyu et al.~\cite{Lei_Xinyu} look at issues in single-factor authentication methods based on a wake-up word, and the lack of a mechanism that can be used to figure out if a user is close-by or not. Using Amazon's Echo device, the authors perform a home burglary attack to manipulate a connected door lock. Likewise, they successfully make a purchase using the compromised device. Authors in \cite{Sugawara2019LightCL} also exploit the lack of proper user authentication and vulnerable microphones to inject voice commands into SPA. By simply modulating the amplitude of laser light, the authors successfully use light-injected voice commands to unlock a connected smart lock integrated with the SPA, and to locate, unlock, and start cars (including Ford and Tesla) provided they are linked with the target's Google account. However, unlike in other classes of attacks where attackers are restricted by distance due to the use of sound for signal injection, attackers here are only limited by their capabilities to carefully aim the laser beam on the devices' microphones. Since light does not accurately penetrate through an opaque object, this attack requires a line of sight to the targeted SPA devices.

\begin{landscape} 
\begin{table*}[t]
\renewcommand{\arraystretch}{1.7}
\scriptsize 
\centering
\Description[Attack table of related research on SPA]{A table showing authors, attack points, and the vulnerabilities they exploit in their attack against the SPA. Many authors utilize the wake-up word and always listening vulnerabilities. The most targeted attack point is the point where the users interact with the SPA devices. }
\caption{categorization of attacks found in previous studies based on vulnerabilities exploited and attack point.}
\label{table:1}
\centering 
\begin{tabular}{|>{\centering\arraybackslash}m{1.35cm}|>{\centering\arraybackslash}m{1.4cm}|>{\centering\arraybackslash}m{0.8cm}|>{\centering\arraybackslash}m{0.96cm}|>{\centering\arraybackslash}m{1.0cm}|>{\centering\arraybackslash}m{0.8cm}|>{\centering\arraybackslash}m{1.0cm}|>{\centering\arraybackslash}m{0.99cm}|>{\centering\arraybackslash}m{1.0cm}|>{\centering\arraybackslash}m{0.7cm}|>{\centering\arraybackslash}m{0.8cm}|>{\centering\arraybackslash}m{0.8cm}|>{\centering\arraybackslash}m{0.55cm}|>{\centering\arraybackslash}m{0.7cm}|>{\centering\arraybackslash}m{0.95cm}|>{\centering\arraybackslash}m{0.7cm}|}
\hline

 \multirow{2}{*}{\begin{tabular}[c]{@{}c@{}}Attack \\ Class\end{tabular}}   & \multirow{2}{*}{Studies}                                         & \multicolumn{3}{c|}{Weak Authentication}             & \multicolumn{3}{c|}{Weak Authorization}                              & \multicolumn{2}{c|}{Profiling}          & \multicolumn{2}{c|}{Adversarial AI} & \multicolumn{3}{c|}{Integrated Techs.} & \multirow{2}{*}{\begin{tabular}[c]{@{}c@{}}Attack\\ Points\end{tabular}} \\ \cline{3-15}
                                                                &                                                                    & Wakeup Word & Always Listening & Synthesized Speech & Payment Auth. & Multiuser Environ.     & External Party & Traffic Analysis   & Uncont. infer. &   Adv ML & NLP Vul  & Skills                 & Cloud            & Smart Devices       &                                                                 \\ \hline
\multirow{4}{*}{\begin{tabular}[c]{@{}c@{}}Side \\ Channel \end{tabular}}                                            & Lei Xinyu et al.~\cite{Lei_Xinyu}                               &  \checkmark   & \checkmark       &                    & \checkmark    &                    &            &                    &            &             &              &                                   &                  & \checkmark          & 1                                                                        \\ \cline{2-16} 
& Zhang et al.~\cite{zhang_yan_ji_zhang_zhang_xu_2017}                                                                                                                              & \checkmark   & \checkmark       & \checkmark         &               &                    &            &                    &            &             &              &                                     &        &                            & 1                                                                        \\  \cline{2-16}

& Segawara et al.~\cite{Sugawara2019LightCL}                                                                                                                               & \checkmark   & \checkmark       & \checkmark         &   \checkmark             &                    &            &                    &            &             &                                              &                  &       &              & 1                                                                        \\ \cline{2-16}

& Roy et al.~\cite{211283}                                                                                                                               & \checkmark   & \checkmark       & \checkmark         &               &                    &            &                    &            &             &                                              &                  &       &              & 1                                                                        \\ \hline
 {\begin{tabular}[c]{@{}c@{}}Behavioral \\ Profiling\end{tabular}}                                              &  Apthorpe et. al.~\cite{DBLP:journals/corr/ApthorpeRF17}         &             &                  &                    &               &                    &            &   \checkmark      &             &              &                                     &        &          &                     & 2                                                                        \\ \hline 

 \multirow{5}{*}{\begin{tabular}[c]{@{}c@{}}Attacks \\ on \\ Voice \\ Models \\ using \\ Adversarial \\ samples\end{tabular}} & Gong \& Poellabaeur~\cite{DBLP:journals/corr/abs-1711-03280}    & \checkmark   & \checkmark       &                    &               &                    &            &                          &             &           \checkmark     &         &                  &  &                   & 1, 3                                                                     \\ \cline{2-16} 
& Sch{\"{o}}nherr et al. \cite{DBLP:journals/corr/abs-1808-05665}                                                                                                                              & \checkmark   & \checkmark       &                    &               &                    &            &                    &            &                  \checkmark      &                    &        &          &                     & 1, 3                                                                     \\ \cline{2-16} 
& Carlini and Wagner~\cite{DBLP:conf/sp/Carlini018}                                                                                                                              & \checkmark   & \checkmark       &                    &               &                    &            &                    &                 & \checkmark       &                           &                  &          &           & 1, 3                                                                     \\ \cline{2-16} 
& Vaidya et al.~\cite{191968}                                              & \checkmark   & \checkmark       & \checkmark         &               &                    &            &                    &            &               \checkmark       &          &                 &                  &                     & 3                                                                        \\  \cline{2-16} 
& Carlini et al.~\cite{197215}                                                                                                                                 & \checkmark   & \checkmark       & \checkmark         &               &                    &            &                    &            &          \checkmark          &                          &                  &           &          & 3                                                                        \\ \hline
 \multirow{2}{*}{\begin{tabular}[c]{@{}c@{}}Skill \\ Squatting \& \\ Masquerading \end{tabular}}                   & Zhang et al.~\cite{2018arXiv180501525Z}                         &              &                  &                    &               &                     &                    &            &              &   \checkmark & \checkmark  & \checkmark   &            &                     & 3, 6                                                                     \\ \cline{2-16} 
& Kumar et al. \cite{217575}                          &              &                  &                    &               &                    &                    &            &               &      \checkmark    & \checkmark    &   \checkmark             &                &                     & 3, 6                                                                     \\ \cline{2-16}
& Security Research Labs \cite{SRLabs_Squatting}                          &              &                  &                    &               &           &                    &            &                &      &     &  \checkmark     &           &                     & 3, 6                                                                     \\ \hline
\end{tabular}
\end{table*}
\end{landscape} 

The non-linearity in the Micro-Electro-Mechanical Systems (MEMS) microphone over ultrasound is exploited by Zhang et al.~\cite{zhang_yan_ji_zhang_zhang_xu_2017}. Non-linearity is described as hardware features that cause signals with high-frequency triggers at high power to be shifted to low frequencies by microphones (and speakers)~\cite{211283}. Even though microphones are designed to be a linear system, they exhibit non-linearity in higher frequencies. By synthesizing high-frequency sounds that are not within the human hearing range but are still intelligible to SPA devices, the authors are able to activate and control the voice of the SPA. This technique is called the dolphin attack as it uses ultrasonic frequencies like what Dolphins use to communicate among themselves. This attack was confirmed on seven popular voice intelligent assistants (Siri, Cortana, Huawei Hi Voice, Google Now, Samsung S Voice, and Alexa) over a range of different voice platform. On the downside, this attack cannot be conducted above a distance of 5ft from the targeted device. Likewise, it requires specialized hardware to synthesize and play the ultrasonic signal, making it unrealistic for a real-world attack. 

In a different study, Roy et al.~\cite{211283} develop a long-range version of the dolphin attack. They achieved a range of 25ft from their target. By exploiting the non-linearity inside the microphone, like in~\cite{zhang_yan_ji_zhang_zhang_xu_2017}, they generated long-range high-frequency signals that are inaudible to human but intelligible to SPA. As in the previous study, they control and issue commands to SPA devices with the assumption that the adversary can synthesize a legitimate voice signal. However, rather than using a single ultrasound speaker as done in~\cite{zhang_yan_ji_zhang_zhang_xu_2017} to play the synthesized signal, the authors used multiple speakers that are physically separated in space. They employ spectrum splicing to optimally slice voice command frequencies and play each slice on independent speakers in a way that the total speaker output is inaudible. Nevertheless, the attack is only feasible in an open environment. This is because high frequencies are more susceptible to interference, which is a limiting factor to the distance~\cite{Texas_Instrument_2013}. Likewise, this attack requires multiple ultrasound speakers, making it more challenging to implement in a real-world attack.

\subsection{Behavioral Profiling}

At point \#2 of the architecture where SPA devices exchange information with the SPA cloud provider, authors in~\cite{DBLP:journals/corr/ApthorpeRF17} identify privacy vulnerabilities with SPA by passively analyzing encrypted smart home traffic. Their study indicates that encryption alone does not offer all the necessary privacy protection requirements. The authors profile users' interaction with Amazon Echo devices by plotting send/receive rates of the stream even with encrypted traffic. This poses a severe privacy implication to smart home users as an attacker can use this to infer their lifestyle and the best time to conduct an attack undetected, as discussed in Section~\ref{sec:issues:en-route}. However, the method used in this study might not apply to a situation where different IoT devices communicate with the same domain because of the difficulty of labeling streams by device type.

\subsection{Attacks on Voice Models using Adversarial Samples}
Here, we discuss attacks on speech recognition and processing system using adversarial inputs. 

Looking at where data-driven ML models operate, authors in~\cite{DBLP:journals/corr/abs-1711-03280} show a new end-to-end scheme that creates adversarial inputs by perturbing the raw waveform of an audio recording. With their end-to-end perturbation scheme, the authors crafted adversarial inputs that mislead the ML model. Note that this is widely used in para-linguistic applications. Their adversarial perturbation has a negligible effect on the audio quality and leads to a vital drop in the efficiency of the state-of-the-art deep neural network approach. On the downside, such an attack needs to be embedded in a legitimate audio signal to make them truly obscure. While this attack was not evaluated on a real SPA, it was successful against paralinguistic tasks which are clearly relevant to SPA. In particular, speaker recognition task for performing voice matching~\cite{Google_CNN,Alexa_Voice_profile} to predict the identity of the speaker.

More recently, Sch{\"{o}}nherr et al.~\cite{DBLP:journals/corr/abs-1808-05665} have proposed an adversarial example based on psychoacoustic hiding to exploit the characteristics of Deep Neural Network (DNN) based ASR systems. 
The attack extended the initial DNN analysis process by adding a back-propagation step to study the level of freedom of an adversarial perturbation in the input signal. It uses forced alignment to identify the best temporal fitting alignment between the maliciously intended transcription and the benign audio sample. It is also used to reduce the perceptibility of the perturbations. The attack is performed against Kaldi\footnote{A widely adopted open-source toolkit written in C++ which offer a wide range of modern algorithms for ASR.}, where it obtained up to 98\% success rate with a computational effort for a 10-secs sound file in less than 2-mins. However, like in~\cite{DBLP:journals/corr/abs-1711-03280}, this attack also needs to be embedded in another audio file, which significantly influences the quality of the adversarial example.

Another important study conducted by Carlini and Wagner in~\cite{DBLP:conf/sp/Carlini018} proposes an attack on speech recognition systems using Connectionist Temporal Classification (CTC) loss. They demonstrated how a carefully designed loss function could be used to generate a better lower-distortion adversarial input. This attack works with a gradient-descent based optimization~\cite{goodfellow2018making} and replaces the loss function with the CTC-loss, which is optimized for time sequences. However, the audio adversarial examples generated when played over-the-air cease to be adversarial, making it unrealistic for a real-world attack.

Similarly, Vaidya et al. \cite{191968} perform an attack on speech recognition systems using unintelligible sound. This is done by modifying the Mel-Frequency Cepstral Coefficients (MFCC) --- feature of the voice command. The attack is performed in two steps: first, altering the input voice signal through feature extraction with adjusted MFCC parameters, and then regenerating an audio signal by applying a reverse MFCC to the extracted features. When put together, this attack is able to craft a well designed adversarial input. The MFCC values are selected in a way that they can create a distorted audio output with least sufficient acoustic information. This audio output can still achieve the desired classification outcome and is correctly interpreted by the SPA while unintelligible to human listeners. Although this attack successfully exploits the differences between how computers and humans decode speech, it could, however, be detected if a user is in proximity --- provided that they hear unsolicited SPA responses. The attack presented by Vaidya et al.~\cite{191968} is extended in the work of Carlini et al.~\cite{197215}, where the authors test the attack effectiveness under a more realistic scenario and craft an adversarial example completely imperceptible to humans by leveraging the knowledge of the target speech recognition system.
 
\subsection{Skill Squatting and Masquerading Attacks}
In this section, we discuss attacks that exploit how skills are invoked and the way skills interact with each other.

Authors in~\cite{2018arXiv180501525Z} target the interaction between third-party skills and the SPA service. Specifically, they analyze two basic threats in Amazon's Alexa and Google's Assistant SPA services: voice squatting and voice masquerading. Voice squatting allows an attacker to use a malicious skill with longest matching skill name, similar phonemes, or paraphrased name to hijack the voice command of another skill as described in section \ref{sec:issues:NLP}. In five randomly sampled vulnerable target skills, the authors successfully ``hijacked'' the skill name of over 50\% of them. 
The feasibility of this type of attack is high, particularly in SPA, such as Alexa that allows multiple skills with the same invocation name. This attack can be used to damage the reputation of a legitimate skill as any poor service of the malicious skill will be blamed on it. 

Equally, in voice masquerading attack, a malicious skill pretends to invoke another skill or fake termination. Then, the skill keeps recording the user's utterances. This attack could be used to snoop on the conversations of the user. While voice squatting attacks exploit the weaknesses in the skill's invocation method, voice masquerading targets user's misconceptions about how SPA skill-switch services work. With some skills requesting for private information, an adversary could use these attacks to obtain sensitive information and cause crucial information leakage to unwanted parties. Voice squatting attack is also shown in the work of Kumar et al.~\cite{217575}. But unlike what was done in~\cite{2018arXiv180501525Z}, Kumar et al. use the intrinsic errors in NLP algorithms and words that are often misinterpreted to craft malicious skills and exploit the ambiguity in the invocation name method.

\section{Countermeasures}
\label{sec:countermeasures}
In mitigating the identified risks and attacks, there have been a number of studies proposing various countermeasures. In this section, we summarise research on countermeasures, highlighting limitations and deficiencies. We give a summary of these in Table~\ref{table:2}. We mapped the proposed countermeasures to the vulnerabilities discussed in Section~\ref{sec:Issues}.
The current mitigation level in the table (last row of Table~\ref{table:2}) aims to provide a quick indication of the extent the issues identified have been resolved by the countermeasures proposed by the existing publications analyzed to date. In some cases, a combination of countermeasures is enough to address a specific concern, while others will require new countermeasures to effectively address them. The table also has a column called ``Usability Impact'' to indicate whether usability is considered or not by the countermeasure. We use the symbol ``!'' where there is ``potential usability impact'' such as where users are required to put on extra wearable devices (sacrificing user convenience) \cite{kepuska_bohouta_2018, feng_fawaz_shin_2017}, or the solution might restrict the SPA capability~\cite{DBLP:journals/corr/abs-1805-10190}, and ``?'' for the rest, which means ``usability not explicitly considered'', as we did not find enough information in the papers to make any claims (positive or negative) about usability. Finally, we also map these countermeasures to the elements of the architecture depicted in Figure~\ref{fig:Capture.png} to describe the points at which the mitigations would be applied. Most countermeasures map to:

\begin{enumerate}
\item { User to SPA device} (\#1): There is a wide range of countermeasures proposed to mitigate attacks at this point of the architecture. In particular, we found many related works mitigating \emph{weak authentication} vulnerabilities.

\item { SPA device to SPA service provider cloud} (\#2): At this point of the infrastructure, we found studies proposing different mitigation techniques to obfuscating traffic between the SPA device and the SPA service provider cloud, with the aim to mitigate \emph{en-route} vulnerabilities within the \emph{profiling} category.

\item { SPA service provider cloud} (\#3): Few of the existing countermeasures also focused on the \emph{Adversarial AI} vulnerabilities that are found at this point of the architecture and recommended measures aim to mitigate the risks associated with them.

\item { New Architecture}: Countermeasures in this category modify to some extent the existing SPA architecture as part of the mitigation and/or mitigate vulnerabilities that cut across multiple points of the infrastructure. We mapped these countermeasures to multiple architecture elements to signal the points where the mitigations apply or the points that would change as part of an architecture modification.
\end{enumerate}

\subsection{Voice Authentication}

One of the defense that has been put in place against weak authentication is voice authentication.
With this defense, the SPA can tell apart individual users when they speak. 
For instance, some SPA such as Google and Amazon perform speaker verification through voice authentication, known as \textit{voice match}~\cite{Google_CNN} and \textit{voice profiles}~\cite{Alexa_Voice_profile} respectively. 
However, none of these mechanisms is enabled by default, and it is left to the users to first realize about their existence and then decide whether

\begin{landscape} 
\begin{table*}[!htb]
\renewcommand{\arraystretch}{1.7}
\scriptsize 
\centering
\Description[Countermeasure table of related research on SPA]{A table showing existing studies and the vulnerabilities they mitigate. Many authors mitigate synthesized speech vulnerabilities when compare with the rest vulnerabilities. Most of the countermeasures are implemented at the point where the users interact with the SPA devices. }
\caption{Categorization of countermeasures found in related studies.} 
\label{table:2}
\begin{tabular}{|>{\centering\arraybackslash}m{1.1cm}|>{\centering\arraybackslash}m{1.6cm}|>{\centering\arraybackslash}m{0.75cm}|>{\centering\arraybackslash}m{0.94cm}|>{\centering\arraybackslash}m{1.0cm}|>{\centering\arraybackslash}m{0.8cm}|>{\centering\arraybackslash}m{0.88cm}|>{\centering\arraybackslash}m{0.8cm}|>{\centering\arraybackslash}m{0.88cm}|>{\centering\arraybackslash}m{0.64cm}|>{\centering\arraybackslash}m{0.54cm}|>{\centering\arraybackslash}m{0.54cm}|>{\centering\arraybackslash}m{0.5cm}|>{\centering\arraybackslash}m{0.5cm}|>{\centering\arraybackslash}m{0.79cm}|>{\centering\arraybackslash}m{0.9cm}|>{\centering\arraybackslash}m{0.9cm}|}
\hline 

 \multirow{2}{*}{Class}                                                                   & \multirow{2}{*}{Studies}                                                     & \multicolumn{3}{c|}{Weak Authentication}             & \multicolumn{3}{c|}{Weak Authorization}                               & \multicolumn{2}{c|}{Profiling}           & \multicolumn{2}{c|}{Adversarial AI} & \multicolumn{3}{c|}{Integrated Techs.} & \multirow{2}{*}{\begin{tabular}[c]{@{}c@{}}Mitigating\\ Point\end{tabular}} & \multirow{2}{*}{\begin{tabular}[c]{@{}c@{}}Usability\\ Impact\end{tabular}} \\ \cline{3-15}
                                                                      &                                                                                          & Wakeup Word & Always Listening & Synthesized Speech & Payment Auth. & Multiuser Environ.  & External Party & Traffic Analysis     & Uncont. Inf. & ML Vul. & NLP Vul. & Skills & Cloud             & Smart Devices      &              &    \\ \hline
\multirow{4}{*}{\begin{tabular}[c]{@{}l@{}}Voice\\ Auth. \end{tabular}}         &  Voice Match / Profiles ~\cite{Google_CNN,Alexa_Voice_profile}         &              &                  & \checkmark         & \checkmark    &                    &             &                    &             &             &              &                              &                &    & 1       &  ?  \\ \cline{2-17} 
& Kepuska and Bohouta~\cite{kepuska_bohouta_2018}                                                                                                  & \checkmark   & \checkmark       & \checkmark         &     &                    &             &                    &                         &              &                &                   &              &       & 1   & !         \\ \cline{2-17} 
& Huan et al.~\cite {feng_fawaz_shin_2017}                                                                & \checkmark   & \checkmark       & \checkmark         & \checkmark    &                    &             &                    &             &             &              &                &                   &                                     &        1   & ! \\ \cline{2-17}
& Chen et al.~\cite{chen_ren_piao_wang_wang_weng_su_mohaisen_2017}                                                                                         &              &                  & \checkmark         &               &                    &                               &             &             &              &                &                   &                    &                & 1     & ? \\ \hline

{\begin{tabular}[c]{@{}c@{}}Location\\ Verification \end{tabular}}& Lei Xinyu et al.~\cite{Lei_Xinyu}                                                                                                                           & \checkmark   & \checkmark       &                    &               &                    &             &                    &             &             &              &                                 &                    &                  & 1      & ? \\ \hline
\multirow{4}{*}{\begin{tabular}[c]{@{}c@{}}Spectral \\ Analysis \& \\ Frequency\\ Filtering \end{tabular}}         & Roy et al.~\cite{211283}                                                                                         &              &                  & \checkmark         &               &                    &             &                    &             &             &              &                                 &                    &                  & 1       & ? \\ \cline{2-17} 
& Zhang et al.~\cite{zhang_yan_ji_zhang_zhang_xu_2017}                                                                                               &              &                  & \checkmark         &               &                    &             &                                &             &              &                &                   &                    &                 & 1     &  ?  \\ \cline{2-17} 

& Lavrentyeva et al.~\cite{DBLP:journals/corr/LavrentyevaNMKK17}                                                                                         &              &                  & \checkmark         &               &                    &             &                    &             &             &              &                               &                    &                & 3     &  ?   \\ \cline{2-17}

& Malik et al. \cite{8695380}                                                                                         &              &                  & \checkmark         &               &                    &             &                    &             &             &              &                &                              &                 & 3      &  ? \\ \hline

\multirow{3}{*}{\begin{tabular}[c]{@{}l@{}}Traffic\\ Shaping\end{tabular}}          &  Liu et al.~\cite{8278156}                                             &             &                  &                   &                    &             &               & \checkmark  &       &      &              &                &                        &                 & 2    & ?   \\ \cline{2-17} 
& Park et al.~\cite{park_basaran_park_son_2014}                                                               &              &                  &                    &               &                    &             & \checkmark  &   &          &                      &        &    &                  & 2     & ?   \\ \cline{2-17} 
& Apthorpe et al.~\cite{apthorpe_huang_reisman_narayanan_feamster_2019}                                                                                         &              &                  &                                  &                    &             &                    & \checkmark  &       &                &                &                   &                    &                & 2      & ? \\ \hline
\multirow{2}{*}{\begin{tabular}[c]{@{}c@{}}Command\\ \& Phonetic \\Analysis\end{tabular}} &  Zhang et al.~\cite{2018arXiv180501525Z}                               &              &                  &                    &               &                       &              &             &             &              &                                  & \checkmark                   &       &          & 3       &  ? \\ \cline{2-17} 
& Kumar et al.~\cite{217575}                                                                                          &              &                  &                    &               &                    &                           &             &                 &   & \checkmark     &   \checkmark         &                    &                  &   3   & ? \\ \hline
\begin{tabular}[c]{@{}c@{}}New \\ Architecture\end{tabular}                              & Coucke et al.~\cite{DBLP:journals/corr/abs-1805-10190}                &  \checkmark   &                  &                    &               &                    & \checkmark  & \checkmark         & \checkmark     &                &        & \checkmark &      \checkmark       &        & New Arch.      & !  \\ \hline

& Current Mitigation Level                                                                                         & \LEFTcircle  & \LEFTcircle      & \CIRCLE            & \LEFTcircle   & \Circle            & \LEFTcircle & \LEFTcircle        & \LEFTcircle  & \LEFTcircle  & \LEFTcircle  & \LEFTcircle    & \LEFTcircle       & \Circle            &      &   \\ \hline

\end{tabular}
\end{table*}
\end{landscape}

\noindent they would like to activate them or not. Even when these mechanisms are activated, they are still open to attack as an attacker can still trick
the system with a collected or synthesized voice sample of the legitimate user~\cite{chen_ren_piao_wang_wang_weng_su_mohaisen_2017}. Collecting voice samples is an easy task since the human voice is open to the public. Unlike passwords that can easily be changed if compromised, a human voice is a feature that is difficult to replace. 

Another important voice authentication method is proposed in~\cite{feng_fawaz_shin_2017}. In this study, the authors present a continuous authentication VAuth system that aims to ensure that the SPA works only on legitimate users' commands. The solution consists of a wearable security token that repeatedly correlates the utterances received by the spa with the body-surface vibrations it acquires from the legitimate user. The solution was said to achieve close to 0.1\% false positive and 97\% detection accuracy and works regardless of differences in accents, languages, and mobility. Even though this system achieves a high detection accuracy, the need to wear devices such as eyeglasses, headset, and necklaces would introduce a potentially unbearable burden and inconvenience to the users.

Kepuska and Bohouta~\cite{kepuska_bohouta_2018} also proposed a multi-modal dialogue system that combines more than one of voice, video, manual gestures, touch, graphics, gaze, and head and body movement for secure SPA authentication. Even though this system might be able to solve the authentication and voice impersonation challenges earlier discussed, the authors have only been able to test the individual components of the system and not the entire system as a whole.

Finally, Chen et al.~\cite{chen_ren_piao_wang_wang_weng_su_mohaisen_2017} propose a software-only impersonation defensive system. The system is developed based on the notion that most synthesized speech needs a loudspeaker to play the sound to an SPA device. As conventional loudspeakers generate a magnetic field when broadcasting a sound, the system monitors the magnetometer reading, which is used to distinguish between voice commands from a human speaker and a loudspeaker. In a situation where the magnetic field emitted is too small to be detected, the system uses the channel size of the sound source to develop a means of authenticating the sound source. However, the effectiveness of the system depends heavily on the environmental magnetic interference. Likewise, the sound source needs to be at a distance of more than 2.3in (6cm) to their system to prevent the magnetic field from interfering with the magnetometer's reading. In addition, the system has a high false acceptance rate when the sound source distance to their system is greater than 4in (10cm) in a situation where the loudspeaker magnetic field is un-shielded and less than about 3in (8cm) when shielded.

\subsection{Location Verification}
Another important measure implemented against weak authentication is presence-based access control system.
This system allows an SPA to verify if a user is truly nearby before accepting any voice commands.
Lei Xinyu et al.~\cite{Lei_Xinyu} propose a solution that uses the channel states information of the router Wi-Fi technology to detect human motions. Interestingly, it eliminates the need for some wearable devices and introduces no added development cost as it uses the existing home Wi-Fi infrastructure. The solution has an advantage over the traditional voice biometrics recognition, i.e.: that becomes ineffective as users age, become tired, or ill. However, the system's effectiveness depends on selecting the best location for the Wi-Fi devices and setting the right parameters for the detection. Besides, it only supports commands that come from the same room where the SPA device is deployed: in their case, an Amazon Echo. Likewise, the system is situational as it works best if there is no structural change to the location where the devices are deployed.

\subsection{Frequency Filtering \& Spectral Analysis}
Another category of countermeasures aims to enhance authentication, particularly protecting the SPA against synthesized speech using frequency filtering and spectral analysis.

In the work of Roy et al.~\cite{211283}, the authors propose a system nicknamed \textit{lip read} that is based on the assumption that some of the features of voice signals--basic frequencies and pitch--is preserved when it passes through non-linearity. It was reported that this system obtains a precision rate of 98\% and a recall rate of 99\% in a situation where the adversary does not influence the attack command. However, there is no formal guarantee of this countermeasure as they are unable to model the frequency and phase responses for general voice commands. Likewise, their defense only considers inaudible voice attack ignoring finding the true trace of non-linearity. Similarly, Zhang et al.~\cite{zhang_yan_ji_zhang_zhang_xu_2017} propose another set of countermeasures against synthesized speech attacks. The authors recommend two hardware-based mitigating measures---the first one aims to enhance the microphones use by the SPA devices. In contrast, the latter hardware-based defense is intended to cancel any unwanted baseband signal. Enhancing the microphone approach entails designing an improved microphone similar to the one found in Apple iPhone 6 plus that can subdue any ultrasonic sound. On the other hand, canceling the unwanted baseband signal of the inaudible voice command solution entails introducing a module before the low pass filter in the subsystem used for voice capturing to identify and cancel the inaudible voice commands baseband signal.
Likewise, the software-based countermeasure relies on the principle that a demodulated attack signal can be distinguished from legitimate ones using a machine-based learning classifier. 

In another study, Malik et al. \cite{8695380} proposed a countermeasure based on higher-order spectral analysis (HOSA) features to detect replay attacks on SPA. The authors show that replay attacks introduce non-linearity, which can be a parameter to detect it. Lavrentyeva et al.~\cite{DBLP:journals/corr/LavrentyevaNMKK17} also explore different countermeasures to defend against voice replay attacks. Even though the countermeasure is implemented at \#3 of the architecture because it needs extensive computational power, it aims to secure \#1. 
The researchers use a reduced version of Light Convolutional Neural Network architecture (LCNN) based on the Max-Feature-Map activation (MFM). The LCNN approach with Fast Fourier Transform (FFT) based features obtained an equal error rate of 7.34\% on the ASVspoof 2017 dataset compared with the spoofing detection method in~\cite{Todisco:2017:CQC:3103639.3103730} with an error rate of 30.74\%. The authors further utilized Support Vector Machine (SVM) classifier to offer valuable input into their system's efficiency. Consequently, their primary system based on systems scores fusion of LCNN (with FFT based features), SVM (i-vector approach), recurrent neural network (RNN), and convolutional neural network (with FFT based features) shows a better equal error rate of 6.73\% on their evaluation dataset.

\subsection{Traffic Shaping}
To defend against profiling, Liu et al.~\cite{8278156} propose a countermeasure to mitigate traffic analysis vulnerabilities (part of the \emph{profiling} category). 
The authors present a solution that protects the smart home against traffic analysis---a community-based ``differential privacy framework''. The framework route traffic between different gateway routers of multiple cooperating smart homes before sending it to the Internet. This masks the source of the traffic with little bandwidth overhead. Nevertheless, this approach requires cooperation from multiple homes, which makes it difficult to implement. In addition, it could result in long network latency if the homes are not geographically close. 

Other approaches can leverage traffic shaping to prevent profiling. 
For instance, in~\cite{park_basaran_park_son_2014}, Park et al. conceal smart home traffic patterns using dummy activities that have a high likelihood of occurrence. 
This is done considering the behavior of the inhabitants of that environment during the time of measurement. 
While this technique is energy efficient and supports low latency transmission of real data, its implementation requires the participation of many devices and can not shape traffic from genuine user activities. 
In another study~\cite{apthorpe_huang_reisman_narayanan_feamster_2019}, Apthorpe et al. propose a traffic shaping algorithm to make it challenging for an adversary to effectively distinguish dummy traffic patterns generated to mimic genuine user activities from the actual genuine traffic. 
However, this method only works against a passive network adversary and protects only traffic rate metadata such as packet times and sizes. This approach needs to be used with other methods to protect the categorical metadata such as protocol, IP address, and DNS hostnames. Likewise, the bandwidth overheads required to reduce the adversary confidence varies with respect to the type of device being protected. In fact, most of the existing traffic shaping techniques depend on effectively mimicking and realistic timing fake user activities.

\subsection{Command and Phonetic Analysis}
Here, we discuss countermeasures aiming at mitigating the issues of malicious skills. 
In particular, the skill vulnerabilities exploiting the interaction between the user and the third-party skill services.

Zhang et al.~\cite{2018arXiv180501525Z} present a system that examines the skill's response and the user's utterance to detect malicious skills that pretend to hand over control to another skill and deceive users into thinking that they are interacting with a different skill. The system relies on a User Intention Classifier (UIC) and a Skills Response Checker (SRC). The SRC semantically analyzes the skill response and compares it against utterances from a black-list of malicious skill responses to flag off any malicious response. While the user UIC, on the other hand, protects the user by checking their utterances to correctly determine their intents of context switches.\footnote{This is, examining the intents of changing from one task to the other.} This is done by matching the meaning of what the user says to the context of the skill the user is presently interacting with and also that of the system commands. They also consider the link between what the user says and the skill that they are currently using. UIC complements the SRC, and their system reports an overall detection precision rate of 95.60\%. Nevertheless, one key shortcoming of this system is the difficulty in implementing a generic UIC due to variation in Natural language-based command and how to distinguish legitimate commands. 

In a similar study, Kumar et al.~\cite{217575} suggests performing phonetic and text analysis for every new skill's invocation name to mitigate voice squatting attacks. They check whether the new skill's invocation name can be mistaken with an existing one, vetting then the creation of the clashing skill. Their solution is similar to what is currently being implemented during domain registration, where registrars do not register domain names that resemble that of popular domains.

\subsection {New Architecture}
In this section, we discuss a countermeasure that proposes a novel architecture for SPA, different from the one described in Section \ref{background-arch}.
In particular, we discuss the work proposed by Coucke et al.~\cite{DBLP:journals/corr/abs-1805-10190}, which proposes changes to the architecture, particularly in terms of the speech recognition functionality. 
Coucke et al.~\cite{DBLP:journals/corr/abs-1805-10190} present a \emph{privacy by design spoken Language Understanding platform} that does not send user queries to the cloud for processing. 
The speech recognition and the intent extraction are done locally on the SPA devices themselves using a partially trained model with \emph{crowd-sourced} data and using \emph{semi-supervised} learning. Many use cases do not need Internet access. However, when the use case requires internet access, such as when data needs to be retrieved or transmitted to an Internet service, then the system processes the data within the SPA device where it was generated rather than in the cloud. This makes it hard for an adversary to perform a mass attack as they can only target a single user or device at once. With such an infrastructure, issues related to \emph{always on always listening}, \emph{cloud}, and \emph{third-party access}, have limited impact since the data is processed locally. 
Besides, it allows personalizing the wake-up word, mitigating the wake-up word vulnerability introduced in Section~\ref{sec:issue-wake-up-words}. However, the platform requires a user to specify the skills on which their assistant will be trained on. Hence, such an assistant can only work within predefined scopes of the selected skills on which their model was trained, thereby restricting their capabilities to only those skills used for their training. It is important to also note that, although this infrastructure modifies the existing SPA architecture so that speech recognition and intent identification is conducted locally, it does not completely eliminate data transmission to other devices or cloud services. The SPA still communicates with other connected devices or cloud services depending on the context of use. This means that attacks like the one described in~\cite{8283484} may still be possible.

\section{Discussion and Open Challenges}
\label{sec:discussion}
Building on the analysis and categorization of the related literature studied in the previous sections, we then offer a synthesis and summary of this review and suggest future research areas.

One can easily observe in Table~\ref{table:1} that vulnerabilities related to weak authentication are the most exploited flaws. The \emph{wake-up word} and the \emph{always listening features} are typically combined and can be described as the gateway of synthesized speech attacks. No related works currently exploit the multiuser environment and external party access. We also observed that the majority of the attacks target point \#1 of the architecture: the point of interaction between the users and the SPA devices as it requires an attacker with lower capabilities. Although few attacks exploit more than one point of the architecture --- e.g.~\cite{2018arXiv180501525Z,217575,DBLP:journals/corr/abs-1711-03280}, none is observed at point \#5, point \#7 and \#8 even though attacks targeting those architectural elements seem possible as discussed in Section~\ref{sec:Attacks}. Similarly, Table~\ref{table:2} shows that countermeasures for \emph{weak authentication} vulnerabilities, and in particular countermeasures towards mitigating synthesized speech have received wide attention in the literature. Taking both Table~\ref{table:1} and Table~\ref{table:2}, we can see a concentration of research efforts towards one particular part of the whole SPA architecture, the direct interaction between the user and the smart speaker --- or point \#1 of the architecture. While indeed, this is an important part of the architecture, SPA should consider security in a holistic manner. 
This shows that despite the growing research efforts in security and privacy in SPA, we, as a community, also need to recognise and tackle SPA problems that go beyond that point of the architecture.
Based on our findings, we suggest a number of open challenges in SPA. 
These include: 
i) a practical evaluation of existing attacks and countermeasures, 
ii) making authentication and authorization stronger as well as smarter,
iii) building secure and privacy-aware speech recognition, 
iv) conducting systematic security and privacy assessments to understand the SPA eco-system and associated risks better, 
v) increasing user awareness and the usability of security and privacy mechanisms in SPA, and 
vi) understanding better profiling risks and potential countermeasures. 
All of which are discussed below in the following subsections.

\subsection{Practical Evaluation of Existing Attacks and Countermeasures}
We observed that many of the attacks target the underlying hardware of the voice infrastructure. For instance,~\cite{211283} and~\cite{zhang_yan_ji_zhang_zhang_xu_2017} use high frequencies signal to attack the non-linearity in SPA devices microphones. While some of these attacks synthesize speech in a way that may be intelligible to humans and easily noticed by users in proximity~\cite{Lei_Xinyu}, other attacks synthesize speech in a way that is unintelligible to the users~\cite{zhang_yan_ji_zhang_zhang_xu_2017,211283}.
Thus, one could argue that the second type of attack is more likely to be successful in practice than the first type. Our study also revealed that many of the attacks require different domain-specific knowledge to be successful, which might not always be available. 
For example, attacks conducted in~\cite{211283,191968,DBLP:journals/corr/abs-1711-03280} need knowledge of the machine classifiers, while the one demonstrated in~\cite{2018arXiv180501525Z} requires the understanding of the SPA skills invocation model.
In some cases, this knowledge is available or can be reverse-engineered from interactions with the SPA and their architecture. 
However, beyond these observations that we can derive from a literature review, some important questions remain unanswered, such as: 1) What is the severity of the existing attacks? 2) What is the likelihood of success of these attacks in practice? 3) What is the cost associated with existing attacks and countermeasures? 4) What is the effectiveness of these countermeasures? and 5) How usable are these countermeasures?

\subsection {Making Authentication Stronger} 
Despite receiving most of the attention in terms of countermeasures, with some of the issues and attacks having a counterpart countermeasure, weak authentication issues have not yet been completely addressed. 
As discussed earlier, many of the attacks targeting the SPA system exploit its weak authentication, especially the \emph{always on, always listening features}.
This attack is usually combined with other vulnerabilities.  Although one could say that the \emph{always on, always listening features} improve the responsiveness of the devices by making resources available to the user before they start uttering commands, the security and privacy risks may outweigh the benefit. Several independent input variables such as voice, video, manual gestures, touch, graphics, gaze, and others like the solution proposed in~\cite{kepuska_bohouta_2018} could be combined to make authentication stronger. 
However, most SPA are designed without environmental sensors. The lack of environmental sensors makes it challenging to implement context-aware authentication systems that could sense the physical environment, and leverage such information to adjust the security parameters accordingly. Also, there may be privacy issues and concerns when using even more personal information (e.g., video). 
Likewise, current authentication mechanisms in integrated technologies like other smart home devices are decentralized. Each integrated technology has its own authentication mechanism. By implementing a centralized mechanism, potentially in an SPA, a user could access multiple integrated technologies by authenticating only once. 
This would enhance usability by lessening the authentication burden on users and improving security as it would ensure consistent authentication across smart home devices. However, this needs to be implemented carefully so as not to create a single point of failure.

Future research can also consider how communication protocols may improve current authentication mechanisms in SPA. There are examples of how these mechanisms can be used in other systems such as remote car unlocking and contactless payment, where they are becoming an effective way to verify users' presence~\cite{Avoine:2018:SDS:3271482.3264628}. Popular among them are the distance-bounding protocols, which can be used to authenticate the user and access their location. These protocols have proven to be practicable especially in a system that is susceptible to distance-based frauds. Distance-bounding protocols are based on timing the delay between when a verifier sends a challenge to the moment the response is received. 
This allows the verifier to detect a third-party interference as any sudden delay in the proper response, which is considered to be the result of a delay due to a long-distance transmissions~\cite{rios2018mob,Avoine:2018:SDS:3271482.3264628}. 
Nevertheless, the effectiveness of this protocol depends on getting the correct propagation time.

\subsection{Enhanced Authorization Models and Mechanisms} 
\label{subsec:discussion-authorization}
More flexible access control and authorization models and mechanisms are needed. 
These mechanisms should be able to dynamically authorize and adapt permissions to users based on the current context and their preferences. 
According to a recent study, users preferred authorization policies in smart homes are affected by some distinct factors~\cite{He:2018:RAC:3277203.3277223}: i) the capabilities within a single device, ii) who is trying to use that capability, iii) and the context of use. Hence, designing authorization models that consider SPA capabilities and the context of use may help create authorization rules that adequately balance security, privacy, and functionality. In fact, similar models have already been implemented successfully in other domains like smartphones~\cite{ 5337017}. Furthermore, we have observed that SPA requires more fine-grained authorization mechanisms. This not only applies to the voice of the user itself, but also to the data that can be obtained from how users interact with the devices. In particular, these interactions can be used to infer, for instance, the sleeping patterns of a user, as discussed earlier.

Novel authorization models and mechanisms for SPA should consider not only single users but also multiple users. However, there are no security and privacy mechanisms for SPA that considers \emph{multi-user environment} issues. This is important, as even if SPA would support multiple accounts, it is a common practice to share accounts between multiple users~\cite{matthews2016she} (especially if one of the accounts has more privileges). The lack of proper authorization can prompt insider misuse, e.g., members of the household spying on their partners~\cite{freed2018stalker}, which can be particularly problematic in the case of intimate partner abuse~\cite{matthews2017security}. 
Moreover, smart home data is relational and it usually refers to a group of people collectively~\cite{databox}, e.g., if there is a way to infer whether there is someone at home or not, this already gives information that can be sensitive to everyone living there. Some general-purpose smart home privacy-enhancing IoT infrastructures like the Databox~\cite{databox} recognize the multiuser problem but no solution has been proposed yet in general for smart homes or in particular for multiuser sharing management in SPA. 
A great deal of research on methods and tools to help users manage data sharing in multiuser and multiparty scenarios have been proposed for social media (see~\cite{such2018multiparty} for a survey), and particular methods for detecting and resolving multiuser data sharing conflicts, such as~\cite{such2016resolving}, could be adapted from there or used to inspire multiuser solutions for the SPA case.

Furthermore, the existing SPA architecture supports only permission-based access control on sensitive data, which is insufficient at controlling how third-party skills use data once they get access. Future research should study how to implement a framework that allows users to pronounce their intended data flow patterns. Similar frameworks ~\cite{197137, inproceedings} have been successfully applied in smartphones for IoT apps. Also, there is a lack of authorization frameworks for data generated during user interactions with a third-party skill, which is one of the personal data assets mentioned in Section~\ref{Subsec:Assets}. Novel authorization mechanisms that allow users to specify, monitor and control what data can be shared with those that have no direct access to the SPA architecture, under what condition should the data be shared (reason), how it should be shared (means) and what it can be used for (purpose) could also help address the issue of external parties.

\subsection{Secure and Privacy-aware Speech Recognition}

NLP and ML models are used in conjunction for speech recognition. Protecting these models against manipulation, e.g., through well-crafted adversarial inputs as pointed out in Section~\ref{sec:issues:ai}, becomes paramount.
It is apparent from Table~\ref{table:1} and Table~\ref{table:2} above that there are many attacks exploiting adversarial ML and NLP issues, and there are substantially more attacks than defenses studied in the related literature. SPA providers need to consider adversarial examples when developing their speech recognition models. However, that is not an easy task, and more research is required in this direction. Some existing countermeasures used in other domains such as adversarial training and distillation could help to develop robust ML models for speech recognition in SPA, but they can be defeated using black-box attacks or attacks that are constructed on iterative optimization~\cite{DBLP:journals/corr/CarliniW16a}. Also, validating the input and reprocessing it to eliminate possible adversarial manipulations before it is fed to the model is a countermeasure that greatly depends on the domain, and is subjected to environmental factors~\cite{papernot2016towards}. Likewise, testing is not enough to secure ML, as an adversary can use a different input from those used for the testing process~\cite{goodfellow2018making}. 

Furthermore, the performance of the current speech recognition system still deserves improvement as shown earlier --- recall that these systems often find it difficult to i) understand words with similar phonemes~\cite{217575}, ii) understand different but similar words, and iii) resolve variation in natural language-based command words~\cite{zhang_yan_ji_zhang_zhang_xu_2017}. Since the word error rate (WER) is the common metric used for evaluating the performance of automatic speech recognition systems~\cite{DBLP:conf/nips/CisseANK17}, it may be easy for an adversary to craft an adversarial input that could maximize the WER of the speech recognition system by exploiting the NLP framework and the ML techniques. This is shown in~\cite{zhang_yan_ji_zhang_zhang_xu_2017}, where the speech recognition system is exploited to manipulate the intent the system understands from the user's command.

Beyond security, obtaining valuable information from big data while still protecting user's privacy has become interesting research in data analysis. While SPA providers let users review and delete their voice recordings, a recent study shows that users are unaware (or do not use) those privacy controls ~\cite{Lau:2018:AYL:3290265.3274371}.
It is also unclear how effective these controls actually are even if used, e.g.: these controls allow the user to delete particular raw utterances but they cannot delete what could be inferred from them (i.e., the model) \cite{alexa-delete}. In light of this, SPA vendors need to understand the privacy challenges of machine learning. For instance, although most existing SPA providers aim to ensure privacy while processing users' voice in the cloud, that is a difficult endeavor with current SPA architectures. With edge computing gradually coming into the limelight, data can now be processed locally, where it is generated, rather than being transmitted to a centralized data processing centre~\cite{roman2018VIS}. This helps to reduce the current dependency on the Internet and eliminate the necessity of putting sensitive data into the cloud. While related work~\cite{DBLP:journals/corr/abs-1805-10190} addresses this direction with a decentralized voice processing platform, it is challenging to build a general-purpose SPA using such platforms.
This is because SPA developed with such platforms can only work within predefined scopes of the selected skills on which their model was trained. Therefore, there is a need for future efforts on how to make voice processing privacy-preserving without hindering SPA's capabilities effectively.

\subsection{AI-based Security and Privacy}
\label{AI-based Security and Privacy}
In addition to using AI techniques for SPA functionality, e.g., speech recognition, they could also be used to make SPA more secure and aid users in managing their privacy as they see fit. AI techniques would include not only data-driven techniques like ML but also knowledge-based techniques such as normative systems and argumentation, which have been successfully used to develop intelligent security and privacy methods in other domains~\cite{such2016intelligent,such2017privacy}. AI techniques could be used to address the issue of \emph{always on always listening} and\emph{ synthesized speech} under the weak authentication vulnerabilities. For instance, it could be applied to detect malicious commands being spoken to the SPA devices (i.e., to make authentication stronger and more resilient to attacks). Likewise, it could be used to solve the issue of \emph{multi-user authorization and over-privileged skills} by applying it to help primary users configure the permissions they grant to other users and third-parties skills respectively. Similar research has already been shown to detect intrusions \cite{corchado2011neural} and to help users in other domains like mobile App permission management~\cite{olejnik2017smarper} and Social Media privacy settings and data sharing~\cite{misra2017pacman}. As for the speech recognition, these ML-based methods need to be engineered considering adversarial cases~\cite{goodfellow2018making}. 

Examples of the use of knowledge-based AI techniques include the use of norms, which have been widely explored in recent years, especially to reduce the autonomy of autonomous and intelligent systems to conform to decent behaviors \cite{criado2011open}. Norms are usually delineated formally using deontic logic to state what is permissible, obligatory, and prohibited, providing a rich framework to express context-dependent policies, e.g., based on Contextual Integrity~\cite{nissenbaum2004privacy}, and they can be defined, verified, and monitored for socio-technical systems like SPA \cite{kafaly2016revani,criado2016selective}. For instance, norms would be beneficial to avoid issues like the case discussed in~\cite{alexa-guardian} where a private conversation is recorded by an Alexa and forwarded to a random contact, as a norm could specify the type of conversations that may or may not be shared with particular contacts and that norm could be verified and monitored for compliance. Another example is norms that govern multiuser interactions with the SPA as discussed in Section~\ref{subsec:discussion-authorization}. Norms for SPA could be elicited automatically as in~\cite{criado2015implicit} or by crowd-sourcing the acceptable flows of information as in~\cite{fogues2017sharing}. Another knowledge-based AI technique like automated negotiation~\cite{baarslag2016negotiation,such2016privacy} could be used to help SPA users navigate the trade-offs and negotiate consent in the very complex SPA ecosystem, including third-party skills and smart devices. For instance, instead of having the user manually inspecting and approving every permission for the many third-party skills that may request them (as it happens now in SPA ecosystems like Amazon Alexa and Google Home), the SPA could automatically negotiate those permissions with the third-party skills. This can be done, however, always in a way in which consent could be revocable and access patterns apparent to the user on-demand, allowing reactive and dynamic data sharing adjustment. Finally, other AI techniques like computational trust~\cite{pinyol2013computational} could be used to choose and only share data with third-party skills and smart devices that are privacy-respecting and trustworthy.

\subsection{Systematic Security and Privacy Assessments}
\label{subsec:Systematic_Security_and_Privacy_Assessments}
SPA are a type of cyber-physical system. Previous research looked at how the assurance techniques and testing methodologies most commonly used in conventional IT systems~\cite{prandini2010towards} apply to cyber-physical systems, including penetration testing, static \& dynamic analysis, fuzzing, and formal verification.
However, it is still unclear how these security testing techniques apply to the SPA system and what are the practices used by third-party developers in this ecosystem. 
Assurance techniques are known to have different cost-effectiveness in practice~\cite{such}, and that cost-effectiveness for one very same assurance technique has been shown to vary across different cyber-physical systems~\cite{asadollah2015survey}, such as Industrial Control Systems~\cite{knowles2015assurance}. 
Therefore, a direction for future research is to study and evaluate how these assurance techniques will perform for the case of SPA and whether or not SPA's unique features like voice recognition and its integration with other technologies like the cloud and other smart devices require novel techniques or methodologies. For instance, the known potential to have composite vulnerabilities that exploit both the physical and the IT part of cyber-physical systems~\cite{ciholascomposite,ciholas2019security} has already been shown to also apply to SPA, e.g.,~\cite{211283}. 
Additionally, authors in~\cite{zhang_yan_ji_zhang_zhang_xu_2017} show that physical properties can be used to compromise the SPA by using high frequencies signals to attack the non-linearity in SPA devices microphones as detailed above in Section~\ref{sec:User-SPA-attack}.
A set of key research questions to answer revolve around which assurance techniques can be used to improve security in SPA systems (see Appendix A in~\cite{knowles2015assurance}). In particular: 1) Can a review of standards and procedures be used to mitigate security risks in SPA systems? 2) Can we run dynamic analysis techniques over components of the SPA architecture? and 3) Can we devise a methodology to provide an independent validation when many components of an SPA system are hosted in the cloud?

Future work should also look at the best and most systematic way to conduct privacy assessments in SPA~\cite{wright2012introduction}. 
However, it remains unclear how many privacy violations there are in the wild of the third-party ecosystem and what is the extent of such violations. 
Measuring privacy violations systematically is particularly challenging as privacy policies are usually unstructured data. Thus, it is hard to infer properties from them automatically. 
Of particular interest might be to study the (extent of) traceability between the actions of the data specified in privacy policies, such as those in the privacy policies of the third-party skills developers in SPA, and the related data operations obvious to users via SPA and/or associated smartphone interfaces, which will also be crucial to help tackle the current \emph{weak authorization} and \emph{profiling} issues of SPA. 
One important research question is whether related works could be adopted to measure policy traceability in the SPA domain.
Methodologies could be adapted from the social media~\cite{anthonysamy2013social} and smartphone apps~\cite{misra2017privacy}, which already showed the extent of traceability in these domains, 
together with methods to help developers automatically map traceability between policies and operationalized controls and maintain it through the development cycle~\cite{anthonysamy2016inferring}. 
As real breaches happen (e.g., \cite{alexa-guardian}), methods to study whether there are gaps in security and privacy policies, such as \cite{kafali2017good} applied to SPA, would also be helpful.
Thus, a systematic study could measure how many privacy policies are complete and how many are broken for the third-party SPA ecosystem. 
Likewise, a longitudinal study is required to comprehend the SPA skill's ecosystem to understand the type of skills available, the capabilities they have, how they are being used, and who is behind them (number of third-party developers, etc.).
This will further ensure a better understanding of the different risks that the ecosystem presents and aid in formulating appropriate security and privacy policies for the users.

\subsection{Increasing User Awareness}
\label{subsec:awareness}
Although implementing a technical defensive measure might go a long way in mitigating some of the identified risks, effective countermeasures will be difficult without better user awareness. Research shows that the lack of awareness about data practices in smart home devices affect users' security and privacy practices~\cite{238321}.
Some SPA users are not very concerned when it comes to the security and privacy issues in SPA~\cite {Zeng:2017:EUS:3235924.3235931}, as they believe they are not valuable targets for attackers~\cite{238321}, or they simply exhibit inaccurate and incomplete mental models of the SPA ecosystem~\cite{Noura_Abdi_2019}. Therefore, it is essential that users understand the risks and threats present in the SPA ecosystem, including the assets that can be compromised and why they need protection for better risk management. Users should be well informed to adopt best practices and even understand what key steps they have to take when either their security or privacy is breached. One crucial way of keeping SPA users informed is to design usable privacy notice that helps them understand and manage their data in SPA, accompanied with usable security and privacy mechanisms (as discussed below in Section \ref{subsec:usability}). 
Privacy notices must be relevant, actionable and understandable as discussed in~\cite{schaub}, and their design should be considered along four main dimensions: 1) timing, when should a privacy notice be presented; 2) channel, how should the privacy notice be delivered; 3) modality, how the information should be conveyed; and 4) control, how choice options are integrated into the privacy notice. Another example would be leveraging the already discussed assessments in Section \ref{subsec:Systematic_Security_and_Privacy_Assessments}, in order to produce a white (or black) list of third-party skills based on the level of security and/or privacy they offer considering the results of the assessments.

\subsection{Usability of SPA Security and Privacy Mechanisms} 
\label{subsec:usability}

While users' awareness is crucial in understanding the system's risks, awareness without usable security and privacy controls mechanisms may not be effective in mitigating these risks. For instance, some SPA users, while aware of some risks, do not know how they can protect themselves~\cite{Noura_Abdi_2019}. In addition to knowing the mechanisms they could use to protect themselves (such as those to achieve a basic level of cyber hygiene~\cite{such2019basic} but in the SPA domain), users should be able to utilize any SPA security and privacy mechanisms in a convenient manner that does not affect usability or functionality of SPA. This is because convenience and connectivity are important concerns for smart home users, influence their perceptions and opinions, and their attitude towards external entities that design and regulate SPA~\cite{Zheng:2018:UPS:3290265.3274469}.
Nonetheless, these measures' primary concern is that they have an important impact on usability, as they clash with the sought ``hands-free'' experience when interacting with SPA. In some other cases where non-technical coping strategies may not be available, SPA users are merely avoiding the SPA functionality they perceive to be risky, e.g., some SPA users only create shopping lists through the SPA but buy the items using the traditional web interface as they perceive buying through the SPA as risky and do not know how to protect themselves~\cite{Noura_Abdi_2019}. 

From all the technical countermeasures that we surveyed in this article (see Section~\ref{sec:countermeasures}), the vast majority of them did not explicitly consider usability. What is worse, there were cases in which some potentially negative usability impacts introduced by the countermeasures were clearly apparent such as where users need to use a wearable device like in \cite{kepuska_bohouta_2018, feng_fawaz_shin_2017}, and where the SPA capability might be restricted~\cite{DBLP:journals/corr/abs-1805-10190}.
Future work should conduct rigorous and systematic studies of the usability of the countermeasures already proposed to assess how usable they really are. Beyond these usability studies of existing countermeasures, future work on SPA security and privacy mechanisms should also consider usability from the onset, not as an afterthought. 
For instance, 
novel SPA security and privacy mechanisms should avoid requiring extensive user involvement. Otherwise, it has been shown they may not be used~\cite{zeng2019understanding}.
A potential avenue to explore as future work regarding this example could be the AI-based techniques discussed in Section~\ref{AI-based Security and Privacy}, which could be leveraged to predict user's preferences and help users set security and privacy controls much easier and with less involvement.

\subsection{Profiling Attacks and Defences} 
Regarding profiling, we can clearly see in Table~\ref{table:1} that few attacks have been reported on this. 
Some of these attacks make some hard assumptions, like having access to all cloud data about a user through their user account. We believe that further research is needed to assess whether other types of more sophisticated profiling could be conducted with access to less information. Furthermore, the community needs to understand whether tracking, which is pervasive across the web~\cite{mayer2012third}, could also apply and be feasible across the SPA ecosystem. In terms of defenses, we can also see in Table~\ref{table:2} the lack of work in this area. 
Some of the challenges we mentioned before would indeed help alleviate profiling such as user awareness and usable controls (Sections~\ref{subsec:awareness} and~\ref{subsec:usability}), systematic privacy assessments (Section~\ref{subsec:Systematic_Security_and_Privacy_Assessments}), and knowledge-based AI techniques to express/verify norms about how data are collected and use of data across the SPA ecosystem (Section~\ref{AI-based Security and Privacy}). However, other open challenges would remain, and profiling-specific countermeasures are also needed. For instance, SPA traffic needs to be properly obfuscated and masked to encode user's interaction with the devices in addition to the existing encryption mechanisms already in place. Note that current encryption mechanisms are not sufficient to avoid traffic profiling as shown in~\cite{DBLP:journals/corr/ApthorpeRF17}. Beyond differential-private approaches like the countermeasure introduced earlier \cite{8278156}, one possible avenue would be to adapt existing mechanisms to the case of SPA, such as traffic morphing techniques~\cite{morphing09} to prevent statistical traffic analysis. This can be done by altering one category of traffic to look like another one. However, this and most other existing traffic analysis countermeasures are vulnerable as they only obfuscate exposed features of the traffic by muffling this features and adding dummy packets. Thus, they are unable to prevent the leakage of many identifying information~\cite{6234422}. Another avenue could be based on mix networks~\cite{Paul_Syverson} and/or onion routing~\cite{Murdoch:2007:STA:1779330.1779341}. However, both of them may also be vulnerable to attack. For instance, mixing is susceptible to long term correlation and sleeper attacks~\cite{Paul_Syverson}, and onion routing is susceptible to an adversary correlating the traffic~\cite{Paul_Syverson_Entropist} and to misconfigured and malicious relays~\cite{DBLP:journals/corr/KadianakisRRW17}.

\section{Conclusions}
\label{sec:conclusions}
This paper analyzes and classifies the security and privacy issues associated with SPA and how a range of malicious actors can exploit them to harm the security and privacy of end-users. 
We have shown that the attack surface of this increasingly popular technology is vast. 
We have noted that the interaction between the users and the SPA devices is currently the weakest link. 
However, we have identified a wide range of attacks that can put users at stake. In as much as there is no single panacea solution for all security issues, the proper understanding of security pitfalls will go a long way in enabling manufacturers, researchers, and developers to design and implement robust security control measures. 
Although there is already very active research on securing intelligent assistants, few of the approaches consider the whole picture of the complex architecture SPA have.
We particularly highlighted open challenges for future research that we deem of critical importance, including making authentication stronger, enhancing authorization models and mechanisms, building secure and privacy-aware speech recognition, conducting systematic security and privacy assessments, developing AI-based security and privacy countermeasures, improving user awareness and usability, and studying further profiling attacks and defenses.

As future work, we would like to keep on expanding our understanding of the different open challenges presented above. While we included all available articles at the time obtained through the method stated earlier, SPA security and privacy is a fast-moving field still in its infancy. We hope this survey serves researchers to help prioritize the most promising areas to improve our understanding of attacks on SPA and to devise usable ways to counter them. Also, most of the literature we found focused on the two most popular SPAs--- Amazon Alexa and Google Assistant. However, there are many other SPAs (e.g., Microsoft Cortana). Even though they may have a similar architecture, there may be specific issues with them not covered in this article, so expanding our current article in this regard would also be an exciting line of future work.

\begin{acks}
The first author would like to thank the Federal Government of Nigeria through Petroleum Technology Development Fund (PTDF) for funding his Ph.D. at King's College London.
\end{acks}

\bibliographystyle{ACM-Reference-Format}
\bibliography{name}


\begin{thebibliography}{157}


\ifx \showCODEN    \undefined \def \showCODEN     #1{\unskip}     \fi
\ifx \showDOI      \undefined \def \showDOI       #1{#1}\fi
\ifx \showISBNx    \undefined \def \showISBNx     #1{\unskip}     \fi
\ifx \showISBNxiii \undefined \def \showISBNxiii  #1{\unskip}     \fi
\ifx \showISSN     \undefined \def \showISSN      #1{\unskip}     \fi
\ifx \showLCCN     \undefined \def \showLCCN      #1{\unskip}     \fi
\ifx \shownote     \undefined \def \shownote      #1{#1}          \fi
\ifx \showarticletitle \undefined \def \showarticletitle #1{#1}   \fi
\ifx \showURL      \undefined \def \showURL       {\relax}        \fi
\providecommand\bibfield[2]{#2}
\providecommand\bibinfo[2]{#2}
\providecommand\natexlab[1]{#1}
\providecommand\showeprint[2][]{arXiv:#2}

\bibitem[\protect\citeauthoryear{Abdi, Ramokapane, and Such}{Abdi
  et~al\mbox{.}}{2019}]%
        {Noura_Abdi_2019}
\bibfield{author}{\bibinfo{person}{Noura Abdi}, \bibinfo{person}{Kopo~M.
  Ramokapane}, {and} \bibinfo{person}{Jose~M. Such}.}
  \bibinfo{year}{2019}\natexlab{}.
\newblock \showarticletitle{More than Smart Speakers: Security and Privacy
  Perceptions of Smart Home Personal Assistants}. In
  \bibinfo{booktitle}{\emph{Fifteenth Symposium on Usable Privacy and Security
  ({SOUPS} 2019)}}. \bibinfo{publisher}{{USENIX} Association},
  \bibinfo{address}{Santa Clara, CA}.
\newblock
\urldef\tempurl%
\url{https://www.usenix.org/conference/soups2019/presentation/abdi}
\showURL{%
\tempurl}


\bibitem[\protect\citeauthoryear{Alepis and Patsakis}{Alepis and
  Patsakis}{2017}]%
        {alepis_patsakis_2017}
\bibfield{author}{\bibinfo{person}{Efthimios Alepis} {and}
  \bibinfo{person}{Constantinos Patsakis}.} \bibinfo{year}{2017}\natexlab{}.
\newblock \showarticletitle{Monkey Says, Monkey Does: Security and Privacy on
  Voice Assistants}.
\newblock \bibinfo{journal}{\emph{IEEE Access}}  \bibinfo{volume}{5}
  (\bibinfo{year}{2017}), \bibinfo{pages}{17841--17851}.
\newblock
\urldef\tempurl%
\url{https://doi.org/10.1109/access.2017.2747626}
\showDOI{\tempurl}


\bibitem[\protect\citeauthoryear{Amazon}{Amazon}{2017}]%
        {Alexa_Voice_profile}
\bibfield{author}{\bibinfo{person}{Amazon}.} \bibinfo{year}{2017}\natexlab{}.
\newblock \bibinfo{title}{{About Alexa Voice Profiles}}.
\newblock
  \bibinfo{howpublished}{\url{https://www.amazon.com/gp/help/customer/display.html?nodeId=202199440}}.
\newblock
\newblock
\shownote{[Online; last accessed 20-February-2019].}


\bibitem[\protect\citeauthoryear{Amazon}{Amazon}{2018}]%
        {alexa_developer}
\bibfield{author}{\bibinfo{person}{Amazon}.} \bibinfo{year}{2018}\natexlab{}.
\newblock \bibinfo{title}{{The Alexa Skill Store for France is a Fast Growing
  Land of Opportunity}}.
\newblock \bibinfo{howpublished}{\url{
  https://developer.amazon.com/docs/ask-overviews/understanding-the-different-types-of-skills.html}}.
\newblock
\newblock
\shownote{[Online; last accessed 29-December-2018].}


\bibitem[\protect\citeauthoryear{Amazon}{Amazon}{2019a}]%
        {alexa-display}
\bibfield{author}{\bibinfo{person}{Amazon}.} \bibinfo{year}{2019}\natexlab{a}.
\newblock \bibinfo{title}{{All-new Echo Show (2nd Gen)}}.
\newblock
  \bibinfo{howpublished}{\url{https://www.amazon.com/All-new-Echo-Show-2nd-Gen/dp/B077SXWSRP}}.
\newblock
\newblock
\shownote{[Online; last accessed 7-January-2019].}


\bibitem[\protect\citeauthoryear{Amazon}{Amazon}{2019b}]%
        {Amazon_Permissions}
\bibfield{author}{\bibinfo{person}{Amazon}.} \bibinfo{year}{2019}\natexlab{b}.
\newblock \bibinfo{title}{{Configure Permissions for Customer Information in
  Your Skill}}.
\newblock \bibinfo{howpublished}{\url{
  https://developer.amazon.com/en-US/docs/alexa/custom-skills/configure-permissions-for-customer-information-in-your-skill.html}}.
\newblock
\newblock
\shownote{[Online; last accessed 21-April-2020].}


\bibitem[\protect\citeauthoryear{Amazon}{Amazon}{nd}]%
        {Understand_Skills}
\bibfield{author}{\bibinfo{person}{Amazon}.} \bibinfo{year}{n.d}\natexlab{}.
\newblock \bibinfo{title}{{Understand How Users Interact with Skills}}.
\newblock
  \bibinfo{howpublished}{\url{https://developer.amazon.com/en-GB/docs/alexa/ask-overviews/understanding-how-users-interact-with-skills.html}}.
\newblock
\newblock
\shownote{[Online; last accessed 21-February-2020].}


\bibitem[\protect\citeauthoryear{Anthonysamy, Edwards, Weichel, and
  Rashid}{Anthonysamy et~al\mbox{.}}{2016}]%
        {anthonysamy2016inferring}
\bibfield{author}{\bibinfo{person}{Pauline Anthonysamy},
  \bibinfo{person}{Matthew Edwards}, \bibinfo{person}{Chris Weichel}, {and}
  \bibinfo{person}{Awais Rashid}.} \bibinfo{year}{2016}\natexlab{}.
\newblock \showarticletitle{Inferring semantic mapping between policies and
  code: the clue is in the language}. In
  \bibinfo{booktitle}{\emph{International Symposium on Engineering Secure
  Software and Systems}}. \bibinfo{publisher}{Springer},
  \bibinfo{pages}{233--250}.
\newblock


\bibitem[\protect\citeauthoryear{Anthonysamy, Greenwood, and
  Rashid}{Anthonysamy et~al\mbox{.}}{2013}]%
        {anthonysamy2013social}
\bibfield{author}{\bibinfo{person}{Pauline Anthonysamy}, \bibinfo{person}{Phil
  Greenwood}, {and} \bibinfo{person}{Awais Rashid}.}
  \bibinfo{year}{2013}\natexlab{}.
\newblock \showarticletitle{Social networking privacy: Understanding the
  disconnect from policy to controls}.
\newblock \bibinfo{journal}{\emph{Computer}} \bibinfo{volume}{46},
  \bibinfo{number}{6} (\bibinfo{year}{2013}), \bibinfo{pages}{60--67}.
\newblock


\bibitem[\protect\citeauthoryear{Apthorpe, Huang, Reisman, Narayanan, and
  Feamster}{Apthorpe et~al\mbox{.}}{2019}]%
        {apthorpe_huang_reisman_narayanan_feamster_2019}
\bibfield{author}{\bibinfo{person}{Noah Apthorpe},
  \bibinfo{person}{Danny~Yuxing Huang}, \bibinfo{person}{Dillon Reisman},
  \bibinfo{person}{Arvind Narayanan}, {and} \bibinfo{person}{Nick Feamster}.}
  \bibinfo{year}{2019}\natexlab{}.
\newblock \showarticletitle{Keeping the Smart Home Private with Smart(er) IoT
  Traffic Shaping}.
\newblock \bibinfo{journal}{\emph{Proceedings on Privacy Enhancing
  Technologies}} \bibinfo{volume}{2019}, \bibinfo{number}{3}
  (\bibinfo{year}{2019}), \bibinfo{pages}{128--148}.
\newblock
\urldef\tempurl%
\url{https://doi.org/10.2478/popets-2019-0040}
\showDOI{\tempurl}


\bibitem[\protect\citeauthoryear{Apthorpe, Reisman, and Feamster}{Apthorpe
  et~al\mbox{.}}{2017}]%
        {DBLP:journals/corr/ApthorpeRF17}
\bibfield{author}{\bibinfo{person}{Noah Apthorpe}, \bibinfo{person}{Dillon
  Reisman}, {and} \bibinfo{person}{Nick Feamster}.}
  \bibinfo{year}{2017}\natexlab{}.
\newblock \showarticletitle{A Smart Home is No Castle: Privacy Vulnerabilities
  of Encrypted IoT Traffic}.
\newblock \bibinfo{journal}{\emph{CoRR}}  \bibinfo{volume}{abs/1705.06805}
  (\bibinfo{year}{2017}).
\newblock
\showeprint[arxiv]{1705.06805}
\urldef\tempurl%
\url{http://arxiv.org/abs/1705.06805}
\showURL{%
\tempurl}


\bibitem[\protect\citeauthoryear{Asadollah, Inam, and Hansson}{Asadollah
  et~al\mbox{.}}{2015}]%
        {asadollah2015survey}
\bibfield{author}{\bibinfo{person}{Sara~Abbaspour Asadollah},
  \bibinfo{person}{Rafia Inam}, {and} \bibinfo{person}{Hans Hansson}.}
  \bibinfo{year}{2015}\natexlab{}.
\newblock \showarticletitle{A survey on testing for cyber physical system}. In
  \bibinfo{booktitle}{\emph{IFIP International Conference on Testing Software
  and Systems}}. \bibinfo{publisher}{Springer}, \bibinfo{pages}{194--207}.
\newblock


\bibitem[\protect\citeauthoryear{{Ava Mutchler}}{{Ava Mutchler}}{2018}]%
        {google_Voicebot.ai}
\bibfield{author}{\bibinfo{person}{{Ava Mutchler}}.}
  \bibinfo{year}{2018}\natexlab{}.
\newblock \bibinfo{title}{Google Assistant App Total Reaches Nearly 2400}.
\newblock \bibinfo{howpublished}{\url{
  https://voicebot.ai/2018/01/24/google-assistant-app-total-reaches-nearly-2400-thats-not-real}}.
\newblock
\newblock
\shownote{[Online; last accessed 22-December-2018].}


\bibitem[\protect\citeauthoryear{Avoine, Bingol, Boureanu, capkun, Hancke,
  Kardas, Kim, Lauradoux, Martin, Munilla, Peinado, Rasmussen, Singelee,
  Tchamkerten, Trujillo-Rasua, and Vaudenay}{Avoine et~al\mbox{.}}{2018}]%
        {Avoine:2018:SDS:3271482.3264628}
\bibfield{author}{\bibinfo{person}{Gildas Avoine},
  \bibinfo{person}{Muhammed~Ali Bingol}, \bibinfo{person}{Ioana Boureanu},
  \bibinfo{person}{Srdjan capkun}, \bibinfo{person}{Gerhard Hancke},
  \bibinfo{person}{Suleyman Kardas}, \bibinfo{person}{Chong~Hee Kim},
  \bibinfo{person}{Cedric Lauradoux}, \bibinfo{person}{Benjamin Martin},
  \bibinfo{person}{Jorge Munilla}, \bibinfo{person}{Alberto Peinado},
  \bibinfo{person}{Kasper~Bonne Rasmussen}, \bibinfo{person}{Dave Singelee},
  \bibinfo{person}{Aslan Tchamkerten}, \bibinfo{person}{Rolando
  Trujillo-Rasua}, {and} \bibinfo{person}{Serge Vaudenay}.}
  \bibinfo{year}{2018}\natexlab{}.
\newblock \showarticletitle{Security of Distance-Bounding: A Survey}.
\newblock \bibinfo{journal}{\emph{ACM Comput. Surv.}} \bibinfo{volume}{51},
  \bibinfo{number}{5}, Article \bibinfo{articleno}{94} (\bibinfo{date}{Sept.}
  \bibinfo{year}{2018}), \bibinfo{numpages}{33}~pages.
\newblock
\showISSN{0360-0300}
\urldef\tempurl%
\url{https://doi.org/10.1145/3264628}
\showDOI{\tempurl}


\bibitem[\protect\citeauthoryear{Baarslag, Alan, Gomer, Liccardi, Marreiros,
  Gerding, et~al\mbox{.}}{Baarslag et~al\mbox{.}}{2016}]%
        {baarslag2016negotiation}
\bibfield{author}{\bibinfo{person}{T. Baarslag}, \bibinfo{person}{A.~T. Alan},
  \bibinfo{person}{R.~C. Gomer}, \bibinfo{person}{I. Liccardi},
  \bibinfo{person}{H. Marreiros}, \bibinfo{person}{E. Gerding},
  {et~al\mbox{.}}} \bibinfo{year}{2016}\natexlab{}.
\newblock \showarticletitle{Negotiation as an interaction mechanism for
  deciding app permissions}. In \bibinfo{booktitle}{\emph{Proc. of CHI Extended
  Abstracts}}. \bibinfo{pages}{2012--2019}.
\newblock


\bibitem[\protect\citeauthoryear{Carlini, Mishra, Vaidya, Zhang, Sherr,
  Shields, Wagner, and Zhou}{Carlini et~al\mbox{.}}{2016}]%
        {197215}
\bibfield{author}{\bibinfo{person}{Nicholas Carlini}, \bibinfo{person}{Pratyush
  Mishra}, \bibinfo{person}{Tavish Vaidya}, \bibinfo{person}{Yuankai Zhang},
  \bibinfo{person}{Micah Sherr}, \bibinfo{person}{Clay Shields},
  \bibinfo{person}{David Wagner}, {and} \bibinfo{person}{Wenchao Zhou}.}
  \bibinfo{year}{2016}\natexlab{}.
\newblock \showarticletitle{Hidden Voice Commands}. In
  \bibinfo{booktitle}{\emph{25th {USENIX} Security Symposium ({USENIX} Security
  16)}}. \bibinfo{publisher}{{USENIX} Association}, \bibinfo{address}{Austin,
  TX}, \bibinfo{pages}{513--530}.
\newblock
\showISBNx{978-1-931971-32-4}
\urldef\tempurl%
\url{www.usenix.org/conference/usenixsecurity16/technical-sessions/presentation/carlini}
\showURL{%
\tempurl}


\bibitem[\protect\citeauthoryear{Carlini and Wagner}{Carlini and
  Wagner}{2016}]%
        {DBLP:journals/corr/CarliniW16a}
\bibfield{author}{\bibinfo{person}{Nicholas Carlini} {and}
  \bibinfo{person}{David~A. Wagner}.} \bibinfo{year}{2016}\natexlab{}.
\newblock \showarticletitle{Towards Evaluating the Robustness of Neural
  Networks}.
\newblock \bibinfo{journal}{\emph{CoRR}}  \bibinfo{volume}{abs/1608.04644}
  (\bibinfo{year}{2016}).
\newblock
\showeprint[arxiv]{1608.04644}
\urldef\tempurl%
\url{http://arxiv.org/abs/1608.04644}
\showURL{%
\tempurl}


\bibitem[\protect\citeauthoryear{Carlini and Wagner}{Carlini and
  Wagner}{2018}]%
        {DBLP:conf/sp/Carlini018}
\bibfield{author}{\bibinfo{person}{Nicholas Carlini} {and}
  \bibinfo{person}{David~A. Wagner}.} \bibinfo{year}{2018}\natexlab{}.
\newblock \showarticletitle{Audio Adversarial Examples: Targeted Attacks on
  Speech-to-Text}. In \bibinfo{booktitle}{\emph{2018 {IEEE} Security and
  Privacy Workshops, {SP} Workshops 2018, San Francisco, CA, USA, May 24,
  2018}}. \bibinfo{pages}{1--7}.
\newblock
\urldef\tempurl%
\url{https://doi.org/10.1109/SPW.2018.00009}
\showURL{%
\tempurl}


\bibitem[\protect\citeauthoryear{Chen, Ren, Piao, Wang, Wang, Weng, Su, and
  Mohaisen}{Chen et~al\mbox{.}}{2017}]%
        {chen_ren_piao_wang_wang_weng_su_mohaisen_2017}
\bibfield{author}{\bibinfo{person}{Si Chen}, \bibinfo{person}{Kui Ren},
  \bibinfo{person}{Sixu Piao}, \bibinfo{person}{Cong Wang},
  \bibinfo{person}{Qian Wang}, \bibinfo{person}{Jian Weng}, \bibinfo{person}{Lu
  Su}, {and} \bibinfo{person}{Aziz Mohaisen}.} \bibinfo{year}{2017}\natexlab{}.
\newblock \showarticletitle{You Can Hear But You Cannot Steal:Defending Against
  Voice Impersonation Attacks on Smartphones}.
\newblock \bibinfo{journal}{\emph{2017 IEEE 37th International Conference on
  Distributed Computing Systems (ICDCS)}} (\bibinfo{year}{2017}).
\newblock
\urldef\tempurl%
\url{https://doi.org/10.1109/icdcs.2017.133}
\showDOI{\tempurl}


\bibitem[\protect\citeauthoryear{Chung and Lee}{Chung and Lee}{2018}]%
        {DBLP:journals/corr/abs-1803-00466}
\bibfield{author}{\bibinfo{person}{Hyunji Chung} {and} \bibinfo{person}{Sangjin
  Lee}.} \bibinfo{year}{2018}\natexlab{}.
\newblock \showarticletitle{Intelligent Virtual Assistant knows Your Life}.
\newblock \bibinfo{journal}{\emph{CoRR}}  \bibinfo{volume}{abs/1803.00466}
  (\bibinfo{year}{2018}).
\newblock
\showeprint[arxiv]{1803.00466}


\bibitem[\protect\citeauthoryear{Chung, Park, and Lee}{Chung
  et~al\mbox{.}}{2017}]%
        {chung_park_lee_2017}
\bibfield{author}{\bibinfo{person}{Hyunji Chung}, \bibinfo{person}{Jungheum
  Park}, {and} \bibinfo{person}{Sangjin Lee}.} \bibinfo{year}{2017}\natexlab{}.
\newblock \showarticletitle{Digital forensic approaches for Amazon Alexa
  ecosystem}.
\newblock \bibinfo{journal}{\emph{Digital Investigation}}  \bibinfo{volume}{22}
  (\bibinfo{year}{2017}), \bibinfo{pages}{S15 to S25}.
\newblock
\urldef\tempurl%
\url{https://doi.org/10.1016/j.diin.2017.06.010}
\showDOI{\tempurl}


\bibitem[\protect\citeauthoryear{Ciholas, Lennie, Sadigova, and Such}{Ciholas
  et~al\mbox{.}}{2019}]%
        {ciholas2019security}
\bibfield{author}{\bibinfo{person}{Pierre Ciholas}, \bibinfo{person}{Aidan
  Lennie}, \bibinfo{person}{Parvin Sadigova}, {and} \bibinfo{person}{Jose~M
  Such}.} \bibinfo{year}{2019}\natexlab{}.
\newblock \showarticletitle{The security of smart buildings: a systematic
  literature review}.
\newblock \bibinfo{journal}{\emph{arXiv preprint arXiv:1901.05837}}
  (\bibinfo{year}{2019}).
\newblock


\bibitem[\protect\citeauthoryear{Ciholas and Such}{Ciholas and Such}{2016}]%
        {ciholascomposite}
\bibfield{author}{\bibinfo{person}{Pierre Ciholas} {and}
  \bibinfo{person}{Jose~M Such}.} \bibinfo{year}{2016}\natexlab{}.
\newblock \showarticletitle{Composite vulnerabilities in Cyber Physical
  Systems}.
\newblock \bibinfo{journal}{\emph{Security and Resilience of Cyber--Physical
  Infrastructures}} (\bibinfo{year}{2016}), \bibinfo{pages}{4}.
\newblock


\bibitem[\protect\citeauthoryear{Cisse, Adi, Neverova, and Keshet}{Cisse
  et~al\mbox{.}}{2017}]%
        {DBLP:conf/nips/CisseANK17}
\bibfield{author}{\bibinfo{person}{Moustapha Cisse}, \bibinfo{person}{Yossi
  Adi}, \bibinfo{person}{Natalia Neverova}, {and} \bibinfo{person}{Joseph
  Keshet}.} \bibinfo{year}{2017}\natexlab{}.
\newblock \showarticletitle{Houdini: Fooling Deep Structured Visual and Speech
  Recognition Models with Adversarial Examples}. In
  \bibinfo{booktitle}{\emph{Advances in Neural Information Processing Systems
  30: Annual Conference on Neural Information Processing Systems 2017, 4-9
  December 2017, Long Beach, CA, {USA}}}. \bibinfo{pages}{6980--6990}.
\newblock


\bibitem[\protect\citeauthoryear{Corchado and Herrero}{Corchado and
  Herrero}{2011}]%
        {corchado2011neural}
\bibfield{author}{\bibinfo{person}{Emilio Corchado} {and}
  \bibinfo{person}{{\'A}lvaro Herrero}.} \bibinfo{year}{2011}\natexlab{}.
\newblock \showarticletitle{Neural visualization of network traffic data for
  intrusion detection}.
\newblock \bibinfo{journal}{\emph{Applied Soft Computing}}
  \bibinfo{volume}{11}, \bibinfo{number}{2} (\bibinfo{year}{2011}),
  \bibinfo{pages}{2042--2056}.
\newblock


\bibitem[\protect\citeauthoryear{Coucke, Saade, Ball, Bluche, Caulier, Leroy,
  Doumouro, Gisselbrecht, Caltagirone, Lavril, Primet, and Dureau}{Coucke
  et~al\mbox{.}}{2018}]%
        {DBLP:journals/corr/abs-1805-10190}
\bibfield{author}{\bibinfo{person}{Alice Coucke}, \bibinfo{person}{Alaa Saade},
  \bibinfo{person}{Adrien Ball}, \bibinfo{person}{Theodore Bluche},
  \bibinfo{person}{Alexandre Caulier}, \bibinfo{person}{David Leroy},
  \bibinfo{person}{Clement Doumouro}, \bibinfo{person}{Thibault Gisselbrecht},
  \bibinfo{person}{Francesco Caltagirone}, \bibinfo{person}{Thibaut Lavril},
  \bibinfo{person}{Mael Primet}, {and} \bibinfo{person}{Joseph Dureau}.}
  \bibinfo{year}{2018}\natexlab{}.
\newblock \showarticletitle{Snips Voice Platform: an embedded Spoken Language
  Understanding system for private-by-design voice interfaces}.
\newblock \bibinfo{journal}{\emph{CoRR}}  \bibinfo{volume}{abs/1805.10190}
  (\bibinfo{year}{2018}).
\newblock


\bibitem[\protect\citeauthoryear{Criado, Argente, and Botti}{Criado
  et~al\mbox{.}}{2011}]%
        {criado2011open}
\bibfield{author}{\bibinfo{person}{Natalia Criado}, \bibinfo{person}{Estefania
  Argente}, {and} \bibinfo{person}{V Botti}.} \bibinfo{year}{2011}\natexlab{}.
\newblock \showarticletitle{Open issues for normative multi-agent systems}.
\newblock \bibinfo{journal}{\emph{AI communications}} \bibinfo{volume}{24},
  \bibinfo{number}{3} (\bibinfo{year}{2011}), \bibinfo{pages}{233--264}.
\newblock


\bibitem[\protect\citeauthoryear{Criado and Such}{Criado and Such}{2015}]%
        {criado2015implicit}
\bibfield{author}{\bibinfo{person}{Natalia Criado} {and}
  \bibinfo{person}{Jose~M Such}.} \bibinfo{year}{2015}\natexlab{}.
\newblock \showarticletitle{Implicit contextual integrity in online social
  networks}.
\newblock \bibinfo{journal}{\emph{Information Sciences}}  \bibinfo{volume}{325}
  (\bibinfo{year}{2015}), \bibinfo{pages}{48--69}.
\newblock


\bibitem[\protect\citeauthoryear{Criado and Such}{Criado and Such}{2016}]%
        {criado2016selective}
\bibfield{author}{\bibinfo{person}{Natalia Criado} {and}
  \bibinfo{person}{Jose~M Such}.} \bibinfo{year}{2016}\natexlab{}.
\newblock \showarticletitle{Selective Norm Monitoring.}. In
  \bibinfo{booktitle}{\emph{IJCAI}}. \bibinfo{pages}{208--214}.
\newblock


\bibitem[\protect\citeauthoryear{Cufoglu}{Cufoglu}{2014}]%
        {Ayse_Cufoglu}
\bibfield{author}{\bibinfo{person}{Ayse Cufoglu}.}
  \bibinfo{year}{2014}\natexlab{}.
\newblock \showarticletitle{User Profiling-A Short Review}.
\newblock \bibinfo{journal}{\emph{International Journal of Computer
  Applications (0975 8887)}} \bibinfo{volume}{108}, \bibinfo{number}{3}
  (\bibinfo{year}{2014}).
\newblock
\urldef\tempurl%
\url{https://research.ijcaonline.org/volume108/number3/pxc3900179.pdf}
\showURL{%
\tempurl}


\bibitem[\protect\citeauthoryear{Denning, Kohno, and Levy}{Denning
  et~al\mbox{.}}{2013}]%
        {Denning:2013:CSM:2398356.2398377}
\bibfield{author}{\bibinfo{person}{Tamara Denning}, \bibinfo{person}{Tadayoshi
  Kohno}, {and} \bibinfo{person}{Henry~M. Levy}.}
  \bibinfo{year}{2013}\natexlab{}.
\newblock \showarticletitle{Computer Security and the Modern Home}.
\newblock \bibinfo{journal}{\emph{Commun. ACM}} \bibinfo{volume}{56},
  \bibinfo{number}{1} (\bibinfo{date}{Jan.} \bibinfo{year}{2013}),
  \bibinfo{pages}{94--103}.
\newblock
\showISSN{0001-0782}
\urldef\tempurl%
\url{https://doi.org/10.1145/2398356.2398377}
\showDOI{\tempurl}


\bibitem[\protect\citeauthoryear{Developer}{Developer}{2019}]%
        {Google-SDK-Preview}
\bibfield{author}{\bibinfo{person}{Google Developer}.}
  \bibinfo{year}{2019}\natexlab{}.
\newblock \bibinfo{title}{{Developer Preview of Local Home SDK}}.
\newblock
  \bibinfo{howpublished}{\url{https://developers.googleblog.com/2019/07/developer-preview-of-local-home-sdk.html}}.
\newblock
\newblock
\shownote{[Online; last accessed 7-January-2020].}


\bibitem[\protect\citeauthoryear{Dyer, Coull, Ristenpart, and Shrimpton}{Dyer
  et~al\mbox{.}}{2012}]%
        {6234422}
\bibfield{author}{\bibinfo{person}{K.~P. Dyer}, \bibinfo{person}{S.~E. Coull},
  \bibinfo{person}{T. Ristenpart}, {and} \bibinfo{person}{T. Shrimpton}.}
  \bibinfo{year}{2012}\natexlab{}.
\newblock \showarticletitle{Peek-a-Boo, I Still See You: Why Efficient Traffic
  Analysis Countermeasures Fail}. In \bibinfo{booktitle}{\emph{2012 IEEE
  Symposium on Security and Privacy}}. \bibinfo{pages}{332--346}.
\newblock
\showISSN{2375-1207}
\urldef\tempurl%
\url{https://doi.org/10.1109/SP.2012.28}
\showDOI{\tempurl}


\bibitem[\protect\citeauthoryear{Easwara~Moorthy and Vu}{Easwara~Moorthy and
  Vu}{2015}]%
        {Easwara_Moorthy_2015}
\bibfield{author}{\bibinfo{person}{Aarthi Easwara~Moorthy} {and}
  \bibinfo{person}{L Vu}.} \bibinfo{year}{2015}\natexlab{}.
\newblock \showarticletitle{Privacy Concerns for Use of Voice Activated
  Personal Assistant in the Public Space}.
\newblock \bibinfo{journal}{\emph{International Journal of Human Computer
  Interaction}} \bibinfo{volume}{31}, \bibinfo{number}{4}
  (\bibinfo{date}{April} \bibinfo{year}{2015}), \bibinfo{pages}{307 to 335}.
\newblock
\urldef\tempurl%
\url{https://doi.org/10.1080/10447318.2014.986642}
\showDOI{\tempurl}


\bibitem[\protect\citeauthoryear{Feng, Fawaz, and Shin}{Feng
  et~al\mbox{.}}{2017}]%
        {feng_fawaz_shin_2017}
\bibfield{author}{\bibinfo{person}{Huan Feng}, \bibinfo{person}{Kassem Fawaz},
  {and} \bibinfo{person}{Kang~G. Shin}.} \bibinfo{year}{2017}\natexlab{}.
\newblock \showarticletitle{Continuous Authentication for Voice Assistants}.
\newblock \bibinfo{journal}{\emph{Proceedings of the 23rd Annual International
  Conference on Mobile Computing and Networking - MobiCom '17}}
  (\bibinfo{year}{2017}).
\newblock
\urldef\tempurl%
\url{https://doi.org/10.1145/3117811.3117823}
\showDOI{\tempurl}


\bibitem[\protect\citeauthoryear{Fernandes, Jung, and Prakash}{Fernandes
  et~al\mbox{.}}{2016a}]%
        {7546527}
\bibfield{author}{\bibinfo{person}{E. Fernandes}, \bibinfo{person}{J. Jung},
  {and} \bibinfo{person}{A. Prakash}.} \bibinfo{year}{2016}\natexlab{a}.
\newblock \showarticletitle{Security Analysis of Emerging Smart Home
  Applications}. In \bibinfo{booktitle}{\emph{2016 IEEE Symposium on Security
  and Privacy (SP)}}. \bibinfo{pages}{636--654}.
\newblock
\showISSN{2375-1207}
\urldef\tempurl%
\url{https://doi.org/10.1109/SP.2016.44}
\showDOI{\tempurl}


\bibitem[\protect\citeauthoryear{Fernandes, Paupore, Rahmati, Simionato, Conti,
  and Prakash}{Fernandes et~al\mbox{.}}{2016b}]%
        {197137}
\bibfield{author}{\bibinfo{person}{Earlence Fernandes}, \bibinfo{person}{Justin
  Paupore}, \bibinfo{person}{Amir Rahmati}, \bibinfo{person}{Daniel Simionato},
  \bibinfo{person}{Mauro Conti}, {and} \bibinfo{person}{Atul Prakash}.}
  \bibinfo{year}{2016}\natexlab{b}.
\newblock \showarticletitle{FlowFence: Practical Data Protection for Emerging
  IoT Application Frameworks}. In \bibinfo{booktitle}{\emph{25th {USENIX}
  Security Symposium ({USENIX} Security 16)}}. \bibinfo{publisher}{{USENIX}
  Association}, \bibinfo{address}{Austin, TX}, \bibinfo{pages}{531--548}.
\newblock
\showISBNx{978-1-931971-32-4}
\urldef\tempurl%
\url{https://www.usenix.org/conference/usenixsecurity16/technical-sessions/presentation/fernandes}
\showURL{%
\tempurl}


\bibitem[\protect\citeauthoryear{Fogues, Murukannaiah, Such, and Singh}{Fogues
  et~al\mbox{.}}{2017}]%
        {fogues2017sharing}
\bibfield{author}{\bibinfo{person}{R. Fogues}, \bibinfo{person}{P.~K.
  Murukannaiah}, \bibinfo{person}{J.~M. Such}, {and} \bibinfo{person}{M.~P.
  Singh}.} \bibinfo{year}{2017}\natexlab{}.
\newblock \showarticletitle{Sharing policies in multiuser privacy scenarios:
  Incorporating context, preferences, and arguments in decision making}.
\newblock \bibinfo{journal}{\emph{ACM TOCHI}} \bibinfo{volume}{24},
  \bibinfo{number}{1} (\bibinfo{year}{2017}), \bibinfo{pages}{5}.
\newblock


\bibitem[\protect\citeauthoryear{Freed, Palmer, Minchala, Levy, Ristenpart, and
  Dell}{Freed et~al\mbox{.}}{2018}]%
        {freed2018stalker}
\bibfield{author}{\bibinfo{person}{Diana Freed}, \bibinfo{person}{Jackeline
  Palmer}, \bibinfo{person}{Diana Minchala}, \bibinfo{person}{Karen Levy},
  \bibinfo{person}{Thomas Ristenpart}, {and} \bibinfo{person}{Nicola Dell}.}
  \bibinfo{year}{2018}\natexlab{}.
\newblock \showarticletitle{A Stalker's Paradise: How Intimate Partner Abusers
  Exploit Technology}. In \bibinfo{booktitle}{\emph{Proceedings of the 2018 CHI
  Conference on Human Factors in Computing Systems}}. \bibinfo{publisher}{ACM},
  \bibinfo{pages}{667}.
\newblock


\bibitem[\protect\citeauthoryear{Fruchter and Liccardi}{Fruchter and
  Liccardi}{2018}]%
        {Fruchter:2018:CAT:3170427.3188448}
\bibfield{author}{\bibinfo{person}{Nathaniel Fruchter} {and}
  \bibinfo{person}{Ilaria Liccardi}.} \bibinfo{year}{2018}\natexlab{}.
\newblock \showarticletitle{Consumer Attitudes Towards Privacy and Security in
  Home Assistants} \emph{(\bibinfo{series}{CHI EA '18})}.
  \bibinfo{publisher}{ACM}, \bibinfo{address}{New York, NY, USA}, Article
  \bibinfo{articleno}{LBW050}, \bibinfo{numpages}{6}~pages.
\newblock
\showISBNx{978-1-4503-5621-3}
\urldef\tempurl%
\url{https://doi.org/10.1145/3170427.3188448}
\showDOI{\tempurl}


\bibitem[\protect\citeauthoryear{Garimella, Mandal, Strom, Hoffmeister,
  Matsoukas, and Parthasarathi}{Garimella et~al\mbox{.}}{2015}]%
        {Garimella2015RobustIB}
\bibfield{author}{\bibinfo{person}{Sri Garimella}, \bibinfo{person}{Arindam
  Mandal}, \bibinfo{person}{Nikko Strom}, \bibinfo{person}{Bj{\"o}rn
  Hoffmeister}, \bibinfo{person}{Spyridon Matsoukas}, {and}
  \bibinfo{person}{Sree Hari~Krishnan Parthasarathi}.}
  \bibinfo{year}{2015}\natexlab{}.
\newblock \showarticletitle{Robust i-vector based adaptation of DNN acoustic
  model for speech recognition}. In \bibinfo{booktitle}{\emph{INTERSPEECH}}.
\newblock


\bibitem[\protect\citeauthoryear{Gong and Poellabauer}{Gong and
  Poellabauer}{2017}]%
        {DBLP:journals/corr/abs-1711-03280}
\bibfield{author}{\bibinfo{person}{Yuan Gong} {and} \bibinfo{person}{Christian
  Poellabauer}.} \bibinfo{year}{2017}\natexlab{}.
\newblock \showarticletitle{Crafting Adversarial Examples For Speech
  Paralinguistics Applications}.
\newblock \bibinfo{journal}{\emph{CoRR}}  \bibinfo{volume}{abs/1711.03280}
  (\bibinfo{year}{2017}).
\newblock
\showeprint[arxiv]{1711.03280}
\urldef\tempurl%
\url{http://arxiv.org/abs/1711.03280}
\showURL{%
\tempurl}


\bibitem[\protect\citeauthoryear{Goodfellow, McDaniel, and Papernot}{Goodfellow
  et~al\mbox{.}}{2018}]%
        {goodfellow2018making}
\bibfield{author}{\bibinfo{person}{Ian Goodfellow}, \bibinfo{person}{Patrick
  McDaniel}, {and} \bibinfo{person}{Nicolas Papernot}.}
  \bibinfo{year}{2018}\natexlab{}.
\newblock \showarticletitle{Making machine learning robust against adversarial
  inputs}.
\newblock \bibinfo{journal}{\emph{Commun. ACM}} \bibinfo{volume}{61},
  \bibinfo{number}{7} (\bibinfo{year}{2018}), \bibinfo{pages}{56--66}.
\newblock


\bibitem[\protect\citeauthoryear{Google}{Google}{2017}]%
        {Google_CNN}
\bibfield{author}{\bibinfo{person}{Google}.} \bibinfo{year}{2017}\natexlab{}.
\newblock \bibinfo{title}{{Set up multiple users for your speaker or smart
  display}}.
\newblock
  \bibinfo{howpublished}{\url{https://support.google.com/assistant/answer/9071681}}.
\newblock
\newblock
\shownote{[Online; last accessed 20-February-2018].}


\bibitem[\protect\citeauthoryear{Google}{Google}{2018a}]%
        {Google_developer}
\bibfield{author}{\bibinfo{person}{Google}.} \bibinfo{year}{2018}\natexlab{a}.
\newblock \bibinfo{title}{{Actions on Google}}.
\newblock \bibinfo{howpublished}{\url{
  https://developers.google.com/actions/samples/}}.
\newblock
\newblock
\shownote{[Online; last accessed 29-December-2018].}


\bibitem[\protect\citeauthoryear{Google}{Google}{2018b}]%
        {GOOGLE}
\bibfield{author}{\bibinfo{person}{Google}.} \bibinfo{year}{2018}\natexlab{b}.
\newblock \bibinfo{title}{{Invocation and Discovery}}.
\newblock
  \bibinfo{howpublished}{\url{https://developers.google.com/actions/sdk/invocation-and-discovery}}.
\newblock
\newblock
\shownote{[Online; last accessed 17-December-2018].}


\bibitem[\protect\citeauthoryear{Haack, Severance, Wallace, and Wohlwend}{Haack
  et~al\mbox{.}}{2017}]%
        {haack_williams}
\bibfield{author}{\bibinfo{person}{William Haack}, \bibinfo{person}{Madeleine
  Severance}, \bibinfo{person}{Michael Wallace}, {and} \bibinfo{person}{Jeremy
  Wohlwend}.} \bibinfo{year}{2017}\natexlab{}.
\newblock \showarticletitle{Security Analysis of Amazon Echo}.
\newblock  (\bibinfo{year}{2017}).
\newblock
\urldef\tempurl%
\url{https://courses.csail.mit.edu/6.857/2017/project/8.pdf}
\showURL{%
\tempurl}


\bibitem[\protect\citeauthoryear{Hannay, Sjoberg, and Dyba}{Hannay
  et~al\mbox{.}}{2007}]%
        {hannay_sjoberg_dyba_2007}
\bibfield{author}{\bibinfo{person}{Jo~E. Hannay}, \bibinfo{person}{Dag~I.K.
  Sjoberg}, {and} \bibinfo{person}{Tore Dyba}.}
  \bibinfo{year}{2007}\natexlab{}.
\newblock \showarticletitle{A Systematic Review of Theory Use in Software
  Engineering Experiments}.
\newblock \bibinfo{journal}{\emph{IEEE Transactions on Software Engineering}}
  \bibinfo{volume}{33}, \bibinfo{number}{2} (\bibinfo{year}{2007}),
  \bibinfo{pages}{87--107}.
\newblock
\urldef\tempurl%
\url{https://doi.org/10.1109/tse.2007.12}
\showDOI{\tempurl}


\bibitem[\protect\citeauthoryear{He, Golla, Padhi, Ofek, Durmuth, Fernandes,
  and Ur}{He et~al\mbox{.}}{2018}]%
        {He:2018:RAC:3277203.3277223}
\bibfield{author}{\bibinfo{person}{Weijia He}, \bibinfo{person}{Maximilian
  Golla}, \bibinfo{person}{Roshni Padhi}, \bibinfo{person}{Jordan Ofek},
  \bibinfo{person}{Markus Durmuth}, \bibinfo{person}{Earlence Fernandes}, {and}
  \bibinfo{person}{Blase Ur}.} \bibinfo{year}{2018}\natexlab{}.
\newblock \showarticletitle{Rethinking Access Control and Authentication for
  the Home Internet of Things (IoT)}. In \bibinfo{booktitle}{\emph{Proceedings
  of the 27th USENIX Conference on Security Symposium}} (Baltimore, MD, USA)
  \emph{(\bibinfo{series}{SEC'18})}. \bibinfo{publisher}{USENIX Association},
  \bibinfo{address}{Berkeley, CA, USA}, \bibinfo{pages}{255--272}.
\newblock
\showISBNx{978-1-931971-46-1}


\bibitem[\protect\citeauthoryear{Hirschberg and Manning}{Hirschberg and
  Manning}{2015}]%
        {hirschberg_manning_2015}
\bibfield{author}{\bibinfo{person}{J. Hirschberg} {and} \bibinfo{person}{C.~D.
  Manning}.} \bibinfo{year}{2015}\natexlab{}.
\newblock \showarticletitle{Advances in natural language processing}.
\newblock \bibinfo{journal}{\emph{Science}} \bibinfo{volume}{349},
  \bibinfo{number}{6245} (\bibinfo{year}{2015}), \bibinfo{pages}{261--266}.
\newblock
\urldef\tempurl%
\url{https://doi.org/10.1126/science.aaa8685}
\showDOI{\tempurl}


\bibitem[\protect\citeauthoryear{Horton}{Horton}{2018}]%
        {The_Telegraph}
\bibfield{author}{\bibinfo{person}{Helena Horton}.}
  \bibinfo{year}{2018}\natexlab{}.
\newblock \bibinfo{title}{{Amazon Alexa recorded owner's conversation and sent
  to 'random' contact, couple complains}}.
\newblock
  \bibinfo{howpublished}{\url{www.telegraph.co.uk/news/2018/05/25/amazon-alexa-recorded-owners-conversation-sent-random-contact/}}.
\newblock
\newblock
\shownote{[Online; last accessed 17-December-2018].}


\bibitem[\protect\citeauthoryear{Hoy}{Hoy}{2018}]%
        {hoy_2018}
\bibfield{author}{\bibinfo{person}{Matthew~B. Hoy}.}
  \bibinfo{year}{2018}\natexlab{}.
\newblock \showarticletitle{Alexa, Siri, Cortana, and More: An Introduction to
  Voice Assistants}.
\newblock \bibinfo{journal}{\emph{Medical Reference Services Quarterly}}
  \bibinfo{volume}{37}, \bibinfo{number}{1} (\bibinfo{year}{2018}),
  \bibinfo{pages}{81--88}.
\newblock
\urldef\tempurl%
\url{https://doi.org/10.1080/02763869.2018.1404391}
\showDOI{\tempurl}


\bibitem[\protect\citeauthoryear{Instruments}{Instruments}{2013}]%
        {Texas_Instrument_2013}
\bibfield{author}{\bibinfo{person}{Texas Instruments}.}
  \bibinfo{year}{2013}\natexlab{}.
\newblock \bibinfo{title}{{AN-1973 Benefits and Challenges of High-Frequency
  Regulators}}.
\newblock
  \bibinfo{howpublished}{\url{http://www.ti.com/lit/an/snva399a/snva399a.pdf}}.
\newblock
\newblock
\shownote{[Online; last accessed 17-December-2018].}


\bibitem[\protect\citeauthoryear{Jia, Chen, Wang, Rahmati, Fernandes, Mao, and
  Prakash}{Jia et~al\mbox{.}}{2017}]%
        {inproceedings}
\bibfield{author}{\bibinfo{person}{Yunhan Jia}, \bibinfo{person}{Qi~Alfred
  Chen}, \bibinfo{person}{Shiqi Wang}, \bibinfo{person}{Amir Rahmati},
  \bibinfo{person}{Earlence Fernandes}, \bibinfo{person}{Zhuoqing Mao}, {and}
  \bibinfo{person}{Atul Prakash}.} \bibinfo{year}{2017}\natexlab{}.
\newblock \showarticletitle{ContexIoT: Towards Providing Contextual Integrity
  to Appified IoT Platforms}.
\newblock


\bibitem[\protect\citeauthoryear{Kadianakis, Roberts, Roberts, and
  Winter}{Kadianakis et~al\mbox{.}}{2017}]%
        {DBLP:journals/corr/KadianakisRRW17}
\bibfield{author}{\bibinfo{person}{George Kadianakis},
  \bibinfo{person}{Claudia~V. Roberts}, \bibinfo{person}{Laura~M. Roberts},
  {and} \bibinfo{person}{Philipp Winter}.} \bibinfo{year}{2017}\natexlab{}.
\newblock \showarticletitle{Anomalous keys in Tor relays}.
\newblock \bibinfo{journal}{\emph{CoRR}}  \bibinfo{volume}{abs/1704.00792}
  (\bibinfo{year}{2017}).
\newblock
\showeprint[arxiv]{1704.00792}
\urldef\tempurl%
\url{http://arxiv.org/abs/1704.00792}
\showURL{%
\tempurl}


\bibitem[\protect\citeauthoryear{Kafali, Ajmeri, and Singh}{Kafali
  et~al\mbox{.}}{2016}]%
        {kafaly2016revani}
\bibfield{author}{\bibinfo{person}{{\"O}zg{\"u}r Kafali},
  \bibinfo{person}{Nirav Ajmeri}, {and} \bibinfo{person}{Munindar~P Singh}.}
  \bibinfo{year}{2016}\natexlab{}.
\newblock \showarticletitle{Revani: Revising and verifying normative
  specifications for privacy}.
\newblock \bibinfo{journal}{\emph{IEEE Intelligent Systems}}
  \bibinfo{volume}{31}, \bibinfo{number}{5} (\bibinfo{year}{2016}),
  \bibinfo{pages}{8--15}.
\newblock


\bibitem[\protect\citeauthoryear{Kafali, Jones, Petruso, Williams, and
  Singh}{Kafali et~al\mbox{.}}{2017}]%
        {kafali2017good}
\bibfield{author}{\bibinfo{person}{{\"O}zg{\"u}r Kafali},
  \bibinfo{person}{Jasmine Jones}, \bibinfo{person}{Megan Petruso},
  \bibinfo{person}{Laurie Williams}, {and} \bibinfo{person}{Munindar~P Singh}.}
  \bibinfo{year}{2017}\natexlab{}.
\newblock \showarticletitle{How good is a security policy against real
  breaches? A HIPAA case study}. In \bibinfo{booktitle}{\emph{2017 IEEE/ACM
  39th International Conference on Software Engineering (ICSE)}}. IEEE,
  \bibinfo{pages}{530--540}.
\newblock


\bibitem[\protect\citeauthoryear{KAMM}{KAMM}{1995}]%
        {Kamm_Candace_1995}
\bibfield{author}{\bibinfo{person}{CANDACE KAMM}.}
  \bibinfo{year}{1995}\natexlab{}.
\newblock \showarticletitle{User interfaces for voice applications}.
\newblock \bibinfo{journal}{\emph{Colloquium Paper}}  \bibinfo{volume}{92}
  (\bibinfo{year}{1995}), \bibinfo{pages}{10031--10037}.
\newblock


\bibitem[\protect\citeauthoryear{Kelly}{Kelly}{2017}]%
        {alexa-CNN}
\bibfield{author}{\bibinfo{person}{Heather Kelly}.}
  \bibinfo{year}{2017}\natexlab{}.
\newblock \bibinfo{title}{{Apple's HomePod is coming. Here's what you need to
  know about smart speakers}}.
\newblock
  \bibinfo{howpublished}{\url{http://money.cnn.com/2017/06/08/technology/gadgets/apple-homepod-smart-speaker-faq/index.html}}.
\newblock
\newblock
\shownote{[Online; last accessed 21-December-2018].}


\bibitem[\protect\citeauthoryear{Kelly and Statt}{Kelly and Statt}{2019}]%
        {alexa-delete}
\bibfield{author}{\bibinfo{person}{Makena Kelly} {and} \bibinfo{person}{Nick
  Statt}.} \bibinfo{year}{2019}\natexlab{}.
\newblock \bibinfo{title}{{Amazon confirms it holds on to Alexa data even if
  you delete audio files}}.
\newblock
  \bibinfo{howpublished}{\url{https://www.theverge.com/2019/7/3/20681423/amazon-alexa-echo-chris-coons-data-transcripts-recording-privacy}}.
\newblock
\newblock
\shownote{[Online; last accessed 21-December-2019].}


\bibitem[\protect\citeauthoryear{Kepuska and Bohouta}{Kepuska and
  Bohouta}{2018}]%
        {kepuska_bohouta_2018}
\bibfield{author}{\bibinfo{person}{Veton Kepuska} {and} \bibinfo{person}{Gamal
  Bohouta}.} \bibinfo{year}{2018}\natexlab{}.
\newblock \showarticletitle{Next-generation of virtual personal assistants
  (Microsoft Cortana, Apple Siri, Amazon Alexa and Google Home)}.
\newblock \bibinfo{journal}{\emph{2018 IEEE 8th Annual Computing and
  Communication Workshop and Conference (CCWC)}} (\bibinfo{year}{2018}),
  \bibinfo{pages}{99--103}.
\newblock
\urldef\tempurl%
\url{https://doi.org/10.1109/ccwc.2018.8301638}
\showDOI{\tempurl}


\bibitem[\protect\citeauthoryear{Kinsella}{Kinsella}{2018a}]%
        {alexa-Voicebot.ai}
\bibfield{author}{\bibinfo{person}{Bret Kinsella}.}
  \bibinfo{year}{2018}\natexlab{a}.
\newblock \bibinfo{title}{{Alexa Skill Store for France is a Fast Growing Land
  of Opportunity}}.
\newblock
  \bibinfo{howpublished}{\url{https://voicebot.ai/2018/11/03/the-alexa-skill-store-for-france-is-a-fast-growing-land-of-opportunity/}}.
\newblock
\newblock
\shownote{[Online; last accessed 22-December-2018].}


\bibitem[\protect\citeauthoryear{Kinsella}{Kinsella}{2018b}]%
        {alexa-Bret_Kinsella}
\bibfield{author}{\bibinfo{person}{Bret Kinsella}.}
  \bibinfo{year}{2018}\natexlab{b}.
\newblock \bibinfo{title}{{Amazon Introduces Skill Connections so Alexa Skills
  Can Work Together}}.
\newblock
  \bibinfo{howpublished}{\url{https://voicebot.ai/2018/10/04/amazon-introduces-skill-connections-so-alexa-skills-can/}}.
\newblock
\newblock
\shownote{[Online; last accessed 24-December-2018].}


\bibitem[\protect\citeauthoryear{Kinsella}{Kinsella}{2018c}]%
        {alexa-Reality}
\bibfield{author}{\bibinfo{person}{Bret Kinsella}.}
  \bibinfo{year}{2018}\natexlab{c}.
\newblock \bibinfo{title}{{The Information Says Alexa Struggles with Voice
  Commerce But Has 50 Million Devices Sold}}.
\newblock
  \bibinfo{howpublished}{\url{https://voicebot.ai/2018/08/06/the-information-says-alexa-struggles-with}}.
\newblock
\newblock
\shownote{[Online; last accessed 7-January-2018].}


\bibitem[\protect\citeauthoryear{Kitchenham, Pearl~Brereton, Budgen, Turner,
  Bailey, and Linkman}{Kitchenham et~al\mbox{.}}{2009}]%
        {kitchenham_pearl}
\bibfield{author}{\bibinfo{person}{Barbara Kitchenham}, \bibinfo{person}{O.
  Pearl~Brereton}, \bibinfo{person}{David Budgen}, \bibinfo{person}{Mark
  Turner}, \bibinfo{person}{John Bailey}, {and} \bibinfo{person}{Stephen
  Linkman}.} \bibinfo{year}{2009}\natexlab{}.
\newblock \showarticletitle{Systematic literature reviews in software
  engineering \- A systematic literature review}.
\newblock \bibinfo{journal}{\emph{Information and Software Technology}}
  \bibinfo{volume}{51}, \bibinfo{number}{1} (\bibinfo{year}{2009}),
  \bibinfo{pages}{7--15}.
\newblock
\urldef\tempurl%
\url{https://doi.org/10.1016/j.infsof.2008.09.009}
\showDOI{\tempurl}


\bibitem[\protect\citeauthoryear{Knowles, Such, Gouglidis, Misra, and
  Rashid}{Knowles et~al\mbox{.}}{2015}]%
        {knowles2015assurance}
\bibfield{author}{\bibinfo{person}{William Knowles}, \bibinfo{person}{Jose~M
  Such}, \bibinfo{person}{Antonios Gouglidis}, \bibinfo{person}{Gaurav Misra},
  {and} \bibinfo{person}{Awais Rashid}.} \bibinfo{year}{2015}\natexlab{}.
\newblock \showarticletitle{Assurance techniques for industrial control systems
  (ics)}. In \bibinfo{booktitle}{\emph{Proceedings of the First ACM Workshop on
  Cyber-Physical Systems-Security and/or PrivaCy}}. \bibinfo{publisher}{ACM},
  \bibinfo{pages}{101--112}.
\newblock


\bibitem[\protect\citeauthoryear{Kumar, Paccagnella, Murley, Hennenfent, Mason,
  Bates, and Bailey}{Kumar et~al\mbox{.}}{2018}]%
        {217575}
\bibfield{author}{\bibinfo{person}{Deepak Kumar}, \bibinfo{person}{Riccardo
  Paccagnella}, \bibinfo{person}{Paul Murley}, \bibinfo{person}{Eric
  Hennenfent}, \bibinfo{person}{Joshua Mason}, \bibinfo{person}{Adam Bates},
  {and} \bibinfo{person}{Michael Bailey}.} \bibinfo{year}{2018}\natexlab{}.
\newblock \showarticletitle{Skill Squatting Attacks on Amazon Alexa}. In
  \bibinfo{booktitle}{\emph{27th {USENIX} Security Symposium ({USENIX} Security
  18)}}. \bibinfo{publisher}{{USENIX} Association},
  \bibinfo{address}{Baltimore, MD}, \bibinfo{pages}{33--47}.
\newblock
\showISBNx{978-1-931971-46-1}


\bibitem[\protect\citeauthoryear{Lai}{Lai}{2018}]%
        {mom.me}
\bibfield{author}{\bibinfo{person}{Angelica Lai}.}
  \bibinfo{year}{2018}\natexlab{}.
\newblock \bibinfo{title}{{Sneaky Kid Orders \$350 Worth of Toys on Her Mom's
  Amazon Account}}.
\newblock
  \bibinfo{howpublished}{\url{https://mom.me/news/271144-sneaky-kid-orders-350-worth-toys-her-moms-amazon-account/}}.
\newblock
\newblock
\shownote{[Online; last accessed 17-December-2018].}


\bibitem[\protect\citeauthoryear{Lau, Zimmerman, and Schaub}{Lau
  et~al\mbox{.}}{2018}]%
        {Lau:2018:AYL:3290265.3274371}
\bibfield{author}{\bibinfo{person}{Josephine Lau}, \bibinfo{person}{Benjamin
  Zimmerman}, {and} \bibinfo{person}{Florian Schaub}.}
  \bibinfo{year}{2018}\natexlab{}.
\newblock \showarticletitle{Alexa, Are You Listening?: Privacy Perceptions,
  Concerns and Privacy-seeking Behaviors with Smart Speakers}.
\newblock \bibinfo{journal}{\emph{Proc. ACM Hum.-Comput. Interact.}}
  \bibinfo{volume}{2}, \bibinfo{number}{CSCW}, Article \bibinfo{articleno}{102}
  (\bibinfo{date}{Nov.} \bibinfo{year}{2018}), \bibinfo{numpages}{31}~pages.
\newblock
\showISSN{2573-0142}
\urldef\tempurl%
\url{https://doi.org/10.1145/3274371}
\showDOI{\tempurl}


\bibitem[\protect\citeauthoryear{Lavrentyeva, Novoselov, Malykh, Kozlov,
  Kudashev, and Shchemelinin}{Lavrentyeva et~al\mbox{.}}{2017}]%
        {DBLP:journals/corr/LavrentyevaNMKK17}
\bibfield{author}{\bibinfo{person}{Galina Lavrentyeva}, \bibinfo{person}{Sergey
  Novoselov}, \bibinfo{person}{Egor Malykh}, \bibinfo{person}{Alexander
  Kozlov}, \bibinfo{person}{Oleg Kudashev}, {and} \bibinfo{person}{Vadim
  Shchemelinin}.} \bibinfo{year}{2017}\natexlab{}.
\newblock \showarticletitle{Audio-replay attack detection countermeasures}.
\newblock \bibinfo{journal}{\emph{CoRR}}  \bibinfo{volume}{abs/1705.08858}
  (\bibinfo{year}{2017}).
\newblock
\showeprint[arxiv]{1705.08858}
\urldef\tempurl%
\url{http://arxiv.org/abs/1705.08858}
\showURL{%
\tempurl}


\bibitem[\protect\citeauthoryear{Lee and Nass}{Lee and Nass}{2005}]%
        {doi:10.1207/S1532785XMEP0701}
\bibfield{author}{\bibinfo{person}{Kwan-Min Lee} {and}
  \bibinfo{person}{Clifford Nass}.} \bibinfo{year}{2005}\natexlab{}.
\newblock \showarticletitle{Social-Psychological Origins of Feelings of
  Presence: Creating Social Presence With Machine Generated Voices}.
\newblock \bibinfo{journal}{\emph{Media Psychology}} \bibinfo{volume}{7},
  \bibinfo{number}{1} (\bibinfo{year}{2005}), \bibinfo{pages}{31--45}.
\newblock
\urldef\tempurl%
\url{https://doi.org/10.1207/S1532785XMEP0701\_2}
\showDOI{\tempurl}


\bibitem[\protect\citeauthoryear{Liu, Zhang, and Fang}{Liu
  et~al\mbox{.}}{2018}]%
        {8278156}
\bibfield{author}{\bibinfo{person}{J. Liu}, \bibinfo{person}{C. Zhang}, {and}
  \bibinfo{person}{Y. Fang}.} \bibinfo{year}{2018}\natexlab{}.
\newblock \showarticletitle{EPIC: A Differential Privacy Framework to Defend
  Smart Homes Against Internet Traffic Analysis}.
\newblock \bibinfo{journal}{\emph{IEEE Internet of Things Journal}}
  \bibinfo{volume}{5}, \bibinfo{number}{2} (\bibinfo{date}{April}
  \bibinfo{year}{2018}), \bibinfo{pages}{1206--1217}.
\newblock
\showISSN{2327-4662}
\urldef\tempurl%
\url{https://doi.org/10.1109/JIOT.2018.2799820}
\showDOI{\tempurl}


\bibitem[\protect\citeauthoryear{Londhe, Ahirwal, and Lodha}{Londhe
  et~al\mbox{.}}{2016}]%
        {londhe_ahirwal_lodha_2016}
\bibfield{author}{\bibinfo{person}{N.~D. Londhe}, \bibinfo{person}{M.~K.
  Ahirwal}, {and} \bibinfo{person}{P. Lodha}.} \bibinfo{year}{2016}\natexlab{}.
\newblock \showarticletitle{Machine learning paradigms for speech recognition
  of an Indian dialect}.
\newblock \bibinfo{journal}{\emph{2016 International Conference on
  Communication and Signal Processing (ICCSP)}} (\bibinfo{year}{2016}).
\newblock
\urldef\tempurl%
\url{https://doi.org/10.1109/iccsp.2016.7754251}
\showDOI{\tempurl}


\bibitem[\protect\citeauthoryear{Luger and Sellen}{Luger and Sellen}{2016}]%
        {Luger:2016:LRB:2858036.2858288}
\bibfield{author}{\bibinfo{person}{Ewa Luger} {and} \bibinfo{person}{Abigail
  Sellen}.} \bibinfo{year}{2016}\natexlab{}.
\newblock \showarticletitle{"Like Having a Really Bad PA": The Gulf Between
  User Expectation and Experience of Conversational Agents}. In
  \bibinfo{booktitle}{\emph{Proceedings of the 2016 CHI Conference on Human
  Factors in Computing Systems}} (San Jose, California, USA)
  \emph{(\bibinfo{series}{CHI '16})}. \bibinfo{publisher}{ACM},
  \bibinfo{address}{New York, NY, USA}, \bibinfo{pages}{5286--5297}.
\newblock
\showISBNx{978-1-4503-3362-7}
\urldef\tempurl%
\url{https://doi.org/10.1145/2858036.2858288}
\showDOI{\tempurl}


\bibitem[\protect\citeauthoryear{Madaan, Ahad, and Sastry}{Madaan
  et~al\mbox{.}}{2018}]%
        {madaan_ahad_sastry_2018}
\bibfield{author}{\bibinfo{person}{Nishtha Madaan}, \bibinfo{person}{Mohd~Abdul
  Ahad}, {and} \bibinfo{person}{Sunil~M. Sastry}.}
  \bibinfo{year}{2018}\natexlab{}.
\newblock \showarticletitle{Data integration in IoT ecosystem: Information
  linkage as a privacy threat}.
\newblock \bibinfo{journal}{\emph{Computer Law and Security Review}}
  \bibinfo{volume}{34}, \bibinfo{number}{1} (\bibinfo{year}{2018}),
  \bibinfo{pages}{125--133}.
\newblock
\urldef\tempurl%
\url{https://doi.org/10.1016/j.clsr.2017.06.007}
\showDOI{\tempurl}


\bibitem[\protect\citeauthoryear{{Malik}, {Malik}, and {Baumann}}{{Malik}
  et~al\mbox{.}}{2019}]%
        {8695380}
\bibfield{author}{\bibinfo{person}{K.~M. {Malik}}, \bibinfo{person}{H.
  {Malik}}, {and} \bibinfo{person}{R. {Baumann}}.}
  \bibinfo{year}{2019}\natexlab{}.
\newblock \showarticletitle{Towards Vulnerability Analysis of Voice-Driven
  Interfaces and Countermeasures for Replay Attacks}. In
  \bibinfo{booktitle}{\emph{2019 IEEE Conference on Multimedia Information
  Processing and Retrieval (MIPR)}}. \bibinfo{pages}{523--528}.
\newblock


\bibitem[\protect\citeauthoryear{Malkin, Deatrick, Tong, Wijesekera, Egelman,
  and Wagner}{Malkin et~al\mbox{.}}{2019}]%
        {malkin2019privacy}
\bibfield{author}{\bibinfo{person}{Nathan Malkin}, \bibinfo{person}{Joe
  Deatrick}, \bibinfo{person}{Allen Tong}, \bibinfo{person}{Primal Wijesekera},
  \bibinfo{person}{Serge Egelman}, {and} \bibinfo{person}{David Wagner}.}
  \bibinfo{year}{2019}\natexlab{}.
\newblock \showarticletitle{Privacy attitudes of smart speaker users}.
\newblock \bibinfo{journal}{\emph{Proceedings on Privacy Enhancing
  Technologies}} \bibinfo{volume}{2019}, \bibinfo{number}{4}
  (\bibinfo{year}{2019}), \bibinfo{pages}{250--271}.
\newblock


\bibitem[\protect\citeauthoryear{Mandal}{Mandal}{2018}]%
        {2800-18}
\bibfield{author}{\bibinfo{person}{Minhua Wu; Sankaran Panchapagesan; Ming Sun;
  Jiacheng Gu; Ryan Thomas; Shiv Naga Prasad Vitaladevuni; Bjorn
  Hoffmeister;~Arindam Mandal}.} \bibinfo{year}{2018}\natexlab{}.
\newblock \showarticletitle{Monophone-Based Background Modeling For Two-Stage
  On-Device Wake Word Detection}.
\newblock  (\bibinfo{year}{2018}).
\newblock
\urldef\tempurl%
\url{http://sigport.org/2800}
\showURL{%
\tempurl}


\bibitem[\protect\citeauthoryear{Martin}{Martin}{2018}]%
        {alexa-Cnet}
\bibfield{author}{\bibinfo{person}{Taylor Martin}.}
  \bibinfo{year}{2018}\natexlab{}.
\newblock \bibinfo{title}{{12 reasons to use Alexa in the kitchen}}.
\newblock
  \bibinfo{howpublished}{\url{https://www.cnet.com/how-to/how-to-use-alexa-in-the-kitchen/}}.
\newblock
\newblock
\shownote{[Online; last accessed 17-December-2018].}


\bibitem[\protect\citeauthoryear{Matthews, Liao, Turner, Berkovich, Reeder, and
  Consolvo}{Matthews et~al\mbox{.}}{2016}]%
        {matthews2016she}
\bibfield{author}{\bibinfo{person}{Tara Matthews}, \bibinfo{person}{Kerwell
  Liao}, \bibinfo{person}{Anna Turner}, \bibinfo{person}{Marianne Berkovich},
  \bibinfo{person}{Robert Reeder}, {and} \bibinfo{person}{Sunny Consolvo}.}
  \bibinfo{year}{2016}\natexlab{}.
\newblock \showarticletitle{She'll just grab any device that's closer: A Study
  of Everyday Device \& Account Sharing in Households}. In
  \bibinfo{booktitle}{\emph{Proceedings of the 2016 CHI Conference on Human
  Factors in Computing Systems}}. \bibinfo{publisher}{ACM},
  \bibinfo{pages}{5921--5932}.
\newblock


\bibitem[\protect\citeauthoryear{Matthews, OLeary, Turner, Sleeper, Woelfer,
  Shelton, Manthorne, Churchill, and Consolvo}{Matthews
  et~al\mbox{.}}{[n.d.]}]%
        {matthews2017security}
\bibfield{author}{\bibinfo{person}{Tara Matthews}, \bibinfo{person}{Kathleen
  OLeary}, \bibinfo{person}{Anna Turner}, \bibinfo{person}{Manya Sleeper},
  \bibinfo{person}{Jill~Palzkill Woelfer}, \bibinfo{person}{Martin Shelton},
  \bibinfo{person}{Cori Manthorne}, \bibinfo{person}{Elizabeth~F Churchill},
  {and} \bibinfo{person}{Sunny Consolvo}.} \bibinfo{year}{[n.d.]}\natexlab{}.
\newblock \showarticletitle{Security and Privacy Experiences and Practices of
  Survivors of Intimate Partner Abuse}.
\newblock \bibinfo{journal}{\emph{IEEE Security \& Privacy}}
  \bibinfo{number}{5} (\bibinfo{year}{[n.\,d.]}), \bibinfo{pages}{76--81}.
\newblock


\bibitem[\protect\citeauthoryear{Mayer and Mitchell}{Mayer and
  Mitchell}{2012}]%
        {mayer2012third}
\bibfield{author}{\bibinfo{person}{Jonathan~R Mayer} {and}
  \bibinfo{person}{John~C Mitchell}.} \bibinfo{year}{2012}\natexlab{}.
\newblock \showarticletitle{Third-party web tracking: Policy and technology}.
  In \bibinfo{booktitle}{\emph{2012 IEEE Symposium on Security and Privacy}}.
  IEEE, \bibinfo{pages}{413--427}.
\newblock


\bibitem[\protect\citeauthoryear{Memon and Anwar}{Memon and Anwar}{2015}]%
        {memon2015colluding}
\bibfield{author}{\bibinfo{person}{Atif~M Memon} {and} \bibinfo{person}{Ali
  Anwar}.} \bibinfo{year}{2015}\natexlab{}.
\newblock \showarticletitle{Colluding apps: Tomorrow's mobile malware threat}.
\newblock \bibinfo{journal}{\emph{IEEE Security \& Privacy}}
  \bibinfo{volume}{13}, \bibinfo{number}{6} (\bibinfo{year}{2015}),
  \bibinfo{pages}{77--81}.
\newblock


\bibitem[\protect\citeauthoryear{Misra and Such}{Misra and Such}{2017}]%
        {misra2017pacman}
\bibfield{author}{\bibinfo{person}{Gaurav Misra} {and} \bibinfo{person}{Jose~M
  Such}.} \bibinfo{year}{2017}\natexlab{}.
\newblock \showarticletitle{PACMAN: Personal Agent for Access Control in Social
  Media}.
\newblock \bibinfo{journal}{\emph{IEEE Internet Computing}}
  \bibinfo{volume}{21}, \bibinfo{number}{6} (\bibinfo{year}{2017}),
  \bibinfo{pages}{18--26}.
\newblock


\bibitem[\protect\citeauthoryear{Misra, Such, and Gill}{Misra
  et~al\mbox{.}}{2017}]%
        {misra2017privacy}
\bibfield{author}{\bibinfo{person}{Gaurav Misra}, \bibinfo{person}{Jose~M
  Such}, {and} \bibinfo{person}{Lauren Gill}.} \bibinfo{year}{2017}\natexlab{}.
\newblock \showarticletitle{A Privacy Assessment of Social Media Aggregators}.
  In \bibinfo{booktitle}{\emph{Proceedings of the 2017 IEEE/ACM International
  Conference on Advances in Social Networks Analysis and Mining 2017}}.
  \bibinfo{publisher}{ACM}, \bibinfo{pages}{561--568}.
\newblock


\bibitem[\protect\citeauthoryear{Modi, Patel, Borisaniya, Patel, and
  Rajarajan}{Modi et~al\mbox{.}}{2012}]%
        {modi_patel_borisaniya_patel_rajarajan_2012}
\bibfield{author}{\bibinfo{person}{Chirag Modi}, \bibinfo{person}{Dhiren
  Patel}, \bibinfo{person}{Bhavesh Borisaniya}, \bibinfo{person}{Avi Patel},
  {and} \bibinfo{person}{Muttukrishnan Rajarajan}.}
  \bibinfo{year}{2012}\natexlab{}.
\newblock \showarticletitle{A survey on security issues and solutions at
  different layers of Cloud computing}.
\newblock \bibinfo{journal}{\emph{The Journal of Supercomputing}}
  \bibinfo{volume}{63}, \bibinfo{number}{2} (\bibinfo{year}{2012}),
  \bibinfo{pages}{561--592}.
\newblock
\urldef\tempurl%
\url{https://doi.org/10.1007/s11227-012-0831-5}
\showDOI{\tempurl}


\bibitem[\protect\citeauthoryear{Mosner, Wu, Raju, Krishnan~Parthasarathi,
  Kumatani, Sundaram, Maas, and Hoffmeister}{Mosner et~al\mbox{.}}{2019}]%
        {mosner_wu_raju_krishnan}
\bibfield{author}{\bibinfo{person}{Ladislav Mosner}, \bibinfo{person}{Minhua
  Wu}, \bibinfo{person}{Anirudh Raju}, \bibinfo{person}{Sree~Hari
  Krishnan~Parthasarathi}, \bibinfo{person}{Kenichi Kumatani},
  \bibinfo{person}{Shiva Sundaram}, \bibinfo{person}{Roland Maas}, {and}
  \bibinfo{person}{Bjorn Hoffmeister}.} \bibinfo{year}{2019}\natexlab{}.
\newblock \showarticletitle{Improving Noise Robustness of Automatic Speech
  Recognition via Parallel Data and Teacher-student Learning}.
\newblock \bibinfo{journal}{\emph{ICASSP 2019 - 2019 IEEE International
  Conference on Acoustics, Speech and Signal Processing (ICASSP)}}
  (\bibinfo{year}{2019}).
\newblock
\urldef\tempurl%
\url{https://doi.org/10.1109/icassp.2019.8683422}
\showDOI{\tempurl}


\bibitem[\protect\citeauthoryear{Murdoch and Zielinski}{Murdoch and
  Zielinski}{2007}]%
        {Murdoch:2007:STA:1779330.1779341}
\bibfield{author}{\bibinfo{person}{Steven~J. Murdoch} {and}
  \bibinfo{person}{Piotr Zielinski}.} \bibinfo{year}{2007}\natexlab{}.
\newblock \showarticletitle{Sampled Traffic Analysis by Internet-exchange-level
  Adversaries}. In \bibinfo{booktitle}{\emph{Proceedings of the 7th
  International Conference on Privacy Enhancing Technologies}} (Ottawa, Canada)
  \emph{(\bibinfo{series}{PET'07})}. \bibinfo{publisher}{Springer-Verlag},
  \bibinfo{address}{Berlin, Heidelberg}, \bibinfo{pages}{167--183}.
\newblock
\urldef\tempurl%
\url{http://dl.acm.org/citation.cfm?id=1779330.1779341}
\showURL{%
\tempurl}


\bibitem[\protect\citeauthoryear{Naik, Gupta, Ge, Lambert, and Sarikaya}{Naik
  et~al\mbox{.}}{2018}]%
        {Naik2018}
\bibfield{author}{\bibinfo{person}{Chetan Naik}, \bibinfo{person}{Arpit Gupta},
  \bibinfo{person}{Hancheng Ge}, \bibinfo{person}{Mathias Lambert}, {and}
  \bibinfo{person}{Ruhi Sarikaya}.} \bibinfo{year}{2018}\natexlab{}.
\newblock \showarticletitle{Contextual Slot Carryover for Disparate Schemas}.
  In \bibinfo{booktitle}{\emph{Proc. Interspeech 2018}}.
  \bibinfo{pages}{596--600}.
\newblock
\urldef\tempurl%
\url{https://doi.org/10.21437/Interspeech.2018-1035}
\showDOI{\tempurl}


\bibitem[\protect\citeauthoryear{Nass, Moon, and Carney}{Nass
  et~al\mbox{.}}{1999}]%
        {doi:10.1111/j.1559-1816.1999.tb00142.x}
\bibfield{author}{\bibinfo{person}{Clifford Nass}, \bibinfo{person}{Youngme
  Moon}, {and} \bibinfo{person}{Paul Carney}.} \bibinfo{year}{1999}\natexlab{}.
\newblock \showarticletitle{Are People Polite to Computers? Responses to
  Computer-Based Interviewing Systems1}.
\newblock \bibinfo{journal}{\emph{Journal of Applied Social Psychology}}
  \bibinfo{volume}{29}, \bibinfo{number}{5} (\bibinfo{year}{1999}),
  \bibinfo{pages}{1093--1109}.
\newblock
\urldef\tempurl%
\url{https://doi.org/10.1111/j.1559-1816.1999.tb00142.x}
\showDOI{\tempurl}


\bibitem[\protect\citeauthoryear{Natatsuka, Iijima, Watanabe, Akiyama, Sakai,
  and Mori}{Natatsuka et~al\mbox{.}}{2019a}]%
        {natatsuka_iijima_watanabe_akiyama_sakai_mori_2019}
\bibfield{author}{\bibinfo{person}{Atsuko Natatsuka}, \bibinfo{person}{Ryo
  Iijima}, \bibinfo{person}{Takuya Watanabe}, \bibinfo{person}{Mitsuaki
  Akiyama}, \bibinfo{person}{Tetsuya Sakai}, {and} \bibinfo{person}{Tatsuya
  Mori}.} \bibinfo{year}{2019}\natexlab{a}.
\newblock \showarticletitle{Poster}.
\newblock \bibinfo{journal}{\emph{Proceedings of the 2019 ACM SIGSAC Conference
  on Computer and Communications Security - CCS '19}} (\bibinfo{year}{2019}).
\newblock
\urldef\tempurl%
\url{https://doi.org/10.1145/3319535.3363274}
\showDOI{\tempurl}


\bibitem[\protect\citeauthoryear{Natatsuka, Iijima, Watanabe, Akiyama, Sakai,
  and Mori}{Natatsuka et~al\mbox{.}}{2019b}]%
        {10.1145/3319535.3363274}
\bibfield{author}{\bibinfo{person}{Atsuko Natatsuka}, \bibinfo{person}{Ryo
  Iijima}, \bibinfo{person}{Takuya Watanabe}, \bibinfo{person}{Mitsuaki
  Akiyama}, \bibinfo{person}{Tetsuya Sakai}, {and} \bibinfo{person}{Tatsuya
  Mori}.} \bibinfo{year}{2019}\natexlab{b}.
\newblock \showarticletitle{Poster: A First Look at the Privacy Risks of Voice
  Assistant Apps}. In \bibinfo{booktitle}{\emph{Proceedings of the 2019 ACM
  SIGSAC Conference on Computer and Communications Security}} (London, United
  Kingdom) \emph{(\bibinfo{series}{CCS ’19})}.
  \bibinfo{publisher}{Association for Computing Machinery},
  \bibinfo{address}{New York, NY, USA}, \bibinfo{pages}{2633–2635}.
\newblock
\showISBNx{9781450367479}
\urldef\tempurl%
\url{https://doi.org/10.1145/3319535.3363274}
\showDOI{\tempurl}


\bibitem[\protect\citeauthoryear{Newman}{Newman}{2018}]%
        {CVE201812716}
\bibfield{author}{\bibinfo{person}{Lily~Hay Newman}.}
  \bibinfo{year}{2018}\natexlab{}.
\newblock \bibinfo{title}{{Millions of Streaming Devices Are Vulnerable to a
  Retro Web Attack}}.
\newblock
  \bibinfo{howpublished}{\url{https://www.wired.com/story/chromecast-roku-sonos-dns-rebinding-vulnerability/}}.
\newblock
\newblock
\shownote{[Online; last accessed 21-April-2020].}


\bibitem[\protect\citeauthoryear{Ni, Yang, Bai, Champion, and Xuan}{Ni
  et~al\mbox{.}}{2009}]%
        {5337017}
\bibfield{author}{\bibinfo{person}{Xudong Ni}, \bibinfo{person}{Zhimin Yang},
  \bibinfo{person}{Xiaole Bai}, \bibinfo{person}{A.~C. Champion}, {and}
  \bibinfo{person}{D. Xuan}.} \bibinfo{year}{2009}\natexlab{}.
\newblock \showarticletitle{DiffUser: Differentiated user access control on
  smartphones}. In \bibinfo{booktitle}{\emph{2009 IEEE 6th International
  Conference on Mobile Adhoc and Sensor Systems}}. \bibinfo{pages}{1012--1017}.
\newblock
\showISSN{2155-6806}
\urldef\tempurl%
\url{https://doi.org/10.1109/MOBHOC.2009.5337017}
\showDOI{\tempurl}


\bibitem[\protect\citeauthoryear{Nissenbaum}{Nissenbaum}{2004}]%
        {nissenbaum2004privacy}
\bibfield{author}{\bibinfo{person}{Helen Nissenbaum}.}
  \bibinfo{year}{2004}\natexlab{}.
\newblock \showarticletitle{Privacy as contextual integrity}.
\newblock \bibinfo{journal}{\emph{Washington Law Review}}  \bibinfo{volume}{79}
  (\bibinfo{year}{2004}), \bibinfo{pages}{119}.
\newblock


\bibitem[\protect\citeauthoryear{Olejnik, Dacosta, Machado, Huguenin, Khan, and
  Hubaux}{Olejnik et~al\mbox{.}}{2017}]%
        {olejnik2017smarper}
\bibfield{author}{\bibinfo{person}{Katarzyna Olejnik}, \bibinfo{person}{Italo
  Dacosta}, \bibinfo{person}{Joana~Soares Machado}, \bibinfo{person}{Kevin
  Huguenin}, \bibinfo{person}{Mohammad~Emtiyaz Khan}, {and}
  \bibinfo{person}{Jean-Pierre Hubaux}.} \bibinfo{year}{2017}\natexlab{}.
\newblock \showarticletitle{Smarper: Context-aware and automatic
  runtime-permissions for mobile devices}. In
  \bibinfo{booktitle}{\emph{Security and Privacy (SP), 2017 IEEE Symposium
  on}}. \bibinfo{publisher}{IEEE}, \bibinfo{pages}{1058--1076}.
\newblock


\bibitem[\protect\citeauthoryear{OVUM}{OVUM}{2017}]%
        {OVUM}
\bibfield{author}{\bibinfo{person}{OVUM}.} \bibinfo{year}{2017}\natexlab{}.
\newblock \showarticletitle{Virtual digital assistants to overtake world
  population by 2021}.
\newblock  (\bibinfo{year}{2017}).
\newblock
\urldef\tempurl%
\url{https://ovum.informa.com/resources/product-content/virtual-digital-assistants-to-overtake-world-population-by-2021}
\showURL{%
\tempurl}


\bibitem[\protect\citeauthoryear{Papayiannis, Amoh, Rozgic, Sundaram, and
  Wang}{Papayiannis et~al\mbox{.}}{2018}]%
        {Papayiannis2018}
\bibfield{author}{\bibinfo{person}{Constantinos Papayiannis},
  \bibinfo{person}{Justice Amoh}, \bibinfo{person}{Viktor Rozgic},
  \bibinfo{person}{Shiva Sundaram}, {and} \bibinfo{person}{Chao Wang}.}
  \bibinfo{year}{2018}\natexlab{}.
\newblock \showarticletitle{Detecting Media Sound Presence in Acoustic Scenes}.
  In \bibinfo{booktitle}{\emph{Proc. Interspeech 2018}}.
  \bibinfo{pages}{1363--1367}.
\newblock
\urldef\tempurl%
\url{https://doi.org/10.21437/Interspeech.2018-2559}
\showDOI{\tempurl}


\bibitem[\protect\citeauthoryear{Papernot, McDaniel, Sinha, and
  Wellman}{Papernot et~al\mbox{.}}{2018}]%
        {papernot2016towards}
\bibfield{author}{\bibinfo{person}{Nicolas Papernot}, \bibinfo{person}{Patrick
  McDaniel}, \bibinfo{person}{Arunesh Sinha}, {and} \bibinfo{person}{Michael
  Wellman}.} \bibinfo{year}{2018}\natexlab{}.
\newblock \showarticletitle{Towards the science of security and privacy in
  machine learning}. In \bibinfo{booktitle}{\emph{3rd IEEE European Symposium
  on Security and Privacy}}.
\newblock


\bibitem[\protect\citeauthoryear{Papernot, McDaniel, and Goodfellow}{Papernot
  et~al\mbox{.}}{2016}]%
        {DBLP:journals/corr/PapernotMG16}
\bibfield{author}{\bibinfo{person}{Nicolas Papernot},
  \bibinfo{person}{Patrick~D. McDaniel}, {and} \bibinfo{person}{Ian~J.
  Goodfellow}.} \bibinfo{year}{2016}\natexlab{}.
\newblock \showarticletitle{Transferability in Machine Learning: from Phenomena
  to Black-Box Attacks using Adversarial Samples}.
\newblock \bibinfo{journal}{\emph{CoRR}}  \bibinfo{volume}{abs/1605.07277}
  (\bibinfo{year}{2016}).
\newblock
\showeprint[arxiv]{1605.07277}
\urldef\tempurl%
\url{http://arxiv.org/abs/1605.07277}
\showURL{%
\tempurl}


\bibitem[\protect\citeauthoryear{Parikh, Tackstrom, Das, and Uszkoreit}{Parikh
  et~al\mbox{.}}{2016}]%
        {45601}
\bibfield{author}{\bibinfo{person}{Ankur~P. Parikh}, \bibinfo{person}{Oscar
  Tackstrom}, \bibinfo{person}{Dipanjan Das}, {and} \bibinfo{person}{Jakob
  Uszkoreit}.} \bibinfo{year}{2016}\natexlab{}.
\newblock \showarticletitle{A Decomposable Attention Model for Natural Language
  Inference}. In \bibinfo{booktitle}{\emph{Proceedings of EMNLP}}.
\newblock
\urldef\tempurl%
\url{https://arxiv.org/abs/1606.01933}
\showURL{%
\tempurl}


\bibitem[\protect\citeauthoryear{Park, Basaran, Park, and Son}{Park
  et~al\mbox{.}}{2014}]%
        {park_basaran_park_son_2014}
\bibfield{author}{\bibinfo{person}{Homin Park}, \bibinfo{person}{Can Basaran},
  \bibinfo{person}{Taejoon Park}, {and} \bibinfo{person}{Sang Son}.}
  \bibinfo{year}{2014}\natexlab{}.
\newblock \showarticletitle{Energy-Efficient Privacy Protection for Smart Home
  Environments Using Behavioral Semantics}.
\newblock \bibinfo{journal}{\emph{Sensors}} \bibinfo{volume}{14},
  \bibinfo{number}{9} (\bibinfo{year}{2014}), \bibinfo{pages}{16235--16257}.
\newblock
\urldef\tempurl%
\url{https://doi.org/10.3390/s140916235}
\showDOI{\tempurl}


\bibitem[\protect\citeauthoryear{Perera, Wakenshaw, Baarslag, Haddadi, Bandara,
  Mortier, Crabtree, Ng, McAuley, and Crowcroft}{Perera et~al\mbox{.}}{2016}]%
        {databox}
\bibfield{author}{\bibinfo{person}{Charith Perera}, \bibinfo{person}{Susan~YL
  Wakenshaw}, \bibinfo{person}{Tim Baarslag}, \bibinfo{person}{Hamed Haddadi},
  \bibinfo{person}{Arosha~K Bandara}, \bibinfo{person}{Richard Mortier},
  \bibinfo{person}{Andy Crabtree}, \bibinfo{person}{Irene~CL Ng},
  \bibinfo{person}{Derek McAuley}, {and} \bibinfo{person}{Jon Crowcroft}.}
  \bibinfo{year}{2016}\natexlab{}.
\newblock \showarticletitle{Valorising the IoT databox: creating value for
  everyone}.
\newblock \bibinfo{journal}{\emph{Transactions on Emerging Telecommunications
  Technologies}} \bibinfo{volume}{28}, \bibinfo{number}{1}
  (\bibinfo{year}{2016}), \bibinfo{pages}{e3125}.
\newblock


\bibitem[\protect\citeauthoryear{Picchi}{Picchi}{2019}]%
        {alexa-CBS-Human}
\bibfield{author}{\bibinfo{person}{Aimee Picchi}.}
  \bibinfo{year}{2019}\natexlab{}.
\newblock \bibinfo{title}{{Amazon workers are listening to what you tell
  Alexa}}.
\newblock
  \bibinfo{howpublished}{\url{https://www.cbsnews.com/news/amazon-workers-are-listening-to-what-you-tell-alexa/}}.
\newblock
\newblock
\shownote{[Online; last accessed 21-February-2020].}


\bibitem[\protect\citeauthoryear{Pinyol and Sabater-Mir}{Pinyol and
  Sabater-Mir}{2013}]%
        {pinyol2013computational}
\bibfield{author}{\bibinfo{person}{I. Pinyol} {and} \bibinfo{person}{J.
  Sabater-Mir}.} \bibinfo{year}{2013}\natexlab{}.
\newblock \showarticletitle{Computational trust and reputation models for open
  multi-agent systems: a review}.
\newblock \bibinfo{journal}{\emph{Artif Intell Rev}} \bibinfo{volume}{40},
  \bibinfo{number}{1} (\bibinfo{year}{2013}), \bibinfo{pages}{1--25}.
\newblock


\bibitem[\protect\citeauthoryear{Prandini and Ramilli}{Prandini and
  Ramilli}{2010}]%
        {prandini2010towards}
\bibfield{author}{\bibinfo{person}{Marco Prandini} {and} \bibinfo{person}{Marco
  Ramilli}.} \bibinfo{year}{2010}\natexlab{}.
\newblock \showarticletitle{Towards a practical and effective security testing
  methodology}. In \bibinfo{booktitle}{\emph{Computers and Communications
  (ISCC), 2010 IEEE Symposium on}}. \bibinfo{publisher}{IEEE},
  \bibinfo{pages}{320--325}.
\newblock


\bibitem[\protect\citeauthoryear{Priyanka and Rajendra}{Priyanka and
  Rajendra}{2016}]%
        {Chouhan_2016}
\bibfield{author}{\bibinfo{person}{Chouhan Priyanka} {and}
  \bibinfo{person}{Singh Rajendra}.} \bibinfo{year}{2016}\natexlab{}.
\newblock \showarticletitle{Security Attacks on Cloud Computing With Possible
  Solution}.
\newblock  \bibinfo{volume}{6}, \bibinfo{number}{1} (\bibinfo{date}{January}
  \bibinfo{year}{2016}).
\newblock
\showISSN{2277 128X}


\bibitem[\protect\citeauthoryear{Purington, Taft, Sannon, Bazarova, and
  Taylor}{Purington et~al\mbox{.}}{2017}]%
        {Purington:2017:AMN:3027063.3053246}
\bibfield{author}{\bibinfo{person}{Amanda Purington},
  \bibinfo{person}{Jessie~G. Taft}, \bibinfo{person}{Shruti Sannon},
  \bibinfo{person}{Natalya~N. Bazarova}, {and} \bibinfo{person}{Samuel~Hardman
  Taylor}.} \bibinfo{year}{2017}\natexlab{}.
\newblock \showarticletitle{"Alexa is My New BFF": Social Roles, User
  Satisfaction, and Personification of the Amazon Echo}. In
  \bibinfo{booktitle}{\emph{Proceedings of the 2017 CHI Conference Extended
  Abstracts on Human Factors in Computing Systems}} (Denver, Colorado, USA)
  \emph{(\bibinfo{series}{CHI EA '17})}. \bibinfo{publisher}{ACM},
  \bibinfo{address}{New York, NY, USA}, \bibinfo{pages}{2853--2859}.
\newblock
\showISBNx{978-1-4503-4656-6}
\urldef\tempurl%
\url{https://doi.org/10.1145/3027063.3053246}
\showDOI{\tempurl}


\bibitem[\protect\citeauthoryear{Ramokapane, Rashid, and Such}{Ramokapane
  et~al\mbox{.}}{2016}]%
        {ramokapane2016assured}
\bibfield{author}{\bibinfo{person}{Kopo~M Ramokapane}, \bibinfo{person}{Awais
  Rashid}, {and} \bibinfo{person}{Jose~M Such}.}
  \bibinfo{year}{2016}\natexlab{}.
\newblock \showarticletitle{Assured deletion in the cloud: requirements,
  challenges and future directions}. In \bibinfo{booktitle}{\emph{Proceedings
  of the 2016 ACM on Cloud Computing Security Workshop}}.
  \bibinfo{pages}{97--108}.
\newblock


\bibitem[\protect\citeauthoryear{Ramokapane, Rashid, and Such}{Ramokapane
  et~al\mbox{.}}{2017}]%
        {ramokapane2017feel}
\bibfield{author}{\bibinfo{person}{Kopo~M. Ramokapane}, \bibinfo{person}{Awais
  Rashid}, {and} \bibinfo{person}{Jose~M. Such}.}
  \bibinfo{year}{2017}\natexlab{}.
\newblock \showarticletitle{``I feel stupid I can't delete...'': A Study of
  Users' Cloud Deletion Practices and Coping Strategies}. In
  \bibinfo{booktitle}{\emph{Thirteenth Symposium on Usable Privacy and Security
  ({SOUPS} 2017)}}. \bibinfo{pages}{241--256}.
\newblock


\bibitem[\protect\citeauthoryear{Reid}{Reid}{2018}]%
        {alexa-Toni_Reid}
\bibfield{author}{\bibinfo{person}{Toni Reid}.}
  \bibinfo{year}{2018}\natexlab{}.
\newblock \bibinfo{title}{{Everything Alexa learned in 2018}}.
\newblock
  \bibinfo{howpublished}{\url{https://blog.aboutamazon.com/devices/everything-alexa-learned-in-2018}}.
\newblock
\newblock
\shownote{[Online; last accessed 4-January-2019].}


\bibitem[\protect\citeauthoryear{Roberts}{Roberts}{2019}]%
        {Assistants-Capecodtoday}
\bibfield{author}{\bibinfo{person}{Mary~Lou Roberts}.}
  \bibinfo{year}{2019}\natexlab{}.
\newblock \bibinfo{title}{{Are Your Voice Assistants Always Listening? The
  simplistic answer is "Yes"...}}
\newblock
  \bibinfo{howpublished}{\url{http://www.capecodtoday.com/article/2019/08/11/248280-Are-Your-Voice-Assistants-Always-Listening}}.
\newblock
\newblock
\shownote{[Online; last accessed 21-February-2020].}


\bibitem[\protect\citeauthoryear{Rodehorst}{Rodehorst}{2019}]%
        {Amazon_Acoustic_cancellation}
\bibfield{author}{\bibinfo{person}{Mike Rodehorst}.}
  \bibinfo{year}{2019}\natexlab{}.
\newblock \bibinfo{title}{{Why Alexa Won't Wake Up When She Hears Her Name in
  Amazon's Super Bowl Ad}}.
\newblock \bibinfo{howpublished}{\url{
  http://web.archive.org/web/20190211063816/https://developer.amazon.com/blogs/alexa/post/37857f29-dd82-4cf4-9ebd-6ebe632f74d3/why-alexa-won-t-wake-up-when-she-hears-her-name-in-amazon-s-super-bowl-ad}}.
\newblock
\newblock
\shownote{[Online; last accessed 21-March-2020].}


\bibitem[\protect\citeauthoryear{Roman, Lopez, and Gritzalis}{Roman
  et~al\mbox{.}}{2018}]%
        {RomanIoT18}
\bibfield{author}{\bibinfo{person}{Rodrigo Roman}, \bibinfo{person}{Javier
  Lopez}, {and} \bibinfo{person}{Stefanos Gritzalis}.}
  \bibinfo{year}{2018}\natexlab{}.
\newblock \showarticletitle{Evolution and Trends in the Security of the
  Internet of Things}.
\newblock \bibinfo{journal}{\emph{IEEE Computer}}  \bibinfo{volume}{51}
  (\bibinfo{date}{07/2018} \bibinfo{year}{2018}), \bibinfo{pages}{16--25}.
\newblock
\showISSN{0018-9162}
\urldef\tempurl%
\url{https://doi.org/10.1109/MC.2018.3011051}
\showDOI{\tempurl}


\bibitem[\protect\citeauthoryear{Roman, Rios, Onieva, and Lopez}{Roman
  et~al\mbox{.}}{ress}]%
        {roman2018VIS}
\bibfield{author}{\bibinfo{person}{Rodrigo Roman}, \bibinfo{person}{Ruben
  Rios}, \bibinfo{person}{Jose~A. Onieva}, {and} \bibinfo{person}{Javier
  Lopez}.} \bibinfo{year}{In Press}\natexlab{}.
\newblock \showarticletitle{Immune System for the Internet of Things using Edge
  Technologies}.
\newblock \bibinfo{journal}{\emph{IEEE Internet of Things Journal}}
  (\bibinfo{year}{In Press}).
\newblock
\showISSN{2327-4662}
\urldef\tempurl%
\url{https://doi.org/10.1109/JIOT.2018.2867613}
\showDOI{\tempurl}


\bibitem[\protect\citeauthoryear{Ronen, Shamir, Weingarten, and Flynn}{Ronen
  et~al\mbox{.}}{2018}]%
        {8283484}
\bibfield{author}{\bibinfo{person}{E. Ronen}, \bibinfo{person}{A. Shamir},
  \bibinfo{person}{A. Weingarten}, {and} \bibinfo{person}{C.~O Flynn}.}
  \bibinfo{year}{2018}\natexlab{}.
\newblock \showarticletitle{IoT Goes Nuclear: Creating a Zigbee Chain
  Reaction}.
\newblock \bibinfo{journal}{\emph{IEEE Security Privacy}} \bibinfo{volume}{16},
  \bibinfo{number}{1} (\bibinfo{date}{January} \bibinfo{year}{2018}),
  \bibinfo{pages}{54 to 62}.
\newblock
\showISSN{1540 to 7993}
\urldef\tempurl%
\url{https://doi.org/10.1109/MSP.2018.1331033}
\showDOI{\tempurl}


\bibitem[\protect\citeauthoryear{Roy, Shen, Hassanieh, and Choudhury}{Roy
  et~al\mbox{.}}{2018}]%
        {211283}
\bibfield{author}{\bibinfo{person}{Nirupam Roy}, \bibinfo{person}{Sheng Shen},
  \bibinfo{person}{Haitham Hassanieh}, {and} \bibinfo{person}{Romit~Roy
  Choudhury}.} \bibinfo{year}{2018}\natexlab{}.
\newblock \showarticletitle{Inaudible Voice Commands: The Long-Range Attack and
  Defense}. In \bibinfo{booktitle}{\emph{15th {USENIX} Symposium on Networked
  Systems Design and Implementation ({NSDI} 18)}}. \bibinfo{publisher}{{USENIX}
  Association}, \bibinfo{address}{Renton, WA}, \bibinfo{pages}{547--560}.
\newblock
\showISBNx{978-1-931971-43-0}


\bibitem[\protect\citeauthoryear{Ruan, Wobbrock, Liou, Ng, and Landay}{Ruan
  et~al\mbox{.}}{2018}]%
        {ruan_wobbrock_liou_ng_landay_2018}
\bibfield{author}{\bibinfo{person}{Sherry Ruan}, \bibinfo{person}{Jacob~O.
  Wobbrock}, \bibinfo{person}{Kenny Liou}, \bibinfo{person}{Andrew Ng}, {and}
  \bibinfo{person}{James~A. Landay}.} \bibinfo{year}{2018}\natexlab{}.
\newblock \showarticletitle{Comparing Speech and Keyboard Text Entry for Short
  Messages in Two Languages on Touchscreen Phones}.
\newblock \bibinfo{journal}{\emph{Proceedings of the ACM on Interactive,
  Mobile, Wearable and Ubiquitous Technologies}} \bibinfo{volume}{1},
  \bibinfo{number}{4} (\bibinfo{year}{2018}), \bibinfo{pages}{1--23}.
\newblock
\urldef\tempurl%
\url{https://doi.org/10.1145/3161187}
\showDOI{\tempurl}


\bibitem[\protect\citeauthoryear{Schaub, Balebako, and Cranor}{Schaub
  et~al\mbox{.}}{2017}]%
        {schaub}
\bibfield{author}{\bibinfo{person}{F. Schaub}, \bibinfo{person}{R. Balebako},
  {and} \bibinfo{person}{L.~F. Cranor}.} \bibinfo{year}{2017}\natexlab{}.
\newblock \showarticletitle{Designing effective privacy notices and controls}.
\newblock \bibinfo{journal}{\emph{IEEE Internet Computing}}
  \bibinfo{volume}{21}, \bibinfo{number}{3} (\bibinfo{year}{2017}),
  \bibinfo{pages}{70--77}.
\newblock


\bibitem[\protect\citeauthoryear{Schonherr, Kohls, Zeiler, Holz, and
  Kolossa}{Schonherr et~al\mbox{.}}{2018}]%
        {DBLP:journals/corr/abs-1808-05665}
\bibfield{author}{\bibinfo{person}{Lea Schonherr}, \bibinfo{person}{Katharina
  Kohls}, \bibinfo{person}{Steffen Zeiler}, \bibinfo{person}{Thorsten Holz},
  {and} \bibinfo{person}{Dorothea Kolossa}.} \bibinfo{year}{2018}\natexlab{}.
\newblock \showarticletitle{Adversarial Attacks Against Automatic Speech
  Recognition Systems via Psychoacoustic Hiding}.
\newblock \bibinfo{journal}{\emph{CoRR}}  \bibinfo{volume}{abs/1808.05665}
  (\bibinfo{year}{2018}).
\newblock
\showeprint[arxiv]{1808.05665}
\urldef\tempurl%
\url{http://arxiv.org/abs/1808.05665}
\showURL{%
\tempurl}


\bibitem[\protect\citeauthoryear{{Shokri}, {Stronati}, {Song}, and
  {Shmatikov}}{{Shokri} et~al\mbox{.}}{2017}]%
        {7958568}
\bibfield{author}{\bibinfo{person}{R. {Shokri}}, \bibinfo{person}{M.
  {Stronati}}, \bibinfo{person}{C. {Song}}, {and} \bibinfo{person}{V.
  {Shmatikov}}.} \bibinfo{year}{2017}\natexlab{}.
\newblock \showarticletitle{Membership Inference Attacks Against Machine
  Learning Models}. In \bibinfo{booktitle}{\emph{2017 IEEE Symposium on
  Security and Privacy (SP)}}. \bibinfo{pages}{3--18}.
\newblock


\bibitem[\protect\citeauthoryear{Singleton}{Singleton}{2017}]%
        {alexa-theverge}
\bibfield{author}{\bibinfo{person}{Micah Singleton}.}
  \bibinfo{year}{2017}\natexlab{}.
\newblock \bibinfo{title}{{Alexa can now set reminders for you}}.
\newblock
  \bibinfo{howpublished}{\url{https://www.theverge.com/circuitbreaker/2017/6/1/15724474/alexa-echo-amazon-reminders-named-timers}}.
\newblock
\newblock
\shownote{[Online; last accessed 21-December-2018].}


\bibitem[\protect\citeauthoryear{Solove}{Solove}{2006}]%
        {solove06}
\bibfield{author}{\bibinfo{person}{D.J. Solove}.}
  \bibinfo{year}{2006}\natexlab{}.
\newblock \showarticletitle{{A taxonomy of privacy}}.
\newblock \bibinfo{journal}{\emph{University of Pennsylvania Law Review}}
  \bibinfo{volume}{154}, \bibinfo{number}{3} (\bibinfo{year}{2006}),
  \bibinfo{pages}{477--560}.
\newblock


\bibitem[\protect\citeauthoryear{SRLabs}{SRLabs}{2019}]%
        {SRLabs_Squatting}
\bibfield{author}{\bibinfo{person}{SRLabs}.} \bibinfo{year}{2019}\natexlab{}.
\newblock \bibinfo{title}{{Smart Spies: Alexa and Google Home expose users to
  vishing and eavesdropping}}.
\newblock \bibinfo{howpublished}{\url{https://srlabs.de/bites/smart-spies/}}.
\newblock
\newblock
\shownote{[Online; last accessed 21-February-2020].}


\bibitem[\protect\citeauthoryear{Statista}{Statista}{2018}]%
        {alexa-Statista}
\bibfield{author}{\bibinfo{person}{Statista}.} \bibinfo{year}{2018}\natexlab{}.
\newblock \bibinfo{title}{{Worldwide intelligent/digital assistant market share
  in 2017 and 2020, by product}}.
\newblock
  \bibinfo{howpublished}{\url{https://www.statista.com/statistics/789633/worldwide-digital-assistant-market-share/}}.
\newblock
\newblock
\shownote{[Online; last accessed 21-December-2018].}


\bibitem[\protect\citeauthoryear{Statt}{Statt}{2019}]%
        {Google-theverge-human}
\bibfield{author}{\bibinfo{person}{Nick Statt}.}
  \bibinfo{year}{2019}\natexlab{}.
\newblock \bibinfo{title}{{Google defends letting human workers listen to
  Assistant voice conversations}}.
\newblock
  \bibinfo{howpublished}{\url{https://www.theverge.com/2019/7/11/20691021/google-assistant-ai-training-controversy-human-workers-listening-privacy}}.
\newblock


\bibitem[\protect\citeauthoryear{Suarez-Tangil, Tapiador, Peris-Lopez, and
  Ribagorda}{Suarez-Tangil et~al\mbox{.}}{2014}]%
        {surveySmart}
\bibfield{author}{\bibinfo{person}{Guillermo Suarez-Tangil},
  \bibinfo{person}{Juan~E Tapiador}, \bibinfo{person}{Pedro Peris-Lopez}, {and}
  \bibinfo{person}{Arturo Ribagorda}.} \bibinfo{year}{2014}\natexlab{}.
\newblock \showarticletitle{Evolution, Detection and Analysis of Malware in
  Smart Devices}.
\newblock \bibinfo{journal}{\emph{IEEE Communications Surveys \& Tutorials}}
  \bibinfo{volume}{16}, \bibinfo{number}{2} (\bibinfo{year}{2014}),
  \bibinfo{pages}{961--987}.
\newblock


\bibitem[\protect\citeauthoryear{Such}{Such}{2017}]%
        {such2017privacy}
\bibfield{author}{\bibinfo{person}{Jose~M Such}.}
  \bibinfo{year}{2017}\natexlab{}.
\newblock \showarticletitle{Privacy and autonomous systems}. In
  \bibinfo{booktitle}{\emph{Proceedings of the 26th International Joint
  Conference on Artificial Intelligence (IJCAI)}}. \bibinfo{publisher}{AAAI
  Press}, \bibinfo{pages}{4761--4767}.
\newblock


\bibitem[\protect\citeauthoryear{Such, Ciholas, Rashid, Vidler, and
  Seabrook}{Such et~al\mbox{.}}{2019}]%
        {such2019basic}
\bibfield{author}{\bibinfo{person}{Jose~M Such}, \bibinfo{person}{Pierre
  Ciholas}, \bibinfo{person}{Awais Rashid}, \bibinfo{person}{John Vidler},
  {and} \bibinfo{person}{Timothy Seabrook}.} \bibinfo{year}{2019}\natexlab{}.
\newblock \showarticletitle{Basic Cyber Hygiene: Does It Work?}
\newblock \bibinfo{journal}{\emph{Computer}} \bibinfo{volume}{52},
  \bibinfo{number}{4} (\bibinfo{year}{2019}), \bibinfo{pages}{21--31}.
\newblock


\bibitem[\protect\citeauthoryear{Such and Criado}{Such and Criado}{2016}]%
        {such2016resolving}
\bibfield{author}{\bibinfo{person}{J.~M. Such} {and} \bibinfo{person}{N.
  Criado}.} \bibinfo{year}{2016}\natexlab{}.
\newblock \showarticletitle{Resolving multi-party privacy conflicts in social
  media}.
\newblock \bibinfo{journal}{\emph{IEEE TKDE}} \bibinfo{volume}{28},
  \bibinfo{number}{7} (\bibinfo{year}{2016}), \bibinfo{pages}{1851--1863}.
\newblock


\bibitem[\protect\citeauthoryear{Such and Criado}{Such and Criado}{2018}]%
        {such2018multiparty}
\bibfield{author}{\bibinfo{person}{J.~M. Such} {and} \bibinfo{person}{N.
  Criado}.} \bibinfo{year}{2018}\natexlab{}.
\newblock \showarticletitle{Multiparty Privacy in Social Media}.
\newblock \bibinfo{journal}{\emph{Commun. ACM}} \bibinfo{volume}{61},
  \bibinfo{number}{8} (\bibinfo{year}{2018}), \bibinfo{pages}{74--81}.
\newblock


\bibitem[\protect\citeauthoryear{Such, Criado, Vercouter, and Rehak}{Such
  et~al\mbox{.}}{2016a}]%
        {such2016intelligent}
\bibfield{author}{\bibinfo{person}{Jose~M Such}, \bibinfo{person}{Natalia
  Criado}, \bibinfo{person}{Laurent Vercouter}, {and} \bibinfo{person}{Martin
  Rehak}.} \bibinfo{year}{2016}\natexlab{a}.
\newblock \showarticletitle{Intelligent Cybersecurity Agents}.
\newblock \bibinfo{journal}{\emph{IEEE Intelligent Systems}}
  \bibinfo{volume}{31}, \bibinfo{number}{5} (\bibinfo{year}{2016}),
  \bibinfo{pages}{3--7}.
\newblock


\bibitem[\protect\citeauthoryear{Such, Gouglidis, Knowles, Gaurav, and
  Awais}{Such et~al\mbox{.}}{2016b}]%
        {such}
\bibfield{author}{\bibinfo{person}{Jose~M. Such}, \bibinfo{person}{Antonios
  Gouglidis}, \bibinfo{person}{William Knowles}, \bibinfo{person}{Misra
  Gaurav}, {and} \bibinfo{person}{Rashid Awais}.}
  \bibinfo{year}{2016}\natexlab{b}.
\newblock \showarticletitle{Information assurance techniques: Perceived cost
  effectiveness}.
\newblock \bibinfo{journal}{\emph{Computers and Security}}
  \bibinfo{volume}{60} (\bibinfo{year}{2016}), \bibinfo{pages}{117--133}.
\newblock
\urldef\tempurl%
\url{https://doi.org/10.1016/j.cose.2016.03.009}
\showDOI{\tempurl}


\bibitem[\protect\citeauthoryear{Such and Rovatsos}{Such and Rovatsos}{2016}]%
        {such2016privacy}
\bibfield{author}{\bibinfo{person}{J.~M. Such} {and} \bibinfo{person}{M.
  Rovatsos}.} \bibinfo{year}{2016}\natexlab{}.
\newblock \showarticletitle{Privacy policy negotiation in social media}.
\newblock \bibinfo{journal}{\emph{ACM Trans. on Autonomous and Adaptive
  Systems}} \bibinfo{volume}{11}, \bibinfo{number}{1} (\bibinfo{year}{2016}),
  \bibinfo{pages}{4}.
\newblock


\bibitem[\protect\citeauthoryear{Sugawara, Cyr, Rampazzi, and Genkin}{Sugawara
  et~al\mbox{.}}{2019}]%
        {Sugawara2019LightCL}
\bibfield{author}{\bibinfo{person}{Takeshi Sugawara}, \bibinfo{person}{Benjamin
  Cyr}, \bibinfo{person}{Sara Rampazzi}, {and} \bibinfo{person}{Daniel
  Genkin}.} \bibinfo{year}{2019}\natexlab{}.
\newblock \showarticletitle{Light Commands: Laser-Based Audio Injection Attacks
  on Voice-Controllable Systems*}.
\newblock


\bibitem[\protect\citeauthoryear{Syverson}{Syverson}{2009}]%
        {Paul_Syverson_Entropist}
\bibfield{author}{\bibinfo{person}{Paul Syverson}.}
  \bibinfo{year}{2009}\natexlab{}.
\newblock \showarticletitle{Why I'm not an Entropist}. In
  \bibinfo{booktitle}{\emph{In the Proceedings of Security Protocols XVII: 17th
  International Workshop}}.
\newblock
\urldef\tempurl%
\url{https://www.freehaven.net/anonbib/cache/entropist.pdf}
\showURL{%
\tempurl}


\bibitem[\protect\citeauthoryear{Syverson}{Syverson}{2011}]%
        {Paul_Syverson}
\bibfield{author}{\bibinfo{person}{Paul Syverson}.}
  \bibinfo{year}{2011}\natexlab{}.
\newblock \showarticletitle{Sleeping dogs lie on a bed of onions but wake when
  mixed}. In \bibinfo{booktitle}{\emph{Proceedings of HotPETS 2011}}.
\newblock
\urldef\tempurl%
\url{https://petsymposium.org/2011/papers/hotpets11-final10Syverson.pdf}
\showURL{%
\tempurl}


\bibitem[\protect\citeauthoryear{Szegedy, Zaremba, Sutskever, Bruna, Erhan,
  Goodfellow, and Fergus}{Szegedy et~al\mbox{.}}{2013}]%
        {DBLP:journals/corr/SzegedyZSBEGF13}
\bibfield{author}{\bibinfo{person}{Christian Szegedy},
  \bibinfo{person}{Wojciech Zaremba}, \bibinfo{person}{Ilya Sutskever},
  \bibinfo{person}{Joan Bruna}, \bibinfo{person}{Dumitru Erhan},
  \bibinfo{person}{Ian~J. Goodfellow}, {and} \bibinfo{person}{Rob Fergus}.}
  \bibinfo{year}{2013}\natexlab{}.
\newblock \showarticletitle{Intriguing properties of neural networks}.
\newblock \bibinfo{journal}{\emph{CoRR}}  \bibinfo{volume}{abs/1312.6199}
  (\bibinfo{year}{2013}).
\newblock


\bibitem[\protect\citeauthoryear{Tabassum, Kosi{\'n}ski, Frik, Malkin,
  Wijesekera, Egelman, and Lipford}{Tabassum et~al\mbox{.}}{2019b}]%
        {tabassum2019investigating}
\bibfield{author}{\bibinfo{person}{Madiha Tabassum}, \bibinfo{person}{Tomasz
  Kosi{\'n}ski}, \bibinfo{person}{Alisa Frik}, \bibinfo{person}{Nathan Malkin},
  \bibinfo{person}{Primal Wijesekera}, \bibinfo{person}{Serge Egelman}, {and}
  \bibinfo{person}{Heather~Richter Lipford}.} \bibinfo{year}{2019}\natexlab{b}.
\newblock \showarticletitle{Investigating Users' Preferences and Expectations
  for Always-Listening Voice Assistants}.
\newblock \bibinfo{journal}{\emph{Proceedings of the ACM on Interactive,
  Mobile, Wearable and Ubiquitous Technologies}} \bibinfo{volume}{3},
  \bibinfo{number}{4} (\bibinfo{year}{2019}), \bibinfo{pages}{1--23}.
\newblock


\bibitem[\protect\citeauthoryear{Tabassum, Kosinski, and Lipford}{Tabassum
  et~al\mbox{.}}{2019a}]%
        {238321}
\bibfield{author}{\bibinfo{person}{Madiha Tabassum}, \bibinfo{person}{Tomasz
  Kosinski}, {and} \bibinfo{person}{Heather~Richter Lipford}.}
  \bibinfo{year}{2019}\natexlab{a}.
\newblock \showarticletitle{"I don{\textquoteright}t own the data": End User
  Perceptions of Smart Home Device Data Practices and Risks}. In
  \bibinfo{booktitle}{\emph{Fifteenth Symposium on Usable Privacy and Security
  ({SOUPS} 2019)}}. \bibinfo{publisher}{{USENIX} Association},
  \bibinfo{address}{Santa Clara, CA}.
\newblock
\urldef\tempurl%
\url{https://www.usenix.org/conference/soups2019/presentation/tabassum}
\showURL{%
\tempurl}


\bibitem[\protect\citeauthoryear{Todisco, Delgado, and Evans}{Todisco
  et~al\mbox{.}}{2017}]%
        {Todisco:2017:CQC:3103639.3103730}
\bibfield{author}{\bibinfo{person}{Massimiliano Todisco},
  \bibinfo{person}{Hctor Delgado}, {and} \bibinfo{person}{Nicholas Evans}.}
  \bibinfo{year}{2017}\natexlab{}.
\newblock \showarticletitle{Constant Q Cepstral Coefficients}.
\newblock \bibinfo{journal}{\emph{Comput. Speech Lang.}} \bibinfo{volume}{45},
  \bibinfo{number}{C} (\bibinfo{date}{Sept.} \bibinfo{year}{2017}),
  \bibinfo{pages}{516--535}.
\newblock
\showISSN{0885-2308}
\urldef\tempurl%
\url{https://doi.org/10.1016/j.csl.2017.01.001}
\showDOI{\tempurl}


\bibitem[\protect\citeauthoryear{Vaidya, Zhang, Sherr, and Shields}{Vaidya
  et~al\mbox{.}}{2015}]%
        {191968}
\bibfield{author}{\bibinfo{person}{Tavish Vaidya}, \bibinfo{person}{Yuankai
  Zhang}, \bibinfo{person}{Micah Sherr}, {and} \bibinfo{person}{Clay Shields}.}
  \bibinfo{year}{2015}\natexlab{}.
\newblock \showarticletitle{Cocaine Noodles: Exploiting the Gap between Human
  and Machine Speech Recognition}. In \bibinfo{booktitle}{\emph{9th {USENIX}
  Workshop on Offensive Technologies ({WOOT} 15)}}.
  \bibinfo{publisher}{{USENIX} Association}, \bibinfo{address}{Washington,
  D.C.}
\newblock


\bibitem[\protect\citeauthoryear{Wang, Hou, Rios, Hallgren, Tippenhauer, and
  Ochoa}{Wang et~al\mbox{.}}{2018}]%
        {rios2018mob}
\bibfield{author}{\bibinfo{person}{Xueou Wang}, \bibinfo{person}{Xiaolu Hou},
  \bibinfo{person}{Ruben Rios}, \bibinfo{person}{Per Hallgren},
  \bibinfo{person}{Nils~Ole Tippenhauer}, {and} \bibinfo{person}{Martin
  Ochoa}.} \bibinfo{year}{2018}\natexlab{}.
\newblock \showarticletitle{Location Proximity Attacks against Mobile Targets}.
  In \bibinfo{booktitle}{\emph{23rd European Symposium on Research in Computer
  Security (ESORICS 2018)}} \emph{(\bibinfo{series}{LNCS},
  Vol.~\bibinfo{volume}{11099})}. Springer, \bibinfo{publisher}{Springer},
  \bibinfo{address}{Barcelona}, \bibinfo{pages}{373--392}.
\newblock
\showISBNx{978-3-319-98988-4}
\urldef\tempurl%
\url{https://doi.org/10.1007/978-3-319-98989-1}
\showDOI{\tempurl}


\bibitem[\protect\citeauthoryear{Watch}{Watch}{2017}]%
        {hrw}
\bibfield{author}{\bibinfo{person}{Human~Right Watch}.}
  \bibinfo{year}{2017}\natexlab{}.
\newblock \showarticletitle{China: Voice Biometric Collection Threatens
  Privacy}.
\newblock  (\bibinfo{year}{2017}).
\newblock
\urldef\tempurl%
\url{https://www.hrw.org/news/2017/10/22/china-voice-biometric-collection-threatens-privacy}
\showURL{%
\tempurl}


\bibitem[\protect\citeauthoryear{White}{White}{2018}]%
        {White:2018:SDV:3289258.3185336}
\bibfield{author}{\bibinfo{person}{Ryen~W. White}.}
  \bibinfo{year}{2018}\natexlab{}.
\newblock \showarticletitle{Skill Discovery in Virtual Assistants}.
\newblock \bibinfo{journal}{\emph{Commun. ACM}} \bibinfo{volume}{61},
  \bibinfo{number}{11} (\bibinfo{date}{Oct.} \bibinfo{year}{2018}),
  \bibinfo{pages}{106--113}.
\newblock
\showISSN{0001-0782}
\urldef\tempurl%
\url{https://doi.org/10.1145/3185336}
\showDOI{\tempurl}


\bibitem[\protect\citeauthoryear{Wolfson}{Wolfson}{2018}]%
        {alexa-guardian}
\bibfield{author}{\bibinfo{person}{Sam Wolfson}.}
  \bibinfo{year}{2018}\natexlab{}.
\newblock \bibinfo{title}{{Amazon's Alexa recorded private conversation and
  sent it to random contact}}.
\newblock
  \bibinfo{howpublished}{\url{www.theguardian.com/technology/2018/may/24/amazon-alexa-recorded-conversation}}.
\newblock
\newblock
\shownote{[Online; last accessed 17-December-2018].}


\bibitem[\protect\citeauthoryear{Wong}{Wong}{2017}]%
        {alexa-CNBC}
\bibfield{author}{\bibinfo{person}{Venessa Wong}.}
  \bibinfo{year}{2017}\natexlab{}.
\newblock \bibinfo{title}{{Burger King's New Ad Will Hijack Your Google Home}}.
\newblock
  \bibinfo{howpublished}{\url{https://www.cnbc.com/2017/04/12/burger-kings-new-ad-will-hijack-your-google-home.html}}.
\newblock
\newblock
\shownote{[Online; last accessed 25-December-2018].}


\bibitem[\protect\citeauthoryear{Wright, Coull, and Monrose}{Wright
  et~al\mbox{.}}{2009}]%
        {morphing09}
\bibfield{author}{\bibinfo{person}{Charles Wright}, \bibinfo{person}{Scott
  Coull}, {and} \bibinfo{person}{Fabian Monrose}.}
  \bibinfo{year}{2009}\natexlab{}.
\newblock \showarticletitle{Traffic Morphing: An efficient defense against
  statistical traffic analysis}. In \bibinfo{booktitle}{\emph{Proceedings of
  the Network and Distributed Security Symposium}}. \bibinfo{publisher}{IEEE}.
\newblock


\bibitem[\protect\citeauthoryear{Wright and De~Hert}{Wright and
  De~Hert}{2012}]%
        {wright2012introduction}
\bibfield{author}{\bibinfo{person}{David Wright} {and} \bibinfo{person}{Paul
  De~Hert}.} \bibinfo{year}{2012}\natexlab{}.
\newblock \showarticletitle{Introduction to privacy impact assessment}.
\newblock In \bibinfo{booktitle}{\emph{Privacy Impact Assessment}}.
  \bibinfo{publisher}{Springer}, \bibinfo{pages}{3--32}.
\newblock


\bibitem[\protect\citeauthoryear{Xinyu, Guan~Hua, Liu, Chi~Yu, and Xie}{Xinyu
  et~al\mbox{.}}{2017}]%
        {Lei_Xinyu}
\bibfield{author}{\bibinfo{person}{Lei Xinyu}, \bibinfo{person}{Tu Guan~Hua},
  \bibinfo{person}{Alex~X.and Liu}, \bibinfo{person}{Li Chi~Yu}, {and}
  \bibinfo{person}{Tian Xie}.} \bibinfo{year}{2017}\natexlab{}.
\newblock \showarticletitle{The Insecurity of Home Digital Voice Assistants:
  Amazon Alexa as a Case Study}.
\newblock  (\bibinfo{year}{2017}).
\newblock
\urldef\tempurl%
\url{https://arxiv.org/pdf/1712.03327.pdf}
\showURL{%
\tempurl}


\bibitem[\protect\citeauthoryear{Yang}{Yang}{2018}]%
        {yang_2018}
\bibfield{author}{\bibinfo{person}{Jun Yang}.} \bibinfo{year}{2018}\natexlab{}.
\newblock \showarticletitle{Multilayer Adaptation Based Complex Echo
  Cancellation and Voice Enhancement}.
\newblock \bibinfo{journal}{\emph{2018 IEEE Int. conference on Acoustics,
  Speech and Signal Processing (ICASSP)}} (\bibinfo{year}{2018}).
\newblock
\urldef\tempurl%
\url{https://doi.org/10.1109/icassp.2018.8461354}
\showDOI{\tempurl}


\bibitem[\protect\citeauthoryear{Zeng, Mare, and Roesner}{Zeng
  et~al\mbox{.}}{2017}]%
        {Zeng:2017:EUS:3235924.3235931}
\bibfield{author}{\bibinfo{person}{Eric Zeng}, \bibinfo{person}{Shrirang Mare},
  {and} \bibinfo{person}{Franziska Roesner}.} \bibinfo{year}{2017}\natexlab{}.
\newblock \showarticletitle{End User Security and Privacy Concerns with Smart
  Homes}.
\newblock  (\bibinfo{year}{2017}), \bibinfo{pages}{65--80}.
\newblock
\showISBNx{978-1-931971-39-3}
\urldef\tempurl%
\url{http://dl.acm.org/citation.cfm?id=3235924.3235931}
\showURL{%
\tempurl}


\bibitem[\protect\citeauthoryear{Zeng and Roesner}{Zeng and Roesner}{2019}]%
        {zeng2019understanding}
\bibfield{author}{\bibinfo{person}{Eric Zeng} {and} \bibinfo{person}{Franziska
  Roesner}.} \bibinfo{year}{2019}\natexlab{}.
\newblock \showarticletitle{Understanding and improving security and privacy in
  multi-user smart homes: a design exploration and in-home user study}. In
  \bibinfo{booktitle}{\emph{28th $\{$USENIX$\}$ Security Symposium
  ($\{$USENIX$\}$ Security 19)}}.
\newblock


\bibitem[\protect\citeauthoryear{Zhang, Yan, Ji, Zhang, Zhang, and Xu}{Zhang
  et~al\mbox{.}}{2017}]%
        {zhang_yan_ji_zhang_zhang_xu_2017}
\bibfield{author}{\bibinfo{person}{Guoming Zhang}, \bibinfo{person}{Chen Yan},
  \bibinfo{person}{Xiaoyu Ji}, \bibinfo{person}{Tianchen Zhang},
  \bibinfo{person}{Taimin Zhang}, {and} \bibinfo{person}{Wenyuan Xu}.}
  \bibinfo{year}{2017}\natexlab{}.
\newblock \showarticletitle{DolphinAttack}.
\newblock \bibinfo{journal}{\emph{Proceedings of the 2017 ACM SIGSAC Conference
  on Computer and Communications Security - CCS '17}} (\bibinfo{year}{2017}).
\newblock
\urldef\tempurl%
\url{https://doi.org/10.1145/3133956.3134052}
\showDOI{\tempurl}


\bibitem[\protect\citeauthoryear{{Zhang}, {Mi}, {Feng}, {Wang}, {Tian}, and
  {Qian}}{{Zhang} et~al\mbox{.}}{2019}]%
        {2018arXiv180501525Z}
\bibfield{author}{\bibinfo{person}{N. {Zhang}}, \bibinfo{person}{X. {Mi}},
  \bibinfo{person}{X. {Feng}}, \bibinfo{person}{X. {Wang}}, \bibinfo{person}{Y.
  {Tian}}, {and} \bibinfo{person}{F. {Qian}}.} \bibinfo{year}{2019}\natexlab{}.
\newblock \showarticletitle{Dangerous Skills: Understanding and Mitigating
  Security Risks of Voice-Controlled Third-Party Functions on Virtual Personal
  Assistant Systems}. In \bibinfo{booktitle}{\emph{2019 IEEE Symposium on
  Security and Privacy (SP)}}. \bibinfo{pages}{1381--1396}.
\newblock


\bibitem[\protect\citeauthoryear{Zhao, Yang, Wang, and Qiu}{Zhao
  et~al\mbox{.}}{2012}]%
        {10.1007/978-3-642-33469-6_30}
\bibfield{author}{\bibinfo{person}{Sendong Zhao}, \bibinfo{person}{Wu Yang},
  \bibinfo{person}{Ding Wang}, {and} \bibinfo{person}{Wenzhen Qiu}.}
  \bibinfo{year}{2012}\natexlab{}.
\newblock \showarticletitle{A New Scheme with Secure Cookie against SSLStrip
  Attack}. In \bibinfo{booktitle}{\emph{Web Information Systems and Mining}},
  \bibfield{editor}{\bibinfo{person}{Fu~Lee Wang}, \bibinfo{person}{Jingsheng
  Lei}, \bibinfo{person}{Zhiguo Gong}, {and} \bibinfo{person}{Xiangfeng Luo}}
  (Eds.). \bibinfo{publisher}{Springer Berlin Heidelberg},
  \bibinfo{address}{Berlin, Heidelberg}, \bibinfo{pages}{214--221}.
\newblock
\showISBNx{978-3-642-33469-6}


\bibitem[\protect\citeauthoryear{Zheng, Apthorpe, Chetty, and Feamster}{Zheng
  et~al\mbox{.}}{2018}]%
        {Zheng:2018:UPS:3290265.3274469}
\bibfield{author}{\bibinfo{person}{Serena Zheng}, \bibinfo{person}{Noah
  Apthorpe}, \bibinfo{person}{Marshini Chetty}, {and} \bibinfo{person}{Nick
  Feamster}.} \bibinfo{year}{2018}\natexlab{}.
\newblock \showarticletitle{User Perceptions of Smart Home IoT Privacy}.
\newblock \bibinfo{journal}{\emph{Proc. ACM Hum.-Comput. Interact.}}
  \bibinfo{volume}{2}, \bibinfo{number}{CSCW}, Article \bibinfo{articleno}{200}
  (\bibinfo{date}{Nov.} \bibinfo{year}{2018}), \bibinfo{numpages}{20}~pages.
\newblock
\showISSN{2573-0142}
\urldef\tempurl%
\url{https://doi.org/10.1145/3274469}
\showDOI{\tempurl}


\end{thebibliography}

\end{document}